\documentclass[%
 aip,
 amsmath,amssymb,
preprint,%
author-numerical,%
floatfix]{revtex4-1}

\usepackage{amsmath,amsfonts,amssymb,bm}
\usepackage{mathptmx}
\usepackage{newtxtext}
\usepackage{newtxmath}
\usepackage{appendix}
\usepackage{enumitem}
\usepackage{subcaption}
\usepackage{lineno}
\usepackage{array}
\usepackage{natbib}
\usepackage{bm}
\usepackage{tikz}
\usetikzlibrary{shapes.geometric}
\usepackage{multirow}

 \pdfoutput=1
\usepackage{graphicx}
\usepackage{dcolumn}
\usepackage[utf8]{inputenc}
\usepackage[T1]{fontenc}
\usepackage{mathptmx}
\usepackage{etoolbox}
\makeatletter
\def\@email#1#2{%
 \endgroup
 \patchcmd{\titleblock@produce}
  {\frontmatter@RRAPformat}
  {\frontmatter@RRAPformat{\produce@RRAP{*#1\href{mailto:#2}{#2}}}\frontmatter@RRAPformat}
  {}{}
}%
\makeatother

\newcommand{\ttau}{\mbox{ $\widetilde{\tau}$} {}}

\newcommand{\tuu}{\mbox{\boldmath $\tilde{u}$} {}}

\newcommand{\tu}{\mbox{$\widetilde{u}$} {}}
\newcommand{\tv}{\mbox{$\widetilde{v}$} {}}
\newcommand{\tw}{\mbox{$\widetilde{w}$} {}}
\newcommand{\thU}{\mbox{${\widetilde{U}}$} {}}
\newcommand{\thu}{\mbox{${\widetilde{u}}$} {}}
\newcommand{\thv}{\mbox{${\widetilde{v}}$} {}}

\begin{document}

\preprint{AIP/123-QED}

\title[]{A dynamic wall modeling approach for Large Eddy Simulation  of offshore wind farms in realistic oceanic conditions}
\author{A. K. Aiyer}
 \email{aaiyer@princeton.edu.}
\affiliation{ 
Department of Mechanical and Aerospace Engineering, Princeton University
}%
\author{L. Deike}%
\affiliation{ 
Department of Mechanical and Aerospace Engineering, Princeton University
}%
\affiliation{%
High Meadows Environmental Institute, Princeton University
}%

\author{M. E. Mueller}
\affiliation{ 
Department of Mechanical and Aerospace Engineering, Princeton University
}%

\date{\today}

\begin{abstract}

Due to the multitude of scales present in realistic oceanic conditions, resolving the surface stress is computationally intensive, motivating modeling approaches. In this work, a dynamic wave drag model is developed for Large Eddy Simulation to quantify the effects of multiscale dynamically rough surfaces on the atmospheric boundary layer. The waves are vertically unresolved, and the total drag due to the horizontally resolved portion of the wave spectrum is computed through a superposition of the force from each mode. As LES can only resolve the horizontal wind-wave interactions to the filter scale $\Delta$, the effects of the horizontally unresolved, subfilter waves are modeled by specifying a roughness length scale characterizing the unresolved wave energy spectrum. This subfilter roughness is set proportional to the subfilter root-mean-square of the height distribution, and the constant of proportionality is evaluated dynamically during the simulation based on the assumption that the total drag force at the wave surface is independent of the filter scale.
The dynamic approach is used to simulate the airflow over a spectrum of moving waves and the results are validated against high-fidelity phase-resolved simulations.  The dynamic approach combined with the wave spectrum drag model is then used to study flow through a fixed-bottom offshore wind farm. The dynamic model accurately adapts to the changing velocity field and accurately predicts the mean velocity profiles and power produced from the offshore wind farm. Further, the effect of the wind-wave interactions on the mean velocity profiles, power production, and kinetic energy budget are quantified.

\end{abstract}

\maketitle


\section{Introduction}
The paradigm shift towards renewable energy coupled with increasing global demand for electricity has made offshore wind energy an important topic of research. The total installed offshore wind energy capacity of the United States is currently $30$ MW and is expected to increase significantly over the next decade, with a planned capacity of $14$ GW by 2030 \cite{osti_1572771}. The offshore environment is generally characterized by stronger wind speeds, weaker turbulence intensities, and larger availability of space, fueling interest in developing offshore wind energy technologies \cite{Esteban2011,Castro-Santos2017}.
The operation of large-scale wind farms in the offshore environment requires accurate characterization of wind speeds and turbulence levels and their modulation by ocean waves. Additionally, as floating structures are gaining popularity and have started being constructed, knowledge of the wave-field dynamics and its impact on the turbine wake and loads are necessary to develop floating offshore wind turbines \cite{Lin2022}. To this end, characterizing the momentum transfer between wind and waves is critical for the future of offshore wind energy.

Limited field measurements for the evolution of turbine wakes in offshore wind farms have focused on wake and power losses through the farm \cite{Christiansen2005,Barthelmie2009}.  However, these large-scale studies are unable to quantify how ocean waves affect a turbine's wake profile and broader wind-wave-wake interactions. Due to these shortcomings, laboratory-scale studies have been conducted with model wind turbines in a wave tank. \citet{Fercak2022} developed an experimental setup combining a wave tank, wind tunnel, and scaled fixed-bottom wind turbine to study wind-wave-wake interactions. The study quantified the effects of the waves on wake meandering and wave-induced stresses. However, experimental studies cannot sample the wide parameter space associated with the offshore environment due to limitations in tank size and the ability to generate waves with varying fetch in addition to the challenges related to aerodynamic scaling.

The ocean surface is composed of a superposition of numerous waves of various lengths and periods. This complex multiscale surface is generally represented using a wave spectrum. The spectrum provides the distribution of wave energy among different wavenumbers on the sea surface. Researchers have developed numerous wave-spectrum models to combine available in-situ and laboratory measurements \cite{hasselmann1973measurements,Toba1973,Donelan1987,fois2014investigation,yurovskaya2013directional,Ryabkova2019}. Computationally representing such an evolving multiscale surface is challenging and requires the development of specialized algorithms. Large Eddy Simulation (LES) has been shown to yield high-fidelity results for flow over complex terrain resolving the large- and intermediate-scale turbulent motions and only requiring modeling of the unresolved subfilter-scale turbulence effects. However, resolving the waves numerically in LES requires a fine grid near the surface to capture the multiscale wave motions. Such phase-resolved simulations accurately describe the near-surface wind-wave interactions but come with a high computational cost \cite{Sullivan2008,Sullivan2014,Hao2019,Hao2020,Wu2022,DESKOS2022109029}. 

Despite their high computational cost, these high-fidelity, phase-resolved LES tools have provided key insights into wind-wave-wake interactions in offshore wind farms.  Computational simulation tools have been developed to resolve the air-sea interface and study the interaction of wind turbines and waves. \citet{Alsam2015} studied the effects of monochromatic fast moving waves on a single wind turbine and found that the higher wave speeds increased the wind velocity, lowered wind shear, and decreased turbulence intensity level. \citet{Yang2014OffshoreW} developed a framework to characterize more realistic wind-wave-wake interactions of an offshore wind farm. The wave surface was modeled using a separate Higher Order Spectral Method (HOSM) coupled to LES of the airflow \cite{Yang2013D}. They found that waves have a significant effect in determining wind farm performance in both the wind-driven regime and in the case of swell. The HOSM relies on potential flow assumptions of the flow being irrotational and inviscid and is limited to non-breaking waves \cite{DUCROZET2017233}. Additionally, the computational cost of the simulation is increased due to the separate wave solver.
 \citet{Xiao2019} and \citet{Yang2022x} studied the effects of swell (waves that transfer momentum to the wind field) on a wind turbine array and found that the swells also affect wake recovery and lead to higher power output.

While providing key insights, these high-fidelity, phase-resolved LES approaches have key shortcomings. To simulate the near-surface wind-wave interactions, a high resolution is required to resolve the multiscale surface in addition to a body-fitted grid. This results in a high computational cost when considering a range of sea states and turbulence parameters associated with the offshore environment. 
Recently, \citet{Aiyer2022} developed a phase-aware wave drag model (WDM) that accurately calculates the wind-wave momentum transfer for LES without the need for a separate solver for the wave field nor a high near-wall resolution. In their model, a hydrodynamic drag that depended on the wave surface gradient and relative wind-wave velocity was imposed to model the effects of the waves. The model relied on horizontally resolving the waves on the computational grid, and the subfilter roughness was set equal to the smooth surface aerodynamic roughness. \citet{Aiyer2022T} applied this wall-modeled approach to offshore wind farms, but only for horizontally resolved monochromatic waves. In this work, a model that calculates the drag for a multiscale moving surface is developed to simulate realistic ocean conditions by modeling the effect of multiple horizontally resolved wave modes.

Furthermore, for a multiscale ocean surface, the effect of waves unresolved by the LES grid must be aggregated into a constant subfilter roughness length $z_0$ or parameterizations that often contain empirical constants that vary under different wind and wave conditions \cite{Charnock1955,Taylor2001,Drennan2003,Deskos2021ReviewLayer}. There is a need for a modeling framework that reduces the computational cost associated with resolving the air-sea interface and concomitantly accounts for subfilter effects without adhoc prescriptions of a constant roughness length. \citet{Yang2013D} employed a dynamic procedure originally developed by \citet{Anderson2011} for stationary surfaces to the wind-wave context. The effect of  subfilter motions  was calculated dynamically during the simulation and the waves were modeled using a separate HOSM solver.
Here, the dynamic approach is extended to the cases where the wave effects are also  modeled rather than computed using the HOSM solver. 
 The dynamic wave spectrum model (Dyn-WaSp) is then applied to study flow through offshore wind farms, where the presence of the wind turbines actively modulates the flow near the wave surface affecting wind-wave momentum transfer and the drag from subfilter waves.

The rest of the paper is organized as follows. A  dynamic wave spectrum (Dyn-WaSp) drag model applicable to a multiscale moving surface is presented in \S \ref{sec:dynamic_model}, and governing equations used for the LES and the wind turbine models are described in  \S \ref{sec:methods}.  The Dyn-WaSp model is then applied to study airflow over a prescribed spectrum of moving waves in  \S \ref{sec:Les_airflow}  and to flow through an offshore wind farm in \S \ref{sec:farm}. Conclusions are drawn in \S \ref{sec:concl}. 

\section{Dynamic wave spectrum drag model}
\label{sec:dynamic_model}


The sea surface is composed of random waves with a wide range of wavelengths and periods.
The height distribution of realistic wave fields generally contains fluctuations over a large range of scales ranging from a few centimeters (capillary waves) to many meters (gravity waves). In general, the total force at the wave surface can be decomposed into a pressure-based form drag and a viscous drag.  
Considering a sloped surface ${\eta}(x,t)$ varying as a function of horizontal coordinate $x$ and time $t$  representing the wave, the average horizontal component of the stress at the surface can be written as \cite{Grare2013}
  \begin{equation}\label{eqn:stress_part}
        \tau_{xz}   = \frac{1}{\rho}\left\langle p_s\eta_x\right\rangle + \left\langle \frac{\tau_{visc,xz}}{\sqrt{1 + \eta_x^2}} \right\rangle,
    \end{equation}
    where $z$ is the vertical direction, $p_s$ is the surface pressure, {$\tau_{visc,xz} = \nu\partial u/\partial z$} is the local viscous surface tangential shear stress, $\eta_x = \partial {\eta}/\partial x$ is the  local interface slope, and the angular brackets
denote averaging over one wavelength. In computational simulations, the accuracy of the pressure and viscous stress relies on the grid resolution discretizing both the surface and the near-wall flow. Direct Numerical Simulations and wall-resolved LES use a high resolution to accurately calculate these stresses. This approach is unfeasible for the atmospheric boundary layer due to the high computational cost.

In LES, the pressure drag due to the wave field is resolved only to the filter scale $\Delta$. The challenge lies in accurately modeling the effect of subfilter motions on the flow. 
The total force at resolution $\Delta$ due to the pressure stress and the subfilter contributions can be written as
\begin{equation}\label{eqn:tot_drag}
    F_{tot,i}^{\Delta} = \iint_A \widetilde{p}^s\widetilde{n}_i\ \mathrm{d}x\mathrm{d}y + \rho_a \iint_A \tau_{ij,\Delta}^{wall} \widetilde{n}_j\ \mathrm{d}x\mathrm{d}y ,
\end{equation}
where $i=x,y$ denotes the horizontal directions, the $(\tilde{..})$ denotes a quantity filtered at scale $\Delta$, $\widetilde{p}^s$ is the filtered surface pressure, $\widetilde{n}$ is the unit normal vector to the filtered surface $\widetilde{\eta}(x,y)$, and $\widetilde{\tau}_{ij,\Delta}^{wall}$ is the subfilter stress due to unresolved wave modes at the wave surface. If the resolved filtered pressure is known, Equation (\ref{eqn:tot_drag}) can be used directly given a model for the subfilter stress. However, for wall-modeled LES, the key is low computational cost, and  models for both terms on the r.h.s of Equation (\ref{eqn:tot_drag}) are required. 

\subsection{Wave Spectrum Model for the Form Drag}
The pressure drag is modeled by extending the approach presented in \citet{Aiyer2022} to a multiscale surface.
The original model was developed for a single wave mode that was horizontally resolved by the LES grid. 
The model is based on applying a hydrodynamic drag force  proportional to the incoming momentum flux onto a frontal surface area. This offers the advantage of incorporating the wave characteristics (surface topology and wave speed) into the formulation without relying on empirical parameterizations.
The model is then expressed as a force per unit volume
\begin{equation}\label{eqn:drag_eqn}
    {F}_{d,i}  = -C_D\frac{\rho_a}{\Delta_z}\widetilde{u}_i U^{\Delta} \left(\widehat{n}_{u,k}\cdot \frac{\partial \widetilde{\eta}}{\partial x_k}\right) {\mathcal{H}\left\{\widehat{n}_{u,k}\cdot \frac{\partial \widetilde{\eta}}{\partial x_k}\right\}}\qquad i =x,y,
\end{equation}
 where  $ak$ is the wave steepness defined as a product of the wave amplitude $a$ and the wavenumber $k$, $C_D =ak/(1+6(ak)^2)$ is the steepness-dependent drag coefficient,  $\rho_a$ is the density of the air,  $\widetilde{u_i}$ is the filtered velocity at the first grid point in the simulation, $\widetilde{\eta}$ is the filtered wave height, $U^{\Delta} = \sqrt{(\widetilde{u} - c_x)^2 + (\widetilde{v}-c_y)^2}$ is the tangential velocity relative to the wave propagation speed, $\bm{\hat{n}_u} = \tuu_{r,c}/U^{\Delta}$ is the velocity unit normal vector, $\tuu_{r,c}$ is the relative velocity between the incoming flow and the wave phase velocity $c = \sqrt{(g/k)}$, and $\mathcal{H}[x]$ is the Heaviside function that ensures that the force is only applied when the flow is incident on the wave frontal area. The drag force formulation incorporates dependence on the wave characteristics based on the relative velocity and the surface gradient. 
 
 In this work, the wave propagation direction is aligned with the mean flow, and further directional effects are not considered. 
The sea surface elevation $\widetilde{\eta}(x,y,t)$ filtered at the LES grid resolution $\Delta$ is constructed as a sum of linear plane waves, each with a random phase $\phi$\cite{Sullivan2014}:
\begin{equation}
    \widetilde{\eta}(x,y,t) = \sum_k a(k) \exp(i[kx - \omega(k)t + \phi],
\end{equation}
where $a^2(k) = 2S(k)dk$ is the amplitude of mode $k$, $S(k)$ is the wave energy spectrum, and the summation is over all the resolved wave modes with $k_{max} = \pi/\Delta$. The random phase model only requires the computation of an initial wave state $\widetilde{\eta}(x,y,t_0)$, and future values can be computed using fast Fourier transforms (FFTs) \cite{Sullivan2014}. The wave surface elevation and total resolvable wave form drag computed is filtered using a Gaussian filter to eliminate contributions from wave modes with wavenumbers larger than the maximum resolvable wavenumber $k_{max}$.
For a multiscale surface with a given wavenumber spectrum, the total wave-induced force is computed as a linear superposition of the effects of individual horizontally resolved wave modes:
\begin{equation}\label{eqn:total_stress}
    F_{d,i}  = -\sum_k C_D(k)\frac{\rho_a}{\Delta_z}\widetilde{u}_i \left((\widetilde{u}_l - c(k))\cdot \frac{\partial \widetilde{\eta}(k)}{\partial x_l}\right){\mathcal{H}\left\{(\widetilde{u}_l - c(k))\cdot \frac{\partial \widetilde{\eta}(k)}{\partial x_l}\right\}}\qquad i =x,y,
\end{equation}
where the summation is carried out over the range of resolvable wavenumbers. Note that, for the case of misaligned wind and waves, the relative velocity in Equation (\ref{eqn:total_stress}) will depend on the local angle of propagation $\theta = \arctan(k_y/k_x)$, but misaligned waves are not considered in this work.
In summary, for a wave propagating in the streamwise direction, each wave mode imposes a hydrodynamic stress proportional to its relative velocity  $\widetilde{{u}} - {c(k)}$ and slope $a(k)k$.

A realistic wave spectrum is a mixture of both wind-waves (slow waves) and swells (fast waves). The wave drag model developed by \citet{Aiyer2022} overestimated the wind-wave momentum flux for cases where locally $\widetilde{u} -c <0$, applicable to very fast waves or swell-like conditions where the momentum transfer is from the waves to the wind. For fast waves,  the drag force decreases linearly as a function of the wave speed as seen from Equation (\ref{eqn:drag_eqn}) and resulted in a larger negative form drag than observed in DNS\cite{Sullivan2000,Yang2010}(see Figure 10b from \citet{Aiyer2022}).  To include these effects in the current framework, a simple parameterization based on high-fidelity phase-resolved numerical simulations of \citet{Cao2021} is proposed. \citet{Cao2021} found that the physical process causing wind-wave momentum flux at high wave age is governed by linear dynamics and can be described by their curvilinear model of the wave boundary layer. The wave growth rate $\beta$ from their model showed good agreement with the phase-resolved LES results (\citet[see Figure 15]{Cao2020}). The averaged (over a wavelength) normalized wave growth rate due to the wind input is related to the surface pressure drag by \citep{Donelan2006,Buckley2020} 
\begin{equation}\label{eqn:norm_growth_rate}
    \beta(k) = \frac{2}{(a(k)k)^2}\frac{1}{\lambda}\int \frac{p(k)}{\rho_au_*^2}\frac{d\eta(k)}{dx}\mathrm{d}x,
\end{equation}
where $\lambda$ is the wavelength, $p(k)$ is the air pressure at the wave surface for mode $k$, and $u_{*}$ is the surface friction velocity. In the current modeling framework, the waves are prescribed and not growing, but the normalized growth rate is merely used to derive a model for the form drag of fast waves. 
The normalized growth rate defined in Equation (\ref{eqn:norm_growth_rate}) can be rewritten in terms of the planar-averaged drag force as 
\begin{equation}
    \beta(k) = \frac{2}{(a(k)k)^2}\frac{<f_d(k)>\Delta_z}{ u_*^2},
\end{equation}
where $f_d = F_d/\rho_a$.
As the wave growth parameterization is calculated from the averaged drag, the swell correction is not phase-aware. 
Note that the normalized growth rate defined in Equation (\ref{eqn:norm_growth_rate}) is implicitly normalized by the square of the wave steepness consistent with wind-wave theory for the form stress \cite{miles_1957,Donelan1990,Wu2022}. Exponents $< 2$ have been proposed by some experimental studies \cite{Grare2013}. In this work, a consistent normalization based on the square of the wave steepness is used while making comparisons to experimental and numerical studies.
For wave age $c/u_* > 25$ a linear fit with respect to the wave age is derived from the data of  \citet{Cao2021}:

\begin{equation}\label{eqn:fit}
    \beta(k) = 25 - \frac{c(k)}{u_*},
\end{equation}
 Using Equation (\ref{eqn:fit}), the modified form drag  can be written as
\begin{equation}\label{eqn:fit_swell}
     F_{d,i}  = -\frac{\rho_a}{\Delta_z}\sum_k \left[\Gamma C_D(k)\widetilde{u}_i \left((\widetilde{u}_l - c(k))\cdot \frac{\partial \widetilde{\eta}(k)}{\partial x_l}\right){\mathcal{H}\left\{(\widetilde{u}_l - c(k))\cdot \frac{\partial \widetilde{\eta}(k)}{\partial x_l}\right\}} + (1-\Gamma) \beta(k)\frac{(aku_*)^2}{2}\right]
\end{equation}
where 

\begin{equation}
\Gamma = \begin{cases}
  1, & u_i-c(k) \geq 0 \\
  0, & u_i-c(k) < 0.
\end{cases}
\end{equation}

In the LES, the wave form drag is calculated using Equation (\ref{eqn:fit_swell}), and the swell parameterization is applied when locally $\widetilde{u} -c < 0$, that is, for momentum transfer from the waves to the wind.

\subsection{ Model for Subfilter Wave Effects}

For multiscale surfaces, the effects of subfilter scales can be recast as quantifying the effective amplitude $\sigma_{\eta}$ of the subfilter waves for wave modes with $k > k_{max}$. 
The model proposed by \citet{Anderson2011} for the subfilter roughness based on the r.m.s of the subfilter height fluctuations is used in the current study. Other candidate models have been proposed based on different wave characteristics such as the steepness and the wave age, for instance; see the review by Yang and co-workers \cite{Yang2013D,Yang2013DA}. However, the results from different models were found to be similar, and the r.m.s height fluctuation model is used for its simplicity. 

The r.m.s of the subfilter height fluctuation for a wave surface $\widetilde{\eta}$  filtered at a scale $\Delta$ is

\begin{equation}
    \sigma_{\eta}^{\Delta} = \left(\widetilde{\eta^2} - \widetilde{\eta}^2\right)^{1/2}.
\end{equation}
For a given wave spectrum $S(k)$, the r.m.s of the subfilter height fluctuation can be written as
\begin{equation}\label{eqn:rms_model}
    \sigma_{\eta}^{\Delta} = \left(2\int_{\pi/\Delta}^{2\pi/\lambda_c} S(k)dk\right)^{1/2},
\end{equation}
where $\lambda_c = 0.05$ m is the critical wavelength that separates gravity and capillary waves. Studies by \citet{Makin1995} have shown that waves smaller than $\lambda_c$ have a negligible contribution to the total sea surface stress and are neglected in the integration. 

The momentum fluxes associated with the subfilter height fluctuations are modeled using an equilibrium log-law model for neutrally stratified flow that expresses the kinematic wall stress nearest to the surface according to

\begin{equation}\label{eqn:MOST}
    \left.\ttau_{iz}\right\vert_{wall} = \left[\frac{\kappa \thU_{avg}}{\log\left(\frac{\Delta_z/2 - \widetilde{\eta}}{z_{0,\Delta}}\right)}\right]^2\frac{\thu_{r}}{\thU_{avg}},
\end{equation}
where $i= x,y$ and $\thU_{avg} = \sqrt{\thu_r^2(x,y,t) + \thv_r^2(x,y,t)}$ is the magnitude of the tangential wind velocity relative to the wave surface, $ \Delta_z/2$ is the first off-wall grid point, and $\widetilde{\eta}(x,y,t)$ is the sea surface elevation filtered at the LES resolution $\Delta$. The relative surface velocities $\thu_r,\thv_r$ are calculated using the velocity at the first grid point and the wave surface orbital velocity  \citep{Yang2013D,Sullivan2014,Yang2014}:
\begin{equation}
    \thu_{r} = \thu(x,y,z_1,t) - \thu_{s}(x,y,t),\qquad \thv_{r} = \thv(x,y,z_1,t) - \thv_{s}(x,y,t),
\end{equation}
where $\thu_s,\thv_s$ are the wave orbital velocities, associated with the corresponding modeled wave mode spectra. This follows previous work in \citet{Aiyer2022T} with the smooth wall aerodynamic roughness supplemented by the subfilter wave roughness.

 The subfilter sea surface roughness is set proportional to the r.m.s of the height fluctuations calculated using Equation (\ref{eqn:rms_model}):
 \begin{equation}
z_{0,\Delta} = \left[z_{0,s}^2 + \left(\alpha_w \sigma_{\eta}^{\Delta}\right)^2\right]^{1/2},
 \end{equation}
where $\alpha_w$ is the roughness parameter that is determined dynamically. Note that the smooth wall roughness \cite{Fairall1996} $z_{0,s} = 0.11\nu/u_*$  has been added for each scale to ensure numerical stability for the case $\alpha_w=0$, which corresponds to a completely resolved spectrum on the LES grid. The dynamic procedure is briefly outlined below  following the procedure in \citet{Anderson2011} and \citet{Yang2013D}.  

\subsection{Dynamic Wave Spectrum Model (Dyn-WaSp)}
To determine the dynamic parameter $\alpha_w$, the total drag is evaluated at the filter scale $\Delta$ and a test- filter scale $\widehat{{\Delta}} = 2\Delta$ (where $\Delta$ is the LES grid resolution) with the assumption that $\alpha_w$ is the same at both scales. A Gaussian filter with a 3-point stencil in physical space is used for the filtering process. The requirement that the total drag at the surface must be independent of the scale chosen leads to the following identity:

\begin{equation}\label{eqn:tot_drag_balance}
    \iint \widetilde{p}^s\widetilde{n}_i\ \mathrm{d}x\mathrm{d}y + \rho_a \iint \tau_{ij,\Delta}^{wall} \widetilde{n}_j\ \mathrm{d}x\mathrm{d}y = \iint \hat{\widetilde{p}}^s\hat{\widetilde{n}}_i\ \mathrm{d}x\mathrm{d}y + \rho_a \iint \tau_{ij,2\Delta}^{wall} \hat{\widetilde{n}}_j\ \mathrm{d}x\mathrm{d}y,
\end{equation}
where the l.h.s represents the total force at scale $\Delta$, the r.h.s represents the total force at scale $\widehat{\Delta}$, and filtering at twice the grid scale is denoted by $\hat{\tilde{(..)}}$. The roughness lengths at the two scales are given by
\begin{equation}
    z_{0,\Delta} = \left[z_{0,s}^2 + \left(\alpha_w \sigma_{\eta}^{\Delta}\right)^2\right]^{1/2}; \quad z_{0,2\Delta} = \left[z_{0,s}^2 + \left(\alpha_w \sigma_{\eta}^{2\Delta}\right)^2\right]^{1/2}.
\end{equation}
For the wind and wave cases considered in the present study, $\alpha_w\sigma_{\eta}^{\Delta}>>z_{0,s}$, and $z_{0,\Delta}$ is primarily determined by the dynamic roughness model.

Substituting the Wave Spectrum (WaSp) model from Equation (\ref{eqn:total_stress}) and the equilibrium wall model from Equation (\ref{eqn:MOST}) at the two scales and further considering an equivalence of total wall stress, the self-consistency condition can be rewritten as 

    \begin{equation}\label{eqn:self_consistency}
    {f^{\Delta}_{d,i}}\Delta_z + \left\langle\left[\frac{\kappa U_{avg}^{\Delta}}{\log\left(\frac{\Delta_z/2 - \widetilde{\eta}}{z_{0,\Delta}}\right)}\right]^2 \frac{\widetilde{u} - \widetilde{u}_{orb}}{U_{avg}^{\Delta}}\right\rangle = {f^{2\Delta}_{d,i}}\Delta_z+ \left\langle\left[\frac{\kappa U_{avg}^{2\Delta}}{\log\left(\frac{\Delta_z/2 - \widehat{\widetilde{\eta}}}{z_{0,2\Delta}}\right)}\right]^2 \frac{\hat{\widetilde{u}} - \hat{\widetilde{u}}_{orb}}{U_{avg}^{2\Delta}}\right\rangle,
    \end{equation}
 where,
\begin{equation}
    {f^{\Delta}_{d,i}}\Delta_z = \sum^{k_{max} = \pi/\Delta} \left[\Gamma C_D(k)\widetilde{u}_i \left((\widetilde{u}_l - c(k))\cdot \frac{\partial \widetilde{\eta}(k)}{\partial x_l}\right){\mathcal{H}\left\{(\widetilde{u}_l - c(k))\cdot \frac{\partial \widetilde{\eta}(k)}{\partial x_l}\right\}} + (1-\Gamma) \beta(k)\frac{(a(k)ku_*)^2}{2}\right]
\end{equation}

\begin{equation}
        {f^{2\Delta}_{d,i}}\Delta_z = \sum^{k_{max} = \pi/2\Delta} \left[\Gamma C_D(k)\widehat{\widetilde{u}}_i \left((\widehat{\widetilde{u}}_l - c(k))\cdot \frac{\partial \widehat{\widetilde{\eta}}(k)}{\partial x_l}\right){\mathcal{H}\left\{(\widehat{\widetilde{u}}_l - c(k))\cdot \frac{\partial \widehat{\widetilde{\eta}}(k)}{\partial x_l}\right\}} + (1-\Gamma) \beta(k)\frac{(a(k)ku_*)^2}{2}\right].
\end{equation}
The angular brackets denote planar averaging. 
As the waves considered here are non-deforming, planar averaging is sufficient.
The unknown parameter $\alpha_w$ enters the equation inside a transcendental function and is challenging to obtain through simple algebraic manipulations. A bisection root-finding algorithm is employed at each time step to solve for $\alpha_w$. This value of $\alpha_w$ is used in Equation (\ref{eqn:MOST}) to calculate the wall stress,  allowing the surface roughness to vary with wind and wave conditions.

In summary, the waves are unresolved in the vertical direction, the effect of waves resolved in the horizontal direction to the filter scale $\Delta$ are modeled, and the subfilter contributions due to unresolved waves are determined dynamically using the  Dyn-WaSp model.

\section{Computational framework}\label{sec:methods}
\subsection{LES Framework}
 LES calculations were performed using NGA, which is a structured, finite difference, low Mach number flow solver \cite{desjardins2008,macart2016}. The wind velocity field is described using the filtered Navier-Stokes equations in the incompressible limit:
\begin{equation}\label{eqn:div}
\nabla \cdot \widetilde{\bm{u}} =0,
\end{equation}
\begin{align}\label{eqn:Navier_stokes}
\frac{\partial \widetilde{\bm{u}}}{\partial t} + \widetilde{\bm{u}} \cdot\nabla\widetilde{\bm{u}} =& -\frac{1}{\rho}\nabla\widetilde{p}+ \nabla\cdot \underline{\underline{\bm{\widetilde{\sigma}}}} + \widetilde{\bm{f}}_d + \widetilde{\bm{f}}_T.
\end{align}
 The equations are discretized on a Cartesian grid $(x,y,z)$.
The variables are staggered on the computational mesh, and the location of each variable is described in detail in Ref. \cite{desjardins2008}.
In Eqs. (\ref{eqn:div}) and (\ref{eqn:Navier_stokes}),  $\tuu = (\tu,\tv,\tw)$ is the velocity vector with the tilde denoting variables filtered on the LES grid; $\widetilde{\bm{f}}_d$ is the drag force applied in the streamwise direction, representing the effects of the waves; $\widetilde{\bm{f}}_T$ is the turbine force calculated by the actuator disk model; and $\widetilde{\sigma}_{ij} = 2\nu \widetilde{S}_{ij} + \ttau^d_{ij}$ is the total deviatoric stress, where $\nu$ is the molecular viscosity, $\widetilde{S}_{ij}$ is the resolved strain rate tensor, and $\ttau^d_{ij}$ is the subfilter stress (SFS) tensor. The SFS tensor is modeled using a Lilly-Smagorinsky type subfilter viscosity model $\ttau_{ij}^d = 2\nu_T\widetilde{S}_{ij}$, where  the subfilter viscosity is computed using the Anisotropic Minimum Dissipation (AMD) model \cite{Rozema2015,Akbar2017}:
\begin{equation}
    \nu_T = \frac{-(\hat{\partial}_k \widetilde{u}_i)(\hat{\partial}_k \widetilde{u}_j)\widetilde{S}_{ij}}{(\partial_l \widetilde{u}_m)(\partial_l \widetilde{u}_m)},
\end{equation}
where $\hat{\partial}_i = \sqrt{C_i}\delta_i \partial_i$ ($i = 1, 2, 3$) is the scaled gradient operator and $C_i = 0.33$ is the modified Poincare constant.

\subsection{Turbine Model and Wave Spectrum }
The wind turbines are represented using an actuator disk model \cite{Burton}. In the present study, the disk averaged velocity is used as the reference velocity rather than the upstream unperturbed velocity and the turbine force in the streamwise direction per unit mass is given by \cite{Calaf2010}

\begin{equation}
    f_T = -\frac{1}{2}C_T^{\prime}\langle u\rangle_d^2 \frac{1}{\Delta x},
\end{equation}
where  $\langle u\rangle_d$ is the local wind velocity spatially averaged over the turbine disk and $C_T^{\prime} = C_T/(1-a_I)^2 = 4/3$ is the effective thrust coefficient, where $C_T$ us the turbine thrust coefficient, and $a_I$ is the induction factor. The turbine force is smoothly distributed across the actuator disk area \cite{Shapiro2019}.

In this work, the height spectrum of wind-generated sea surface waves obtained during the Joint North Sea Wave Observation Project (JONSWAP)\cite{hasselmann1973measurements} is used. The one-dimensional form of the spectrum in angular frequency space is given by

\begin{equation}\label{eqn:Jonswap}
    E_J(\omega) = \frac{\alpha_P g^2}{\omega^5}\exp{\left[-\frac{5}{4}\left(\frac{\omega_p}{\omega}\right)^4\right]}\gamma^{r},
\end{equation}
where
\begin{equation}
    r = \exp\left[-\frac{(\omega-\omega_p)^2}{2\Sigma^2\omega_p^2}\right],
\end{equation}
where $\Sigma$ is the standard deviation given by:
\begin{equation}
\Sigma = \begin{cases}
  0.07, & \omega \leq \omega_p \\
  0.09, & \omega > \omega_p.
\end{cases}
\end{equation}
Here, $\omega_p$ is the angular frequency at the spectrum peak, $\alpha_P = 0.076\left((U_{10}^2/(gF)\right)^{0.22}$ is the Phillips constant \cite{phillips_1985}, $F$ is the wave fetch, $\gamma=3.3$ is the peak enhancement factor, and $U_{10}$ is the velocity at $10$ m height.
Using the gravity wave dispersion relation $\omega^2 = gk$, the  spectrum can be transformed into wavenumber space using $S(k) = (g/(2\omega))E_J(\omega)$.

\subsection{Simulation Setup}

\begin{figure}
    \centering
    \includegraphics[width = 0.7\textwidth,trim={0 0 0 0},clip]{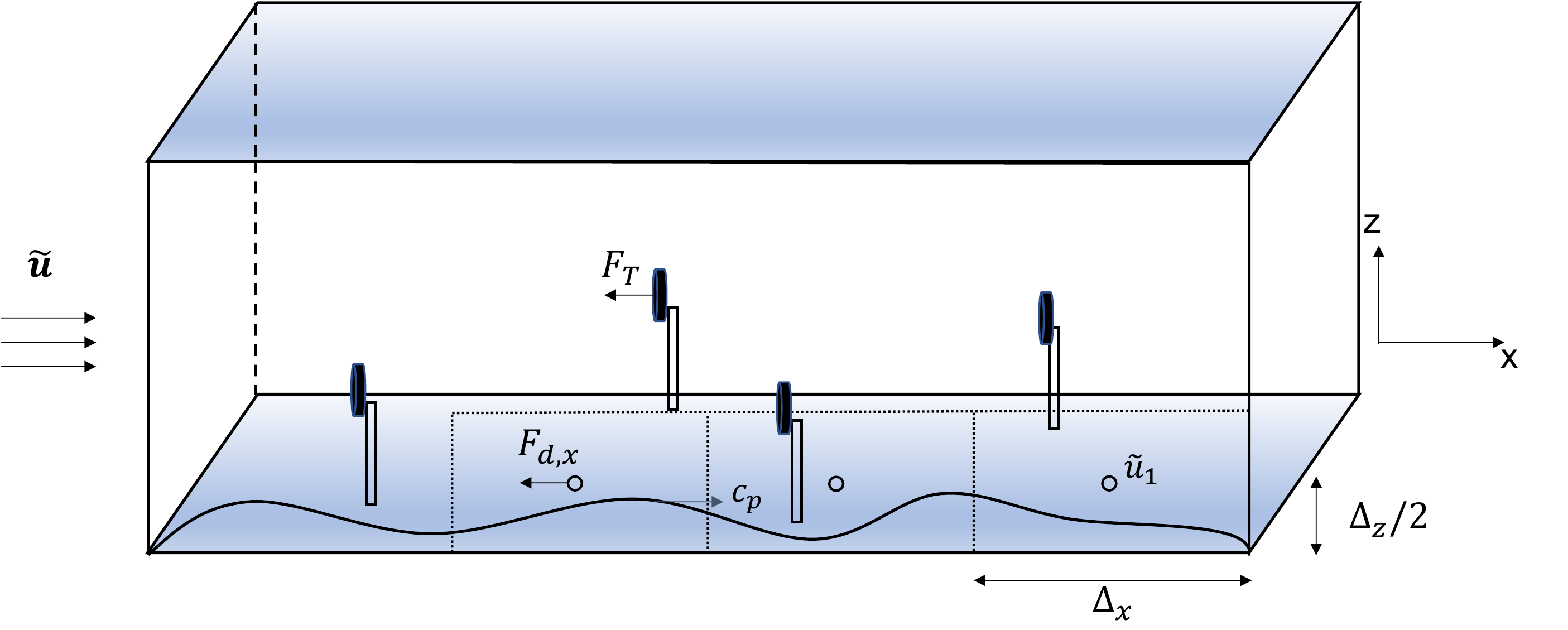}

     \caption{ Sketch of the simulation setup depicting the Dynamic Wave Spectrum (Dyn-WaSp) model  ($F_{d,x}$), which calculates the hydrodynamic wave drag for the horizontally resolved portion of the wave spectrum. The waves are multiscale, and the subfilter contributions are calculated dynamically. The wind turbines are modeled using actuator disks.}
    \label{fig:sketch}
\end{figure}

 \begin{table}
\centering
\begin{ruledtabular}
\begin{tabular}{c c c c c c} 
 Case &$U_{10}$ ($ms^{-1}$)
 & $c_p/u_{\ast}$ & $c_p\ (ms^{-1})$ & $k_p\ (m^{-1})$ & $H_s\ (m)$\\ [5pt] \hline
 CU6&$12$  &  $6$ & $2.66$ & $1.38$ & $0.26$  \\ 
 CU10 &$12$ & $10$ & $4.43$ &$0.5$ & $0.56$\\
CU18 &$12$ & $18$ & $7.99$ & $0.15$  & $1.34$\\
 \hline
\end{tabular}
\end{ruledtabular}

\caption{Parameters for JONSWAP wave spectrum  }
\label{tab:wave_param}
\end{table}

 \begin{table}
\centering
\begin{ruledtabular}
\begin{tabular}{c c c c c c c}
 Case & $\lambda_p (m)$  &$\Delta_x/\lambda_p$ & $H/\lambda_p$ & $N_x \times N_y \times N_z$ & $\langle \alpha_w\rangle_t$ \\ [5pt] \hline
 CU6(C)&$4.55$ & $0.355$ &  $2$ & $64\times 64\times 32$ & $0.278$\\ 
  CU6(F)&$4.55$ & $0.156$ &  $2$ & $128\times 128\times 64$ & $0.304$ \\ 
 CU10(C) &$12.62$ & $0.686$ &  $2$ & $92\times 92\times 44$ & $0.18$\\ 
 CU10(F) & $12.62$ & $0.493$ &  $2$ & $128\times 128\times 44$ & $0.194$ \\ 
CU18(C) &$41$ & $3.2725$ &  $2$ & $64\times 64\times 60$ & $0.037$\\ 
CU18(F) &$41$ & $1.6362$ &  $2$ & $128\times 128\times 60$ & $0.088$\\ 
 \hline
\end{tabular}
\end{ruledtabular}
\caption{Parameters for simulation domain }
\label{tab:sim_param}
\end{table}

In the first set of simulations, airflow over a spectrum of waves (without wind turbines) with different wave characteristics is considered. The wave parameters are chosen to mimic the different cases simulated in phase-resolved simulations of \citet{Yang2013D}. 
The details of the wave characteristics for the different simulations considered are provided in Table \ref{tab:wave_param}. Each case is labeled by the wave age at the spectrum peak, i.e CU6 is the case with $c_p/u_* = 6$ etc. The computational domain size in the horizontal directions is $L_x=L_y = 5\lambda_p$ for each case, where $\lambda_p$ is the wavelength at the spectrum peak. The case with wave age $c/u_* = 6$ corresponds to a wave field with the smallest peak wavelength. A larger domain size with $L_x = 20\lambda_p$ is chosen to ensure the effect of large-scale structures is well captured. For each case, two grid resolutions are run to probe the robustness of the dynamic model. Details of the grid discretization for each case along with the horizontal and vertical resolutions are provided in Table \ref{tab:sim_param}.  The discretization in the vertical direction $N_z$ is chosen to ensure that the maximum wave height lies below the cell center of the first grid point, $a_{max}  < 0.95\Delta_z/2$, and  {that the grid aspect ratio is ideal for wall-modeled LES, i.e, $\Delta_x/\Delta_z \geq 1$ \citep{Piomelli2002,kawai2012}. 
Periodic boundary conditions are applied in the $x$ and $y$ directions, and the flow is driven with a constant external pressure gradient. The pressure gradient results in a friction velocity  (or surface stress) $ u_{\ast}^2 =\rho_a^{-1}(\partial P/\partial x) H$. A free-slip boundary condition is used for the top of the domain. The bottom boundary is bounded by multiscale waves, and the subfilter-scale roughness is calculated dynamically using the procedure described in Section \ref{sec:dynamic_model}. The simulation setup is depicted in Figure 1.

Next, for the flow through an offshore wind farm, the wave characteristics are fixed (shown in Table \ref{tab:wave_paramt}), and the turbine configurations are varied. The different cases considered in this study are summarized in Table \ref{tab:sim_paramt}.
  A $N_{row} \times 3$ wind turbine array is distributed uniformly in the simulation domain of extent ($L_x, L_y, H$) = ($2100$, $1500$, $1000$) m with $N_{row} = 2,4$. The computational domain is discretized uniformly using ($N_x,\ N_y,\ N_z$) = ($128,\ 128,\ 125$) grid points.  Due to the periodic boundary conditions, the setup represents an infinite wind farm under fully developed conditions \cite{Calaf2010,Yang2014}. The wind turbines have a hub height of $H_{hub} = 100$ m and a diameter of $D= 100$ m. The spanwise spacing parameter (normalized distance between two turbines) has a fixed value of $s_y = (L_y/N_{col}) =  5$, and the streamwise spacing parameter is $s_x  = L_x/N_{row} = 10.5$ and $5.25$ for $N_{row} = 2$ and $4$, respectively. These spacings are typical of commercial wind farms and mimic the configuration in \citet{Yang2014OffshoreW}. Additionally, the case with a sinusoidal wave field with the wave characteristics matched to the peak wavelength and significant wave height $H_s$ of the JONSWAP spectrum is considered.    At the initial stage of the simulation, the airflow boundary layer over the prescribed waves is allowed to develop until equilibrium is reached. The wind turbine forcing is then turned on, and the wind farm boundary layer is allowed to develop.

  \begin{center}
\begin{table}
\parbox{.33\linewidth}{
\centering
\begin{ruledtabular}    
\begin{tabular}{c c c}
 $c_p/u_*$&$H_sk_p/2$ & $k_p$ (m) \\
\hline
  $11$ & $0.09$ & $0.1$ \\
\end{tabular}
\end{ruledtabular}
\caption{\label{tab:wave_paramt}Wave parameters for offshore wind farm study}
}
\hfill
\parbox{.57\linewidth}{
\centering
\begin{ruledtabular}    
\begin{tabular}{c c c c }
Case & Description & $s_x\times s_y$ & $<\alpha_w>_t$\\
\hline
 SPW & Wave Spectrum & $--$&  $2.35\times 10 ^{-2}$\\
 SPWJ2 & Wave Spectrum + turbines & $10.5\times 5$  & $2.22\times 10 ^{-3}$\\
SWJ2  & Monochromatic wave + turbines &  $10.5\times 5$ & --\\
SPWJ4 &Wave Spectrum + turbines & $5.25\times 5$ &$2.26\times 10 ^{-3}$\\
\end{tabular}
\end{ruledtabular}
\caption{\label{tab:sim_paramt}Simulation parameters for offshore wind farm study}
}
\end{table}
\end{center}

\section{LES of airflow over a spectrum of waves}
\label{sec:Les_airflow}

In this section, the Dyn-WaSp model is used to simulate airflow over a multiscale ocean surface. The results are validated against phase-resolved simulations performed by \citet{Yang2013D}  for different wave characteristics. The contribution from different wave modes to the total stress is analyzed, and, finally, the performance of the dynamic model for different grid resolutions is demonstrated.

\subsection{Mean Velocity Profiles}

Different wave modes contribute to the overall height distribution of the ocean wave field.  
The one-dimensional wavenumber spectra of surface elevation for the three cases are shown in Figure \ref{fig:spectra}a, along with the location of the maximum resolvable wavenumber for the different grid resolutions considered. 
A transect of the surface for the different cases is shown in Figure  \ref{fig:spectra}b.
In the current simulations, directional spreading is neglected, and the wave propagates in the streamwise direction aligned with the mean flow.  The multiscale nature of the generated surface is visible, with contribution from different wavenumbers based on the corresponding wave spectrum from Figure \ref{fig:spectra}. It is important to note that, although there is less energy in the high wavenumber tail, the contribution to the overall wave form drag is non-negligible. \citet{Hwang2005WaveWaves} found that the intermediate-scale waves with wavelengths less than $6$ m or wavenumbers greater than $1$ m$^{-1}$ are the dominant contributor to the ocean surface drag. As the wave phase speed is inversely proportional to the wavenumber, the high wave number region corresponds to slow waves, that impart a drag to the wind  due to the larger relative velocity between the wind and the wave surface.  Hence, the entire spectrum (resolved and subfilter waves) needs to be accurately represented to determine the sea surface drag.
\begin{figure}
    \centering
    
    \subfloat[Wave Spectra]{\includegraphics[width=0.9\textwidth]{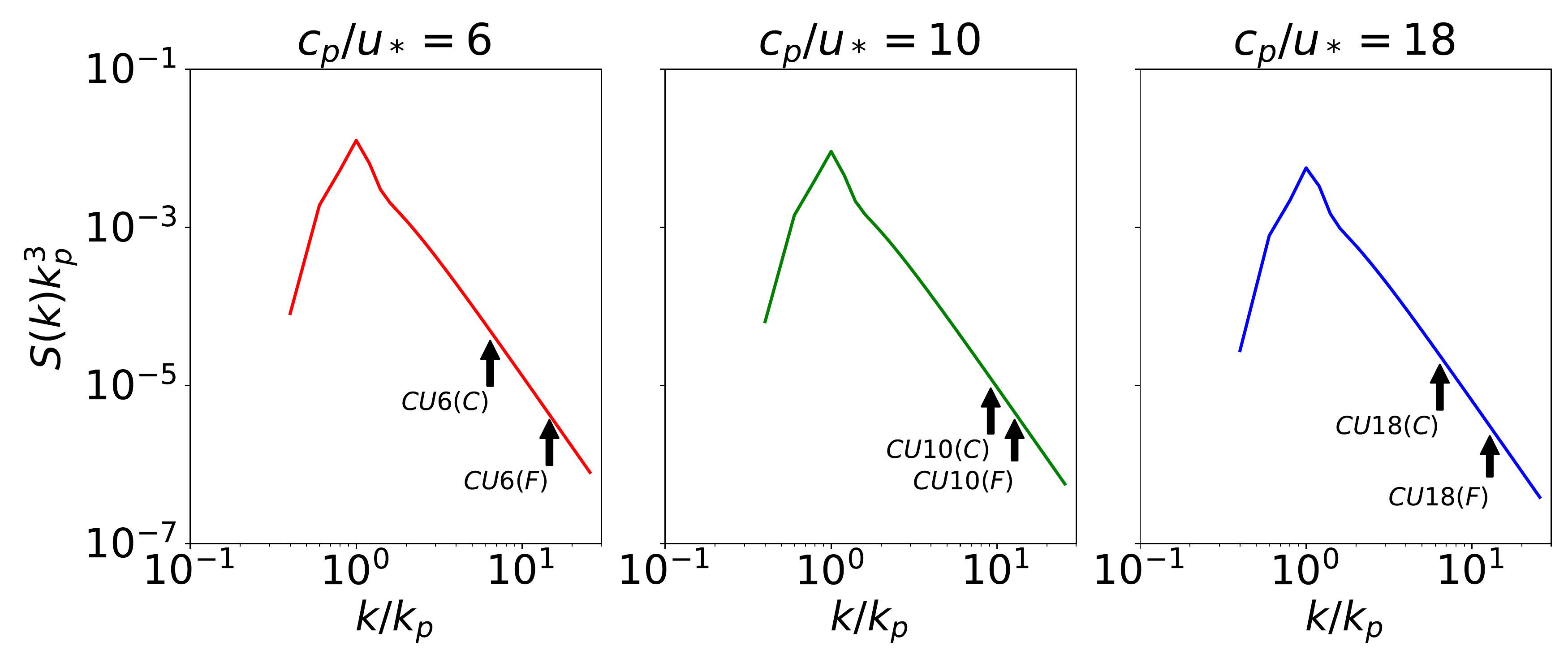}}
    \vfill
   \subfloat[Height Distribution]{\includegraphics[width=0.9\textwidth]{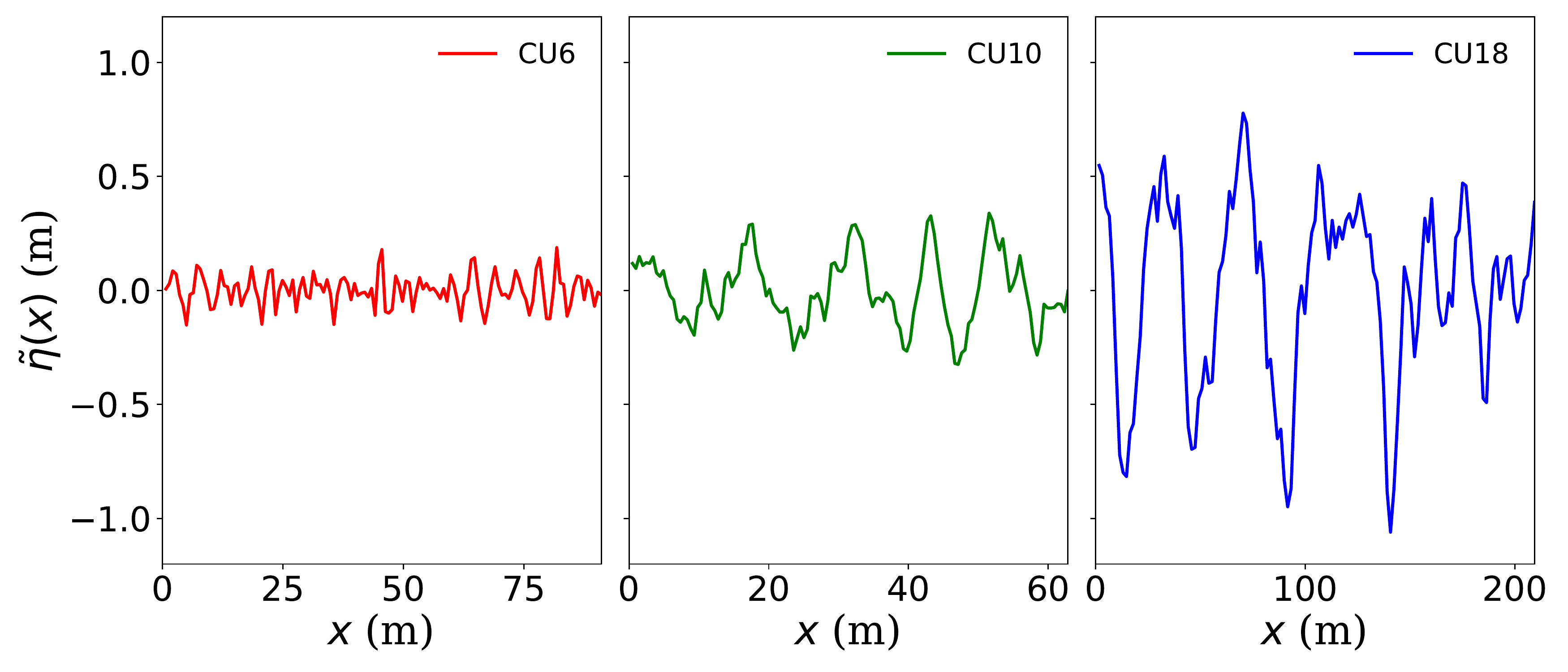}}
    \caption{a) One-dimensional wavenumber spectra of surface elevation normalized by the peak wavenumber $k_p$ and b) resolved wave height distribution as a function of normalized streamwise coordinates prescribed at the bottom boundary for the fine resolution case: $c_p/u_* = 6$ (red line), $c_p/u_* = 10$ (green line), $c_p/u_* = 18$ (blue line). In the current simulations, directional spreading is neglected, and the waves propagate along the streamwise direction. The arrows in each panel correspond to the grid resolution limit of the LES for the different cases in Table \ref{tab:sim_param}.}
    \label{fig:spectra}
\end{figure}

\begin{figure}
    \centering
    
    \subfloat[\label{fig:mean_cu6}Velocity]{\includegraphics[width = 0.45\textwidth]{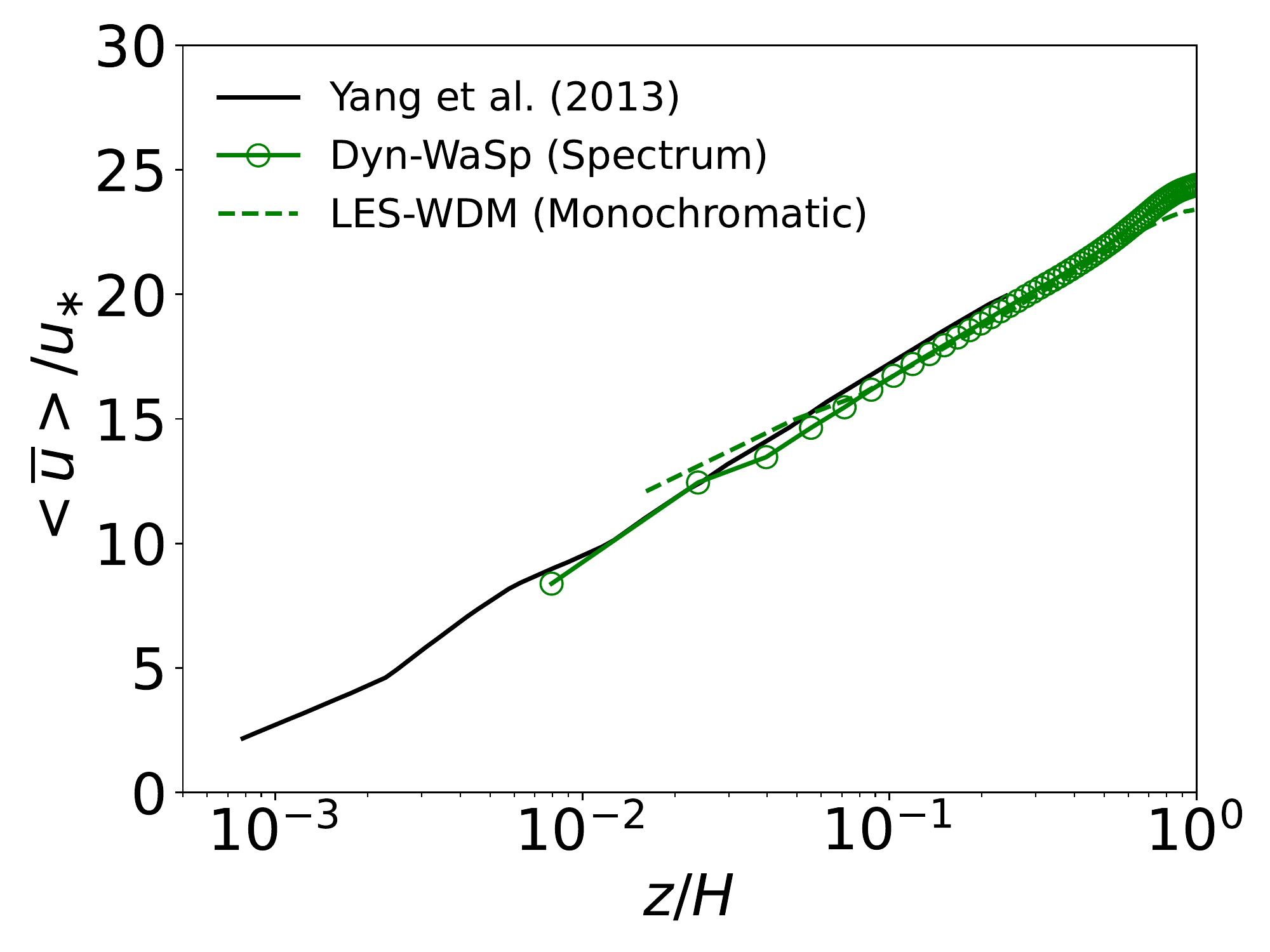}}
    \subfloat[\label{fig:ws_cu6}Growth Rate]{\includegraphics[width = 0.45\textwidth]{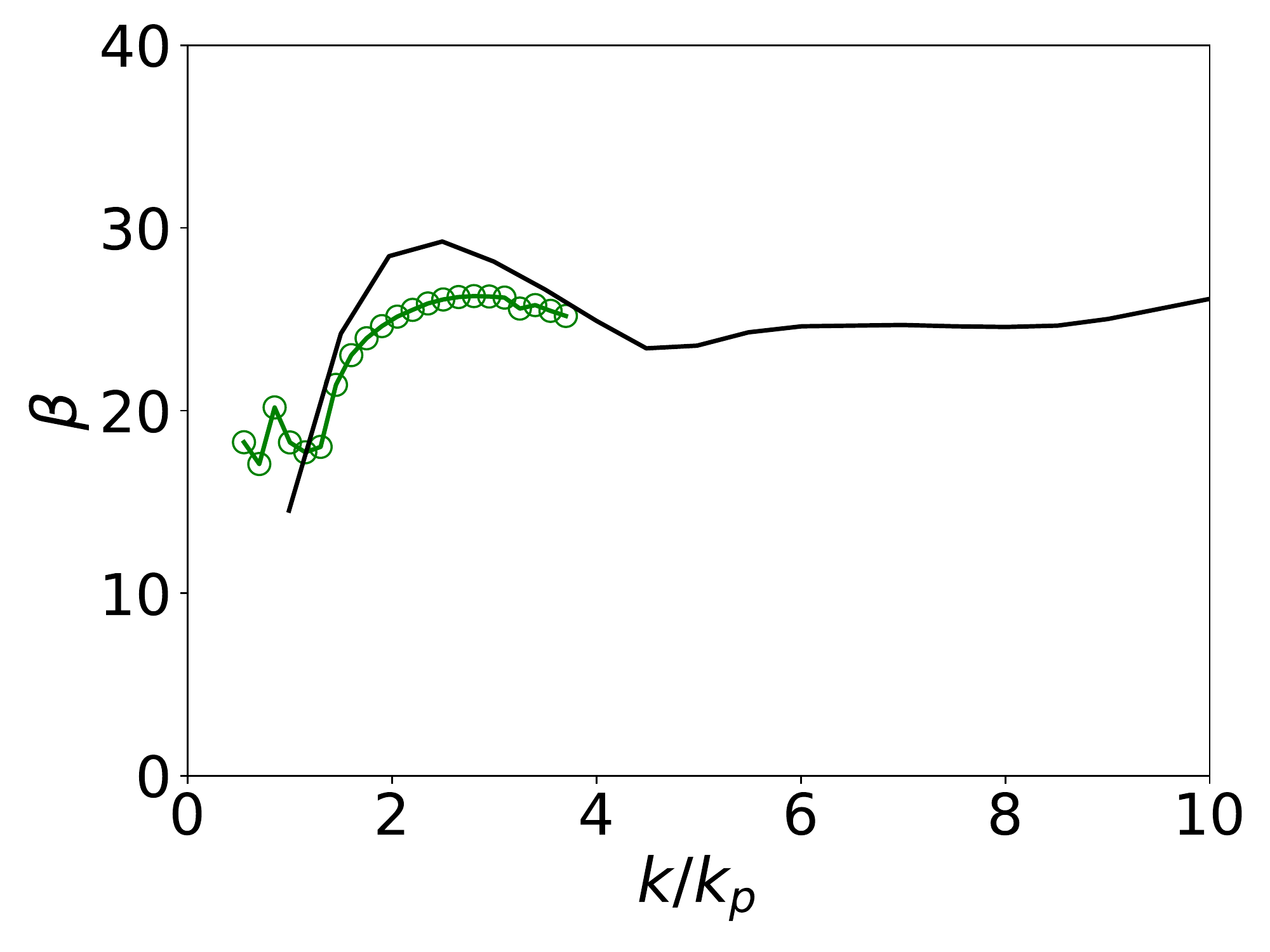}}
\vfill
    \subfloat[\label{fig:CU10_mean} Velocity]{\includegraphics[width = 0.45\textwidth]{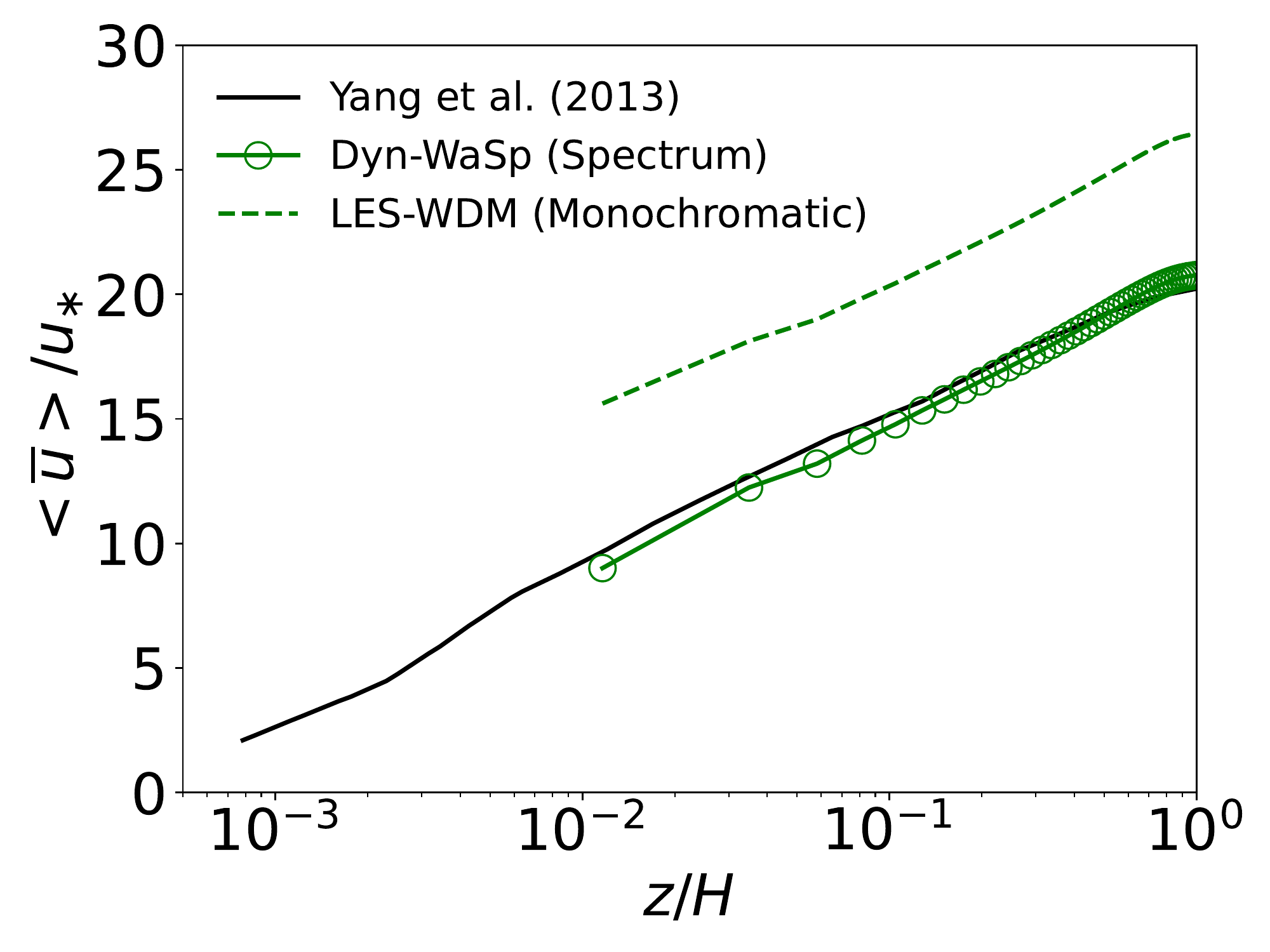}}
        \subfloat[\label{fig:CU10_growth}Growth Rate]{\includegraphics[width = 0.45\textwidth]{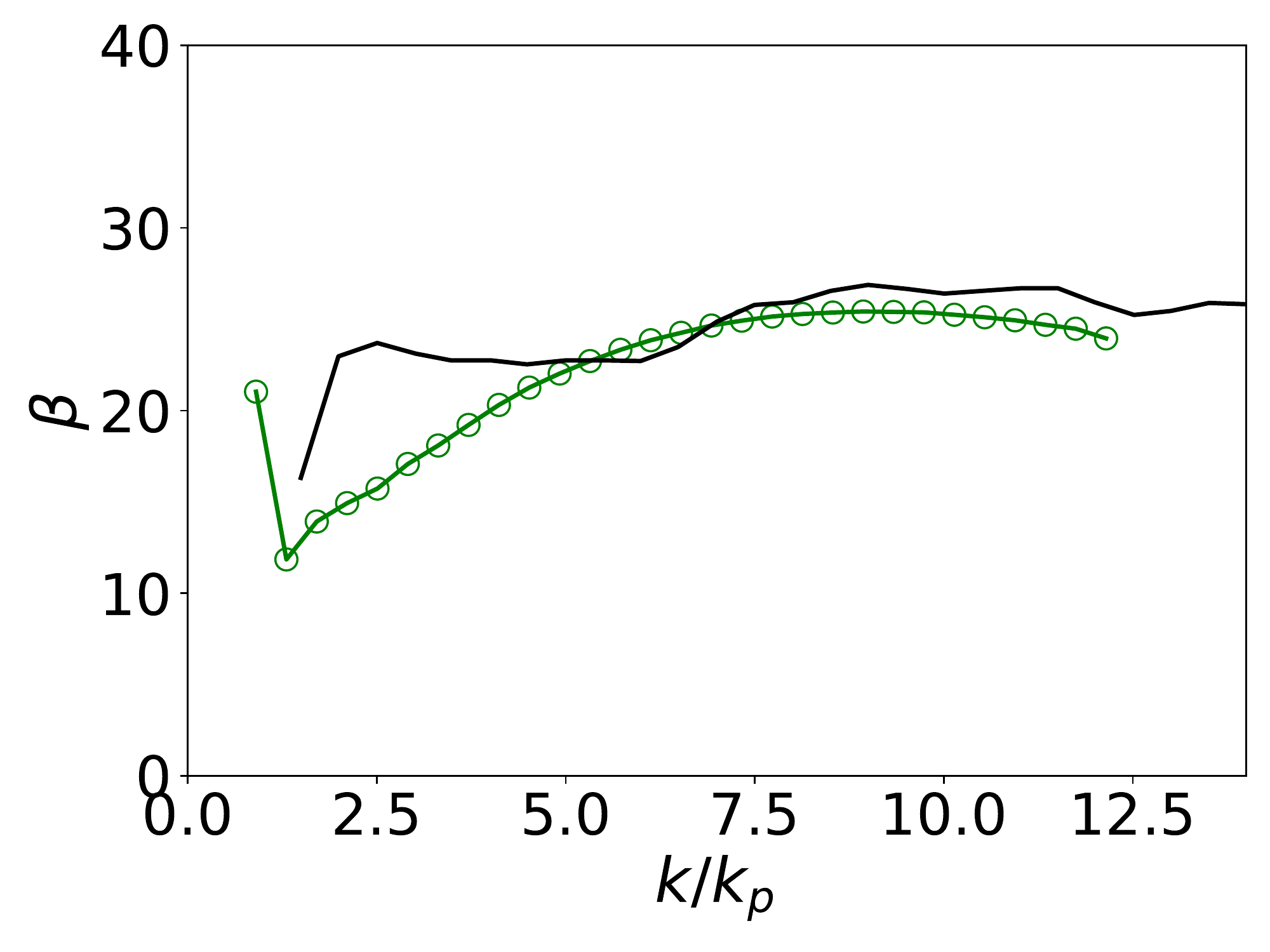}}
        \vfill
            \subfloat[ \label{fig:CU18} Velocity]{\includegraphics[width = 0.45\textwidth]{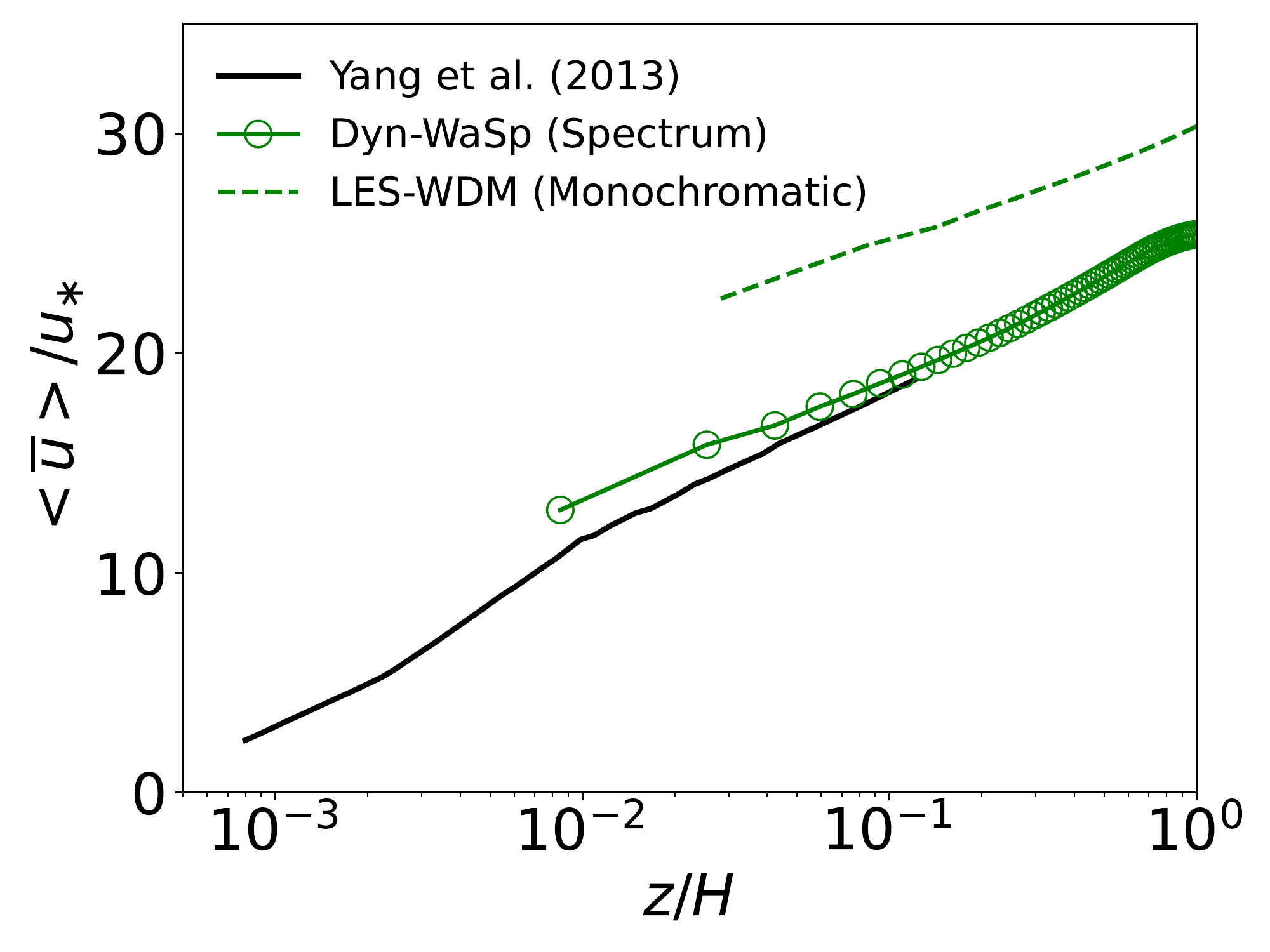}}
            \subfloat[\label{fig:CU18_growth}Growth Rate]{\includegraphics[width = 0.45\textwidth]{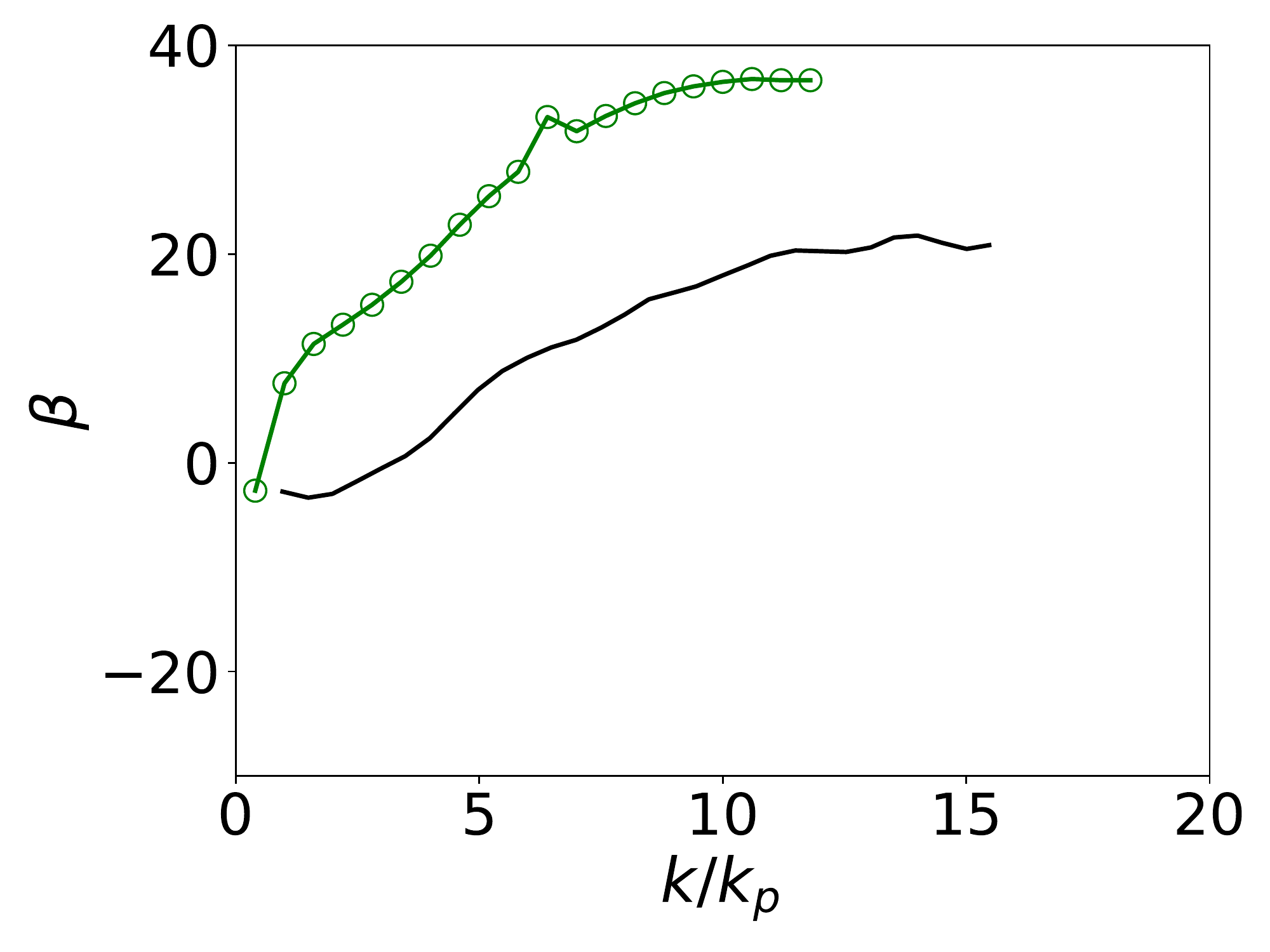}}
            \vfill
    \caption{Temporally and spatially averaged velocity profiles and normalized wave growth rate  for (a,b) $c_p/u_* = 6$, (c,d) $c_p/u_* = 10$, and (e,f) $c_p/u_*=18$, respectively. The results from the LES with the Dyn-WaSp model are shown using the green circle and line, and the phase-resolved LES from \citet{Yang2013D} are shown using the black solid lines. For the velocity profiles, the results from a monochromatic wave train are shown using the dashed green lines. }
    \label{fig:cu6_mean}
\end{figure}





In Figure \ref{fig:mean_cu6}, the temporally and spatially averaged mean velocity profile from the LES is depicted. The results show good agreement with the phase-resolved simulations from \citet{Yang2013D}. Additionally shown are LES results where the wave field is described using a monochromatic wave train with wavenumber set to the peak wavenumber $k_p$ and amplitude calculated using the significant wave height $a = H_s/2$. For this case, the sinusoidal wave train with the model from \citet{Aiyer2022} also shows good agreement with the phase-resolved model. The monochromatic wave train is based on the spectrum peak which also corresponds to the peak drag producing waves. Hence, resolving the peak wavenumber for this case $k_p = 1.38$ m is sufficient to capture the mean behavior of the streamwise velocity. Figure \ref{fig:ws_cu6} shows the distribution of the wave growth rate calculated for the CU6(F) case. The modeled normalized growth rate shows  good qualitative and quantitative agreement with the phase-resolved calculations. Further, the growth rate reaches a constant value for high-wavenumbers consistent with previous studies \cite{Belcher1998,CAVALERI2007603}. Note that, in the current simulations, the pressure-based form stress is modeled, and the waves are prescribed. However, a normalized growth rate can be used to facilitate comparison with the phase-resolved simulations and experiments.

 %
 

The mean profiles for cases with $c_p/u_* = 10$ and $c_p/u_* = 18$ are shown in Figures \ref{fig:CU10_mean} and \ref{fig:CU18}. For these cases, the equivalent monochromatic wave with the matched peak characteristics under-predicts the wave form drag, and the mean velocity profiles are faster than the phase-resolved cases. The peak wavenumber for these two cases is $k_p = 0.5\ m^{-1}$ and $k_p = 0.15\ m^{-1}$, and the intermediate scale waves correspond to $k/k_p > 2$ and $7$, respectively. For the CU10 case, the growth rate shows excellent qualitative and quantitative agreement with the phase-resolved simulations as shown in Figure \ref{fig:CU10_growth}. The growth rate for the CU18 case is higher than that of the phase-resolved simulation. This case's velocity near the surface is slightly higher than the phase-resolved simulations. This leads to an over-prediction of the surface stress as the stress depends on the square of the velocity. Additionally, the normalized growth rate is sensitive to the wave steepness where a 10\% difference in the steepness results in a 20\% difference in the growth rate.
Qualitatively the growth rate is slightly negative for the smallest wavenumber (highest wave speed) and begins to saturate for the higher wavenumbers similar to the phase-resolved simulations. 

In order to highlight the range of dimensional wavenumbers simulated in each case, the normalized growth rate is plotted as a function of dimensional wavenumber (for the high resolution cases) in Figure \ref{fig:stress_growth_dim}. The values of $\beta$ differ at small $k$ and reach a constant value for large $k$ for CU6 and CU10 cases. This behavior mimics the behavior of turbulence where the large-scale motions differ, while universality is preserved in the small scales \cite{Yang2013D}. For the CU18 case, the higher wavenumbers are not resolved, and the saturation is not reached.  In order to facilitate comparison with other numerical and experimental studies, Figure \ref{fig:stress_growth_all} depicts the normalized growth rate as a function of the wave age $c/u_*$. The results from the current LES with  the Dyn-WaSp model are shown with solid symbols. Additionally shown are lab-scale experimental data from wave tanks \cite{Grare2013,Buckley2020}, DNS of flow over a monochromatic wave train \cite{Sullivan2000,Kihara2007,Yang2010}, fully coupled DNS \cite{Wu2022} and data from LES coupled with a HOSM wave solver \cite{Liu2010,Yang2013D,Hao2019}. The normalized growth rate has been shown to depend on the Reynolds number of the flow, and domain size and measurements from experiments and simulations show significant scatter \cite{plant1982,Grare2013,Hao2019,Buckley2020}. The normalized growth rate calculated with the Dyn-WaSp model fall in the range of existing experiments, computation, and theory.

In summary, if the intermediate scale waves are resolved in the simulation, using a representative sinusoidal wave train matched to the peak wave characteristics is sufficient to capture the surface roughness.  However, in general, the performance of the Dyn-WaSp model is superior and accurately predicts both the mean velocity profiles and the form stress and does not rely on resolving the intermediate scale waves. The dynamic approach will be advantageous, especially for simulations of marine atmospheric boundary layer flows relevant to offshore wind farms where typical grid resolutions are of the order O(10) m, and intermediate scale waves would remain unresolved.

\begin{figure}
    \centering
    \subfloat[\label{fig:stress_growth_dim}] {\includegraphics[width = 0.5\textwidth]{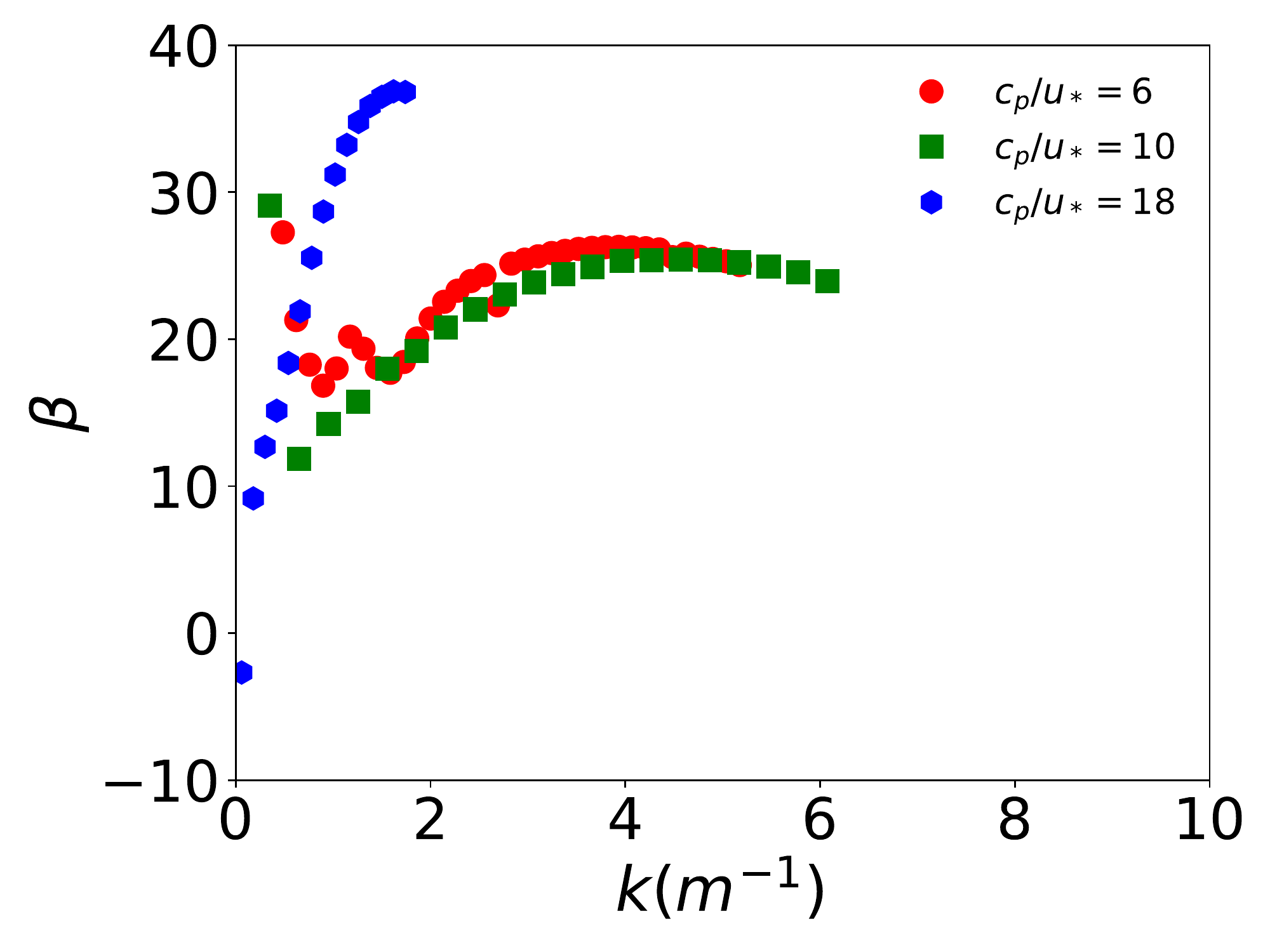}}
    \subfloat[\label{fig:stress_growth_all}]{\includegraphics[width = 0.5\textwidth]{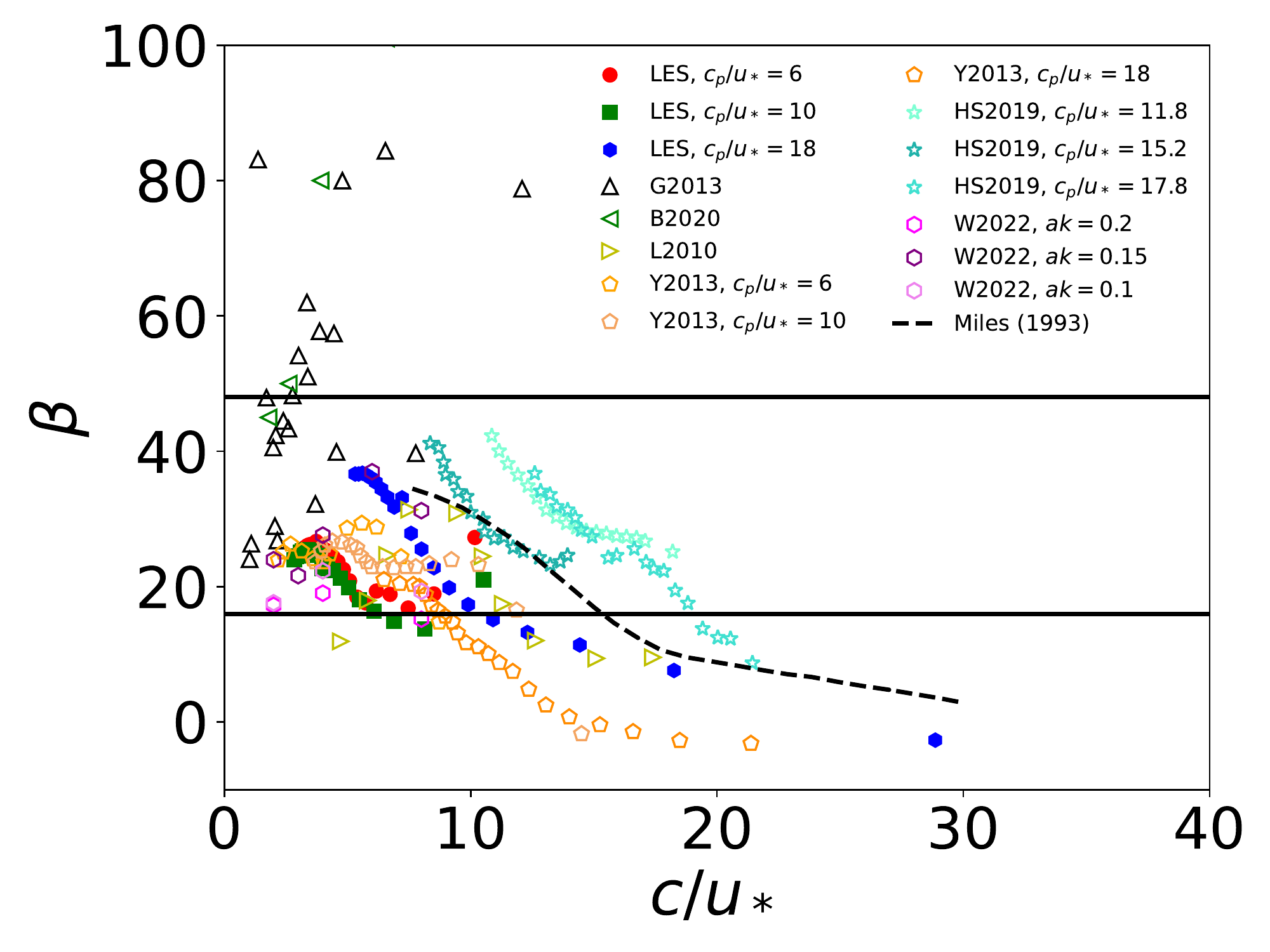}}

    \caption{ a) Normalized wave growth rate as a function of dimensional wavenumber for the different cases considered and b) wave growth rate as a function of wave age for current LES with Dyn-WaSp model (Solid symbols). Additionally shown are measurements:  \citet{Grare2013} (G2013) (black triangles), and \citet{Buckley2020} (green left facing triangles); DNS: \citet{Wu2022} (W2022) (violet, purple, and magenta hexagons); and LES: \citet{Liu2010} (L2010) (Yellow right facing triangles), \citet{Yang2013D} (Y2013) (orange pentagons), and \citet{Hao2019} (H2019) (turquoise stars).  The solid line at $\beta = 48$ and $\beta = 16$ corresponds to the empirical formula from \citet{plant1982}, and the dashed lines are from the critical-layer theory of \citet{Miles1993}.  }
    \label{fig:stress_growth}
\end{figure}


\subsection{Dynamic Roughness Parameter}
\begin{figure}
    \centering
    \subfloat[\label{fig:alphat}]{\includegraphics[width = 0.5\textwidth]{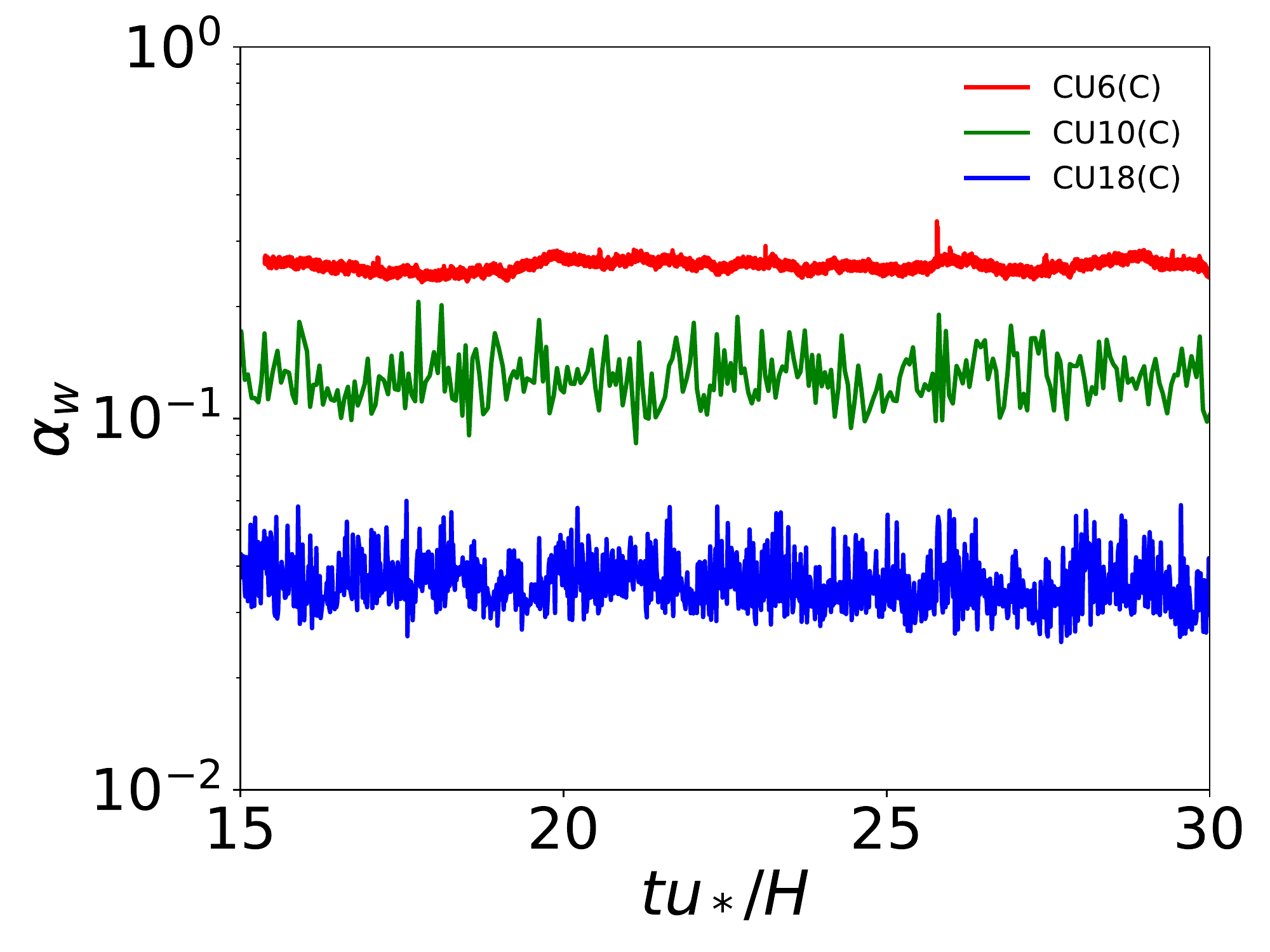}}
        \subfloat[\label{fig:alpha_av}]{\includegraphics[width = 0.5\textwidth]{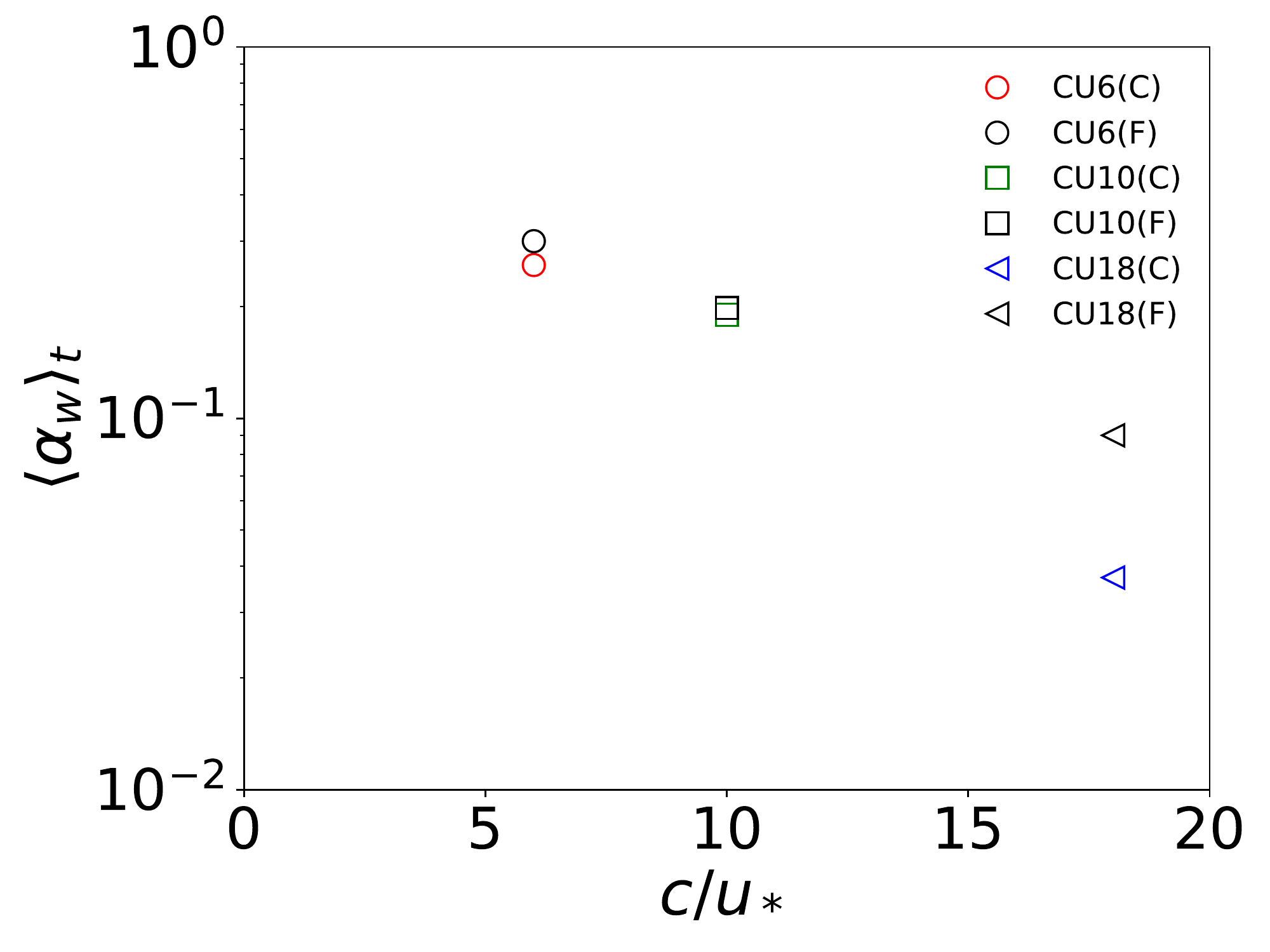}}

      \caption{a) Evolution of the dynamic coefficient $\alpha_w$ as a function of time and b) the time-averaged value for $c_p/u_* = 6$ (circles), $c_p/u_* = 10$ (squares), and $c_p/u_* = 18$ (triangles).}
    \label{fig:alph_characeristics}
\end{figure}

\begin{figure}
    \centering
    
    \includegraphics[width = 0.6\textwidth]{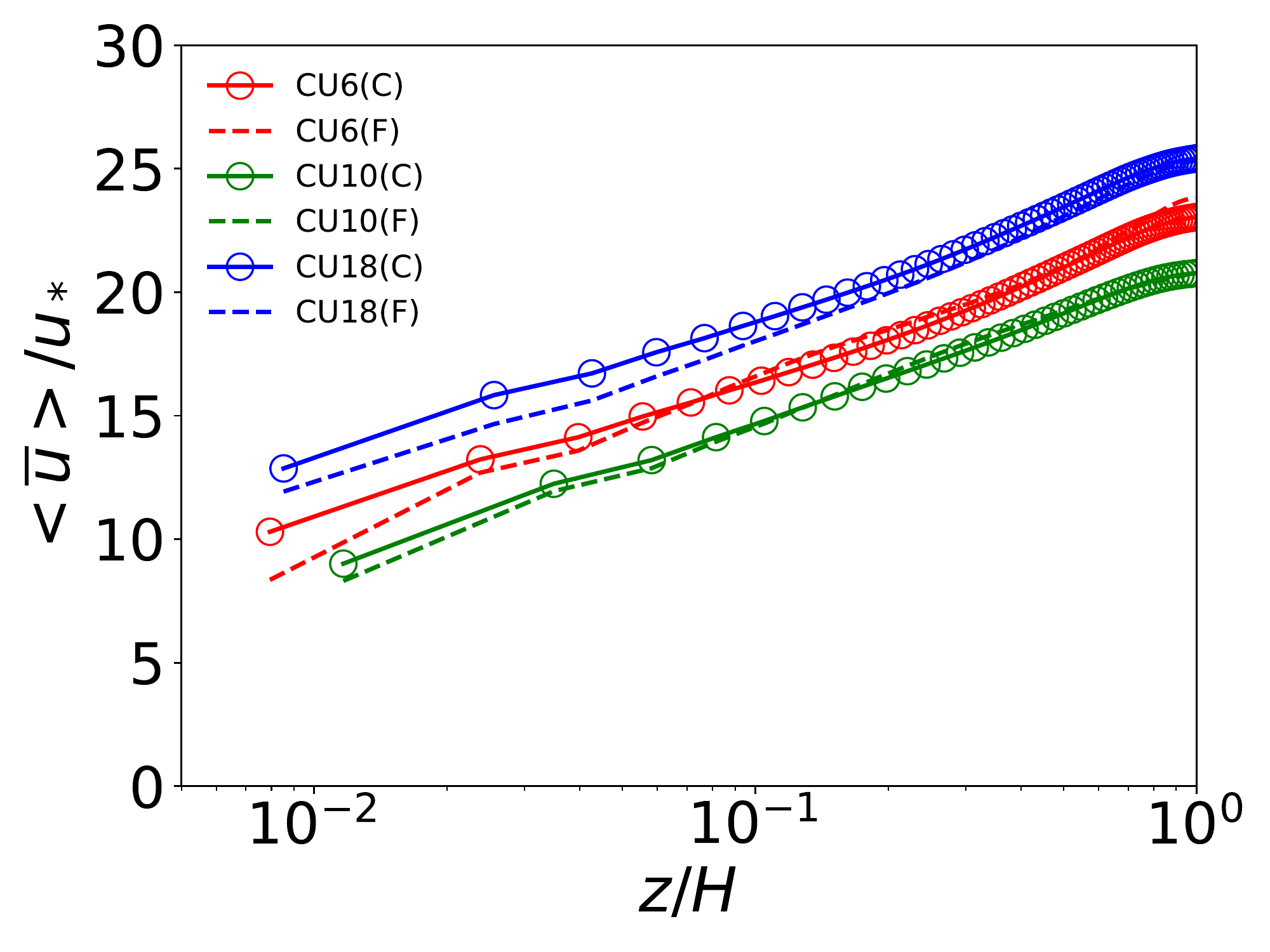}

    \caption{Non-dimensional mean streamwise velocity profiles in semi-logarithmic scale for flow over multiscale waves for $c/u_* = 6$, $c/u_* = 10$, and $c/u_* = 18$ using the Dyn-WaSp model.}
    \label{fig:Grid_compare}
\end{figure}
The dynamic procedure calculates the parameter $\alpha_w$ to ensure the total surface drag force at the filter scale $\Delta$ is equal to the total force at the test- filter scale $2\Delta$. Filtering at a larger test- filter scale reduces the resolved form drag as the surface is now smoother at the larger scale. However, since more fluctuations are subfilter, the unresolved roughness scale is larger leading to a correspondingly higher subfilter drag. The model relies on the existence of a unique value of $\alpha_w$ that equates the total force at both scales. 

Figure \ref{fig:alphat} shows the time history of the dynamic parameter $\alpha_w$. The value fluctuates around a steady state similar to that observed in \citet{Anderson2011} and \citet{Yang2013D}. A unique value of $\alpha_w$ is obtained in each case, without undesirable instabilities.  The time-averaged value of the dynamic parameter is shown as a function of wave age in Figure \ref{fig:alpha_av}. Here, $\alpha_w$ is calculated at two grid resolutions for each case. The time-averaged value of $\alpha_w$ is independent of filter size for the low wave age cases. The filter and the test filter scales for these cases lie in the saturation region of the normalized wave growth rate, leading to a scale-independent parameter. For the highest wave age $c_p/u_* = 18$, fewer drag producing waves are resolved, and the motion of the subfilter waves is non-negligible. Hence, the r.m.s model does not yield a value of the dynamic parameter that is independent of resolution. Different subfilter models that account for the wave phase speed or a scale-dependent filter could be used for the high wave-age cases and will be the subject of future studies. 
The mean velocity profiles for the different resolutions are shown in Figure \ref{fig:Grid_compare}. It is evident that the profiles are  independent of grid resolutions. Even for the case with $c_p/u_* = 18$ where the dynamic parameter showed a dependence on filter width, the dynamic model predicts velocity profiles that are grid independent.

In summary, the Dyn-WaSp model provides an accurate description of the airflow over the multiscale rough surface. The approach shows minimal grid dependence and benefits from a reduced computational cost of O(100), based on grid resolution, compared to phase-resolved simulations, making it ideal for large-scale marine atmospheric boundary layer simulations.

\section{LES of offshore wind farm}
\label{sec:farm}

In this section simulations of flow through an offshore wind farm with the bottom boundary prescribed as a multiscale wave field are presented. The Dyn-WaSp model is coupled with an actuator disk model for the wind turbines. The mean velocity profiles, bottom friction velocity, normalized farm power, and kinetic energy budgets are quantified.



    \subsection{Results and Discussions}

 In this section, statistics for the airflow flow over a spectrum of waves for different turbine configurations are presented.
\subsubsection{Velocity Profiles and Surface Roughness}
Temporally and spatially averaged mean streamwise velocity profiles for the monochromatic wave and spectrum case are shown in Figure \ref{fig:wind_mean}.  Phase-averaged LES over waves usually assume a constant value for the surface roughness $z_0 = 2\times 10^{-4}$ m
shown as the black dashed line in Figure \ref{fig:wind_mean} \cite{Goit2022}. Such a description is clearly insufficient to quantify the correct wave form stress or mean velocity profile and does not generalize to different wave conditions. For the sinusoidal case, the total drag exerted by the waves is dependent on the relative wind-wave velocity quantified by the wave age $c/u_*$, and the wave steepness $ak$. With the inclusion of the full wave spectrum, the flow field is reduced significantly due to the inclusion of slower wave modes resulting in a higher drag. 

With the addition of wind turbines, the flow slows down as energy is extracted from the flow by the wind farm. Figure \ref{fig:contour_plots}a shows the instantaneous $x-z$ contours of the streamwise velocity normalized by the friction velocity $u_*$ at the y-midplane for case SPWJ2.  There exists a large variability of instantaneous velocity around different turbines due to complex interaction between the wakes \cite{Wu2011,Yang2014,Porte-Agel2020}.
Figure \ref{fig:contour_plots}b depicts the time-averaged streamwise velocity where a wake region  due to the actuator disk forcing is observed behind the turbine extending to approximately $6$ D downstream. The variance of streamwise velocity $\overline{u^{\prime 2}}$ is shown in Figure \ref{fig:contour_plots}c. At the edge of the wake region, a shear layer is generated resulting in two high Reynolds stress regions starting from the rotor tip. 
\begin{figure}
    \centering
    \includegraphics[width = 0.7\textwidth]{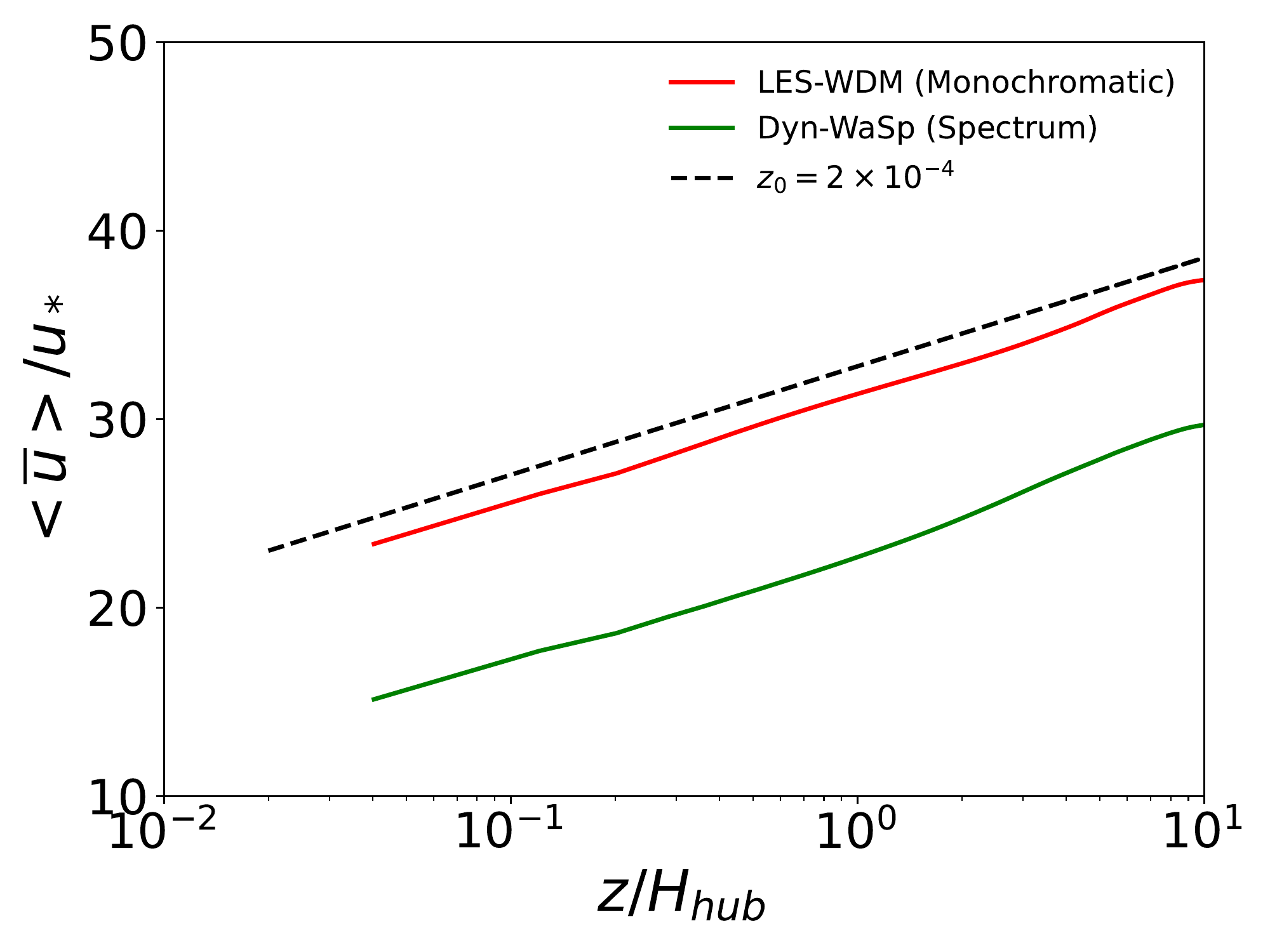}
        \caption{Temporally and spatially averaged mean streamwise velocity from the LES with Dyn-WaSp (Full Spectrum) model and LES with the Wave Drag Model (Monochromatic). Resolving more wave modes results in an increased surface roughness and slower velocity.   }
    \label{fig:wind_mean}
\end{figure}

\begin{figure}
    \centering  
    \subfloat[Instantaneous Streamwise Velocity]{\includegraphics[trim={0 3cm 0 3.5cm},clip,width =0.7\textwidth]{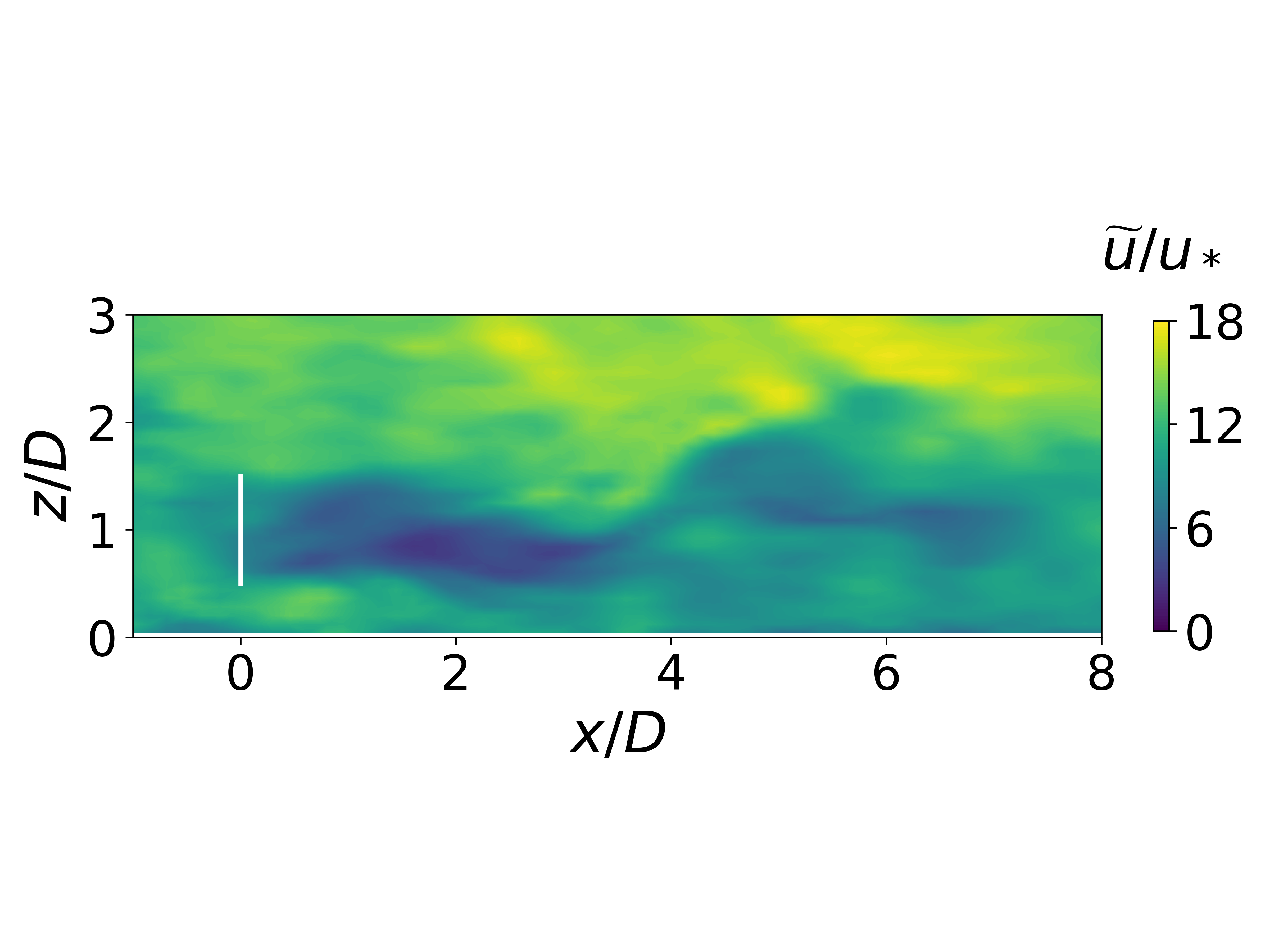}}
    \hfill
    \subfloat[Averaged Streamwise Velocity]{\includegraphics[trim={0 3cm 0 3.5cm},clip,width =0.7\textwidth]{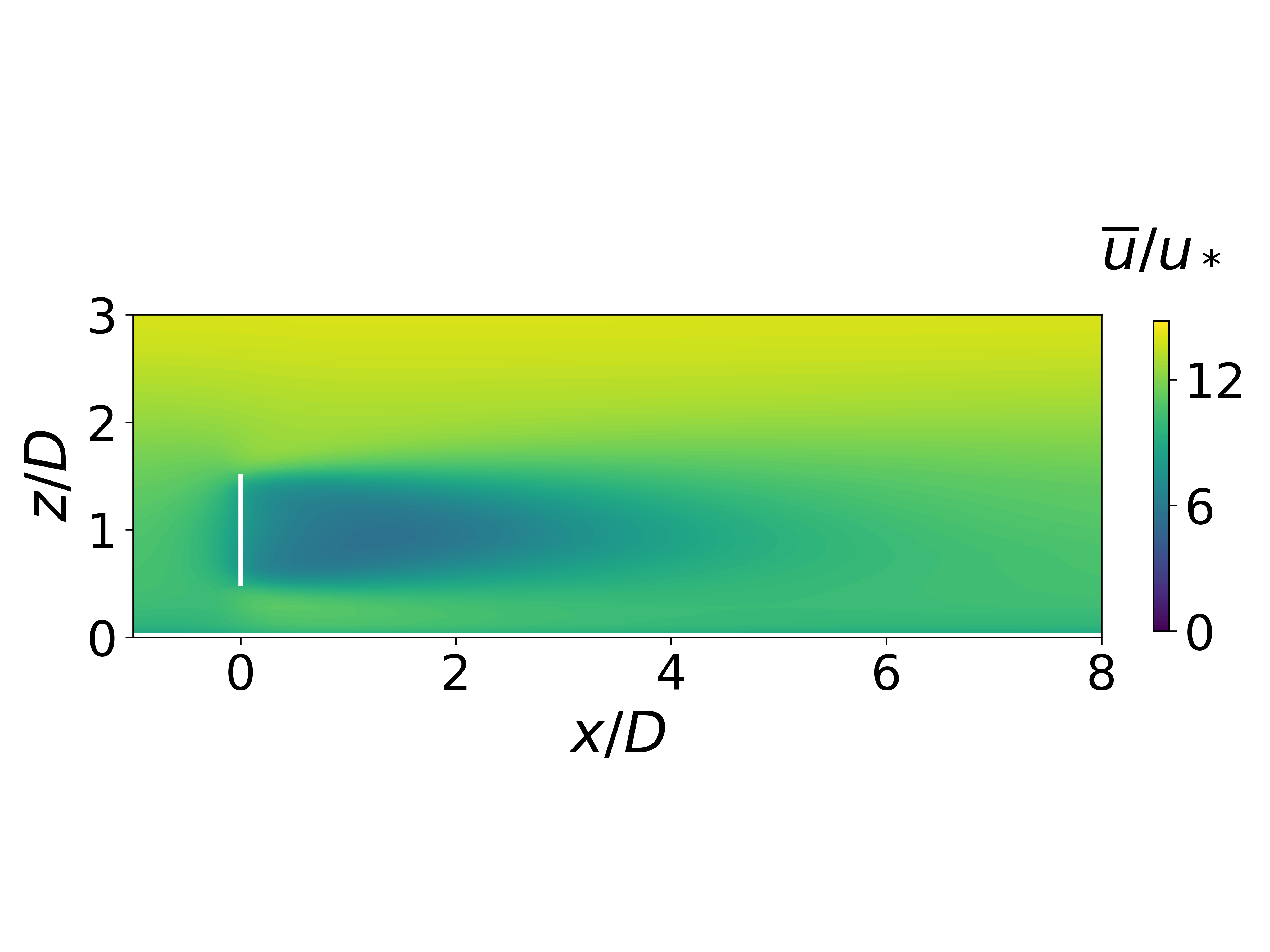}}
    \hfill
    \subfloat[Streamwise Velocity Variance]{\includegraphics[trim={0 3cm 0 3cm},clip,width =0.7\textwidth]{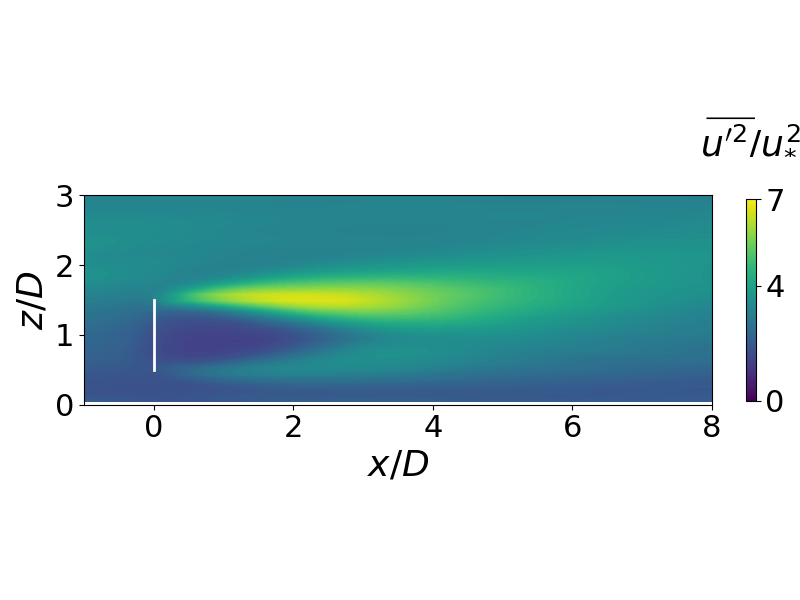}}
        \caption{x-z contours at the y midplane across the center of the actuator disk for the a) instantaneous streamwise velocity $\tilde{u}$, b) time-averaged streamwise velocity $\overline{\tilde{u}}$, and c) time-averaged variance of the streamwise velocity $\overline{u^{' 2}}$.}
    \label{fig:contour_plots}
\end{figure}

\begin{figure}
    \centering
    \includegraphics[width = 0.7\textwidth]{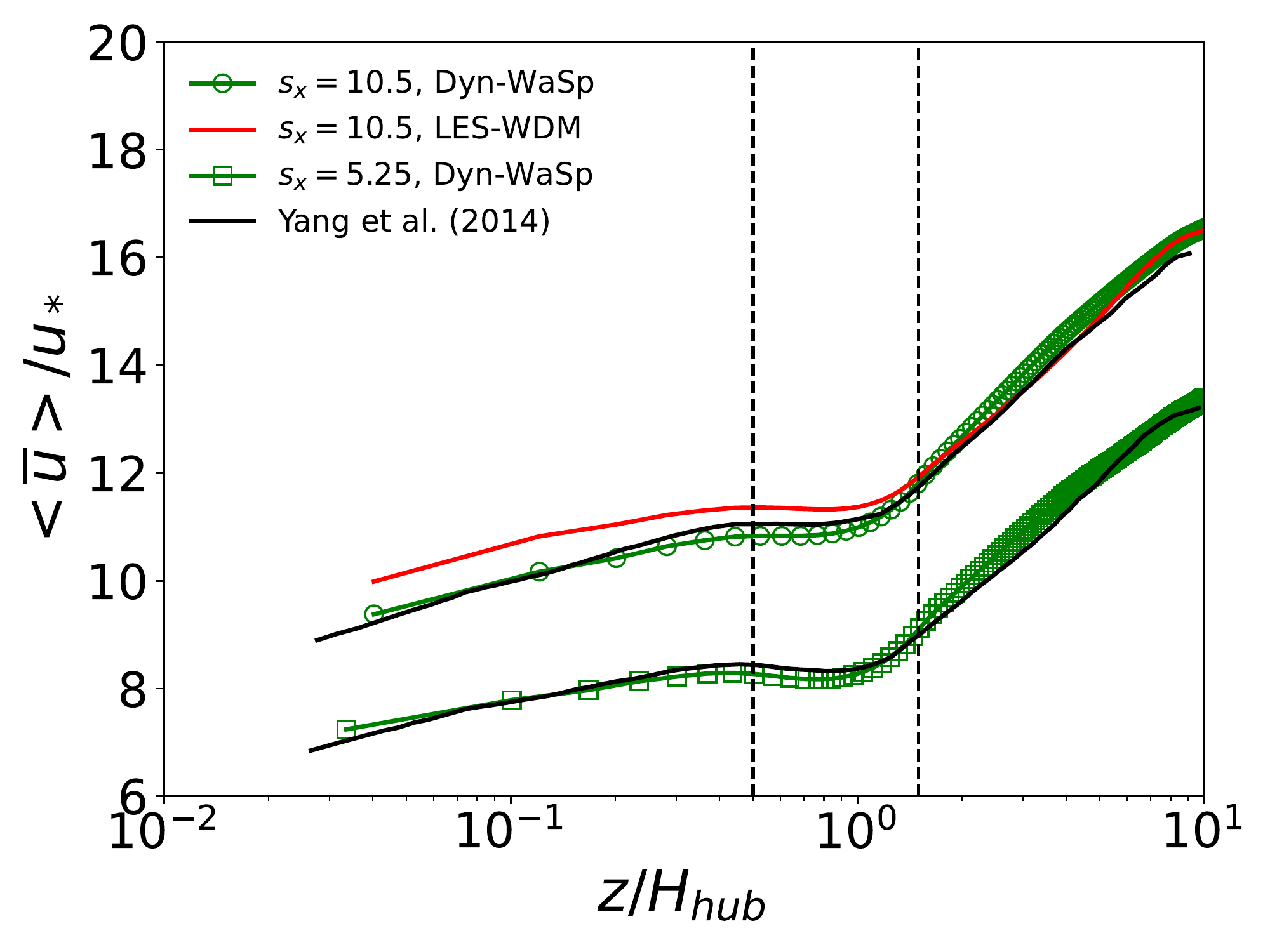}
        \caption{Temporally and spatially averaged velocity profiles for $c_p/u_* = 11$ from LES with the Dyn-WaSp model with $s_x = 10.5$ (green line with circles), $s_x = 5.25$ (green line with squares), LES with a monochromatic wave (red line), and phase-resolved LES \cite{Yang2014} (black lines). The turbine hub region is depicted by the vertical dashed black lines. }
    \label{fig:wind_turb_mean}
\end{figure}

The temporally and spatially- averaged streamwise velocity profiles from the different wind farm configurations, are depicted in Figure \ref{fig:wind_turb_mean}.  Additionally shown is the profile for a monochromatic wave. The profiles predicted from the Dyn-WaSp model for both cases show good agreement both below the wind turbine region and above the turbine layer. 
For a broad range of wave modes, such as observed in the ocean, characterizing the effect of the unresolved roughness scales, in addition to the effects of wave steepness and wave age is important. The Dyn-WaSp model accounts for the wave spectrum and does not rely on adhoc prescription of the surface roughness.

The velocity profiles in Figure \ref{fig:wind_turb_mean} depict two constant stress layers that emerge in the averaged streamwise velocity below and above the wind turbine layer \cite{Frandsen2006,Calaf2010}. The velocity profile in the layer below the wind turbines is controlled by the total surface stress $ \tau_{total}$ (both resolved and subfilter) or equivalently, the bottom friction velocity $u_{*,bot}=\sqrt{\tau_{total}/\rho}$ that depends on the waves and the wind turbines. Predicting this quantity will determine the accuracy in predicting the turbine velocity at hub height.
The values of the total surface stress as well as the bottom friction velocity are shown in Table \ref{tab:power_fric_vel} and compared to the phase-resolved simulations \cite{Yang2014OffshoreW}. The values obtained from the LES with the Dyn-WaSp model show good quantitative agreement with the phase-resolved simulations with a less than $5\%$  difference for the friction velocity and $10\%$ difference for the total stress.

\subsubsection{Wave form stress}
Figure \ref{fig:wind_turb_stress} shows the distribution of the resolved form stress $D_p = f_d(x,y,\Delta_z/2)\Delta_z$ normalized by the surface friction velocity $u_{*,bot}$ for the case without turbines and the two cases with turbines. For the case without turbines $u_{*,bot} = u_*$, and the waves that produce the most drag are the waves with the largest energy with wavenumber $k = k_p$. The total normalized resolved form stress is $18\%$ for the case without turbines. The flow decelerates as energy is extracted by the turbines resulting in a decreased friction velocity at the wave surface. The effect of the waves on the airflow is dependent on the difference between the airflow velocity and the wave phase speed, or equivalently the wave age $c/u_*$. In the presence of the turbines, the effective wave age $c/u_{*,bot}$ is larger compared to the case without turbines. As the drag force is proportional to the difference between the wind and the wave velocity (see Equation (\ref{eqn:fit_swell})), this higher wave age corresponds to a smaller drag. This is evident in the middle and right panels of Figure \ref{fig:wind_turb_stress} where the peak of resolved stress distribution shifts towards higher wavenumbers.  The maximum resolvable wavenumber $k_{max}\approx 2 k_p$, corresponds to a minimum resolvable wave age of $c_{min}/u_{*,bot}$. For the case without the wind turbines this corresponds to a minimum wave age of $\approx 7.7$, and slower waves are accounted for using the dynamic subfilter model. With the addition of turbines, the minimum wave age increases to $\approx 19$, and a larger fraction of the drag producing waves are subfilter. Without the dynamic approach, the effects of the waves would not be accounted for in the simulation resulting in an incorrect prediction of the surface stress. Additionally, in wind farms, the main drag producing waves spatially evolve depending on the changes in airflow, and the model gives a more accurate picture of the varying surface roughness.


\subsubsection{Power Density and Kinetic Energy Budget}

The power extracted from the flow by the wind turbines can be calculated based on the turbine induced force and the wind velocity. The power extracted by a single turbine can be written as 
\begin{equation}\label{eqn:pturb}
    P_{turb} = - \left(\frac{\pi D^2}{4} f_T \langle u\rangle_d \right). 
\end{equation}
Equation (\ref{eqn:pturb}) assumes that the effective power coefficient $C_p^{\prime} = C_T^{\prime}$, and should only be used for an inter- comparison between cases and not as an absolute prediction of power output.

The power density can be defined based on the streamwise and spanwise turbine spacings and the rotor diameter as
\begin{equation}
    \mathcal{P} = \frac{P_{turb}}{s_x s_y D^2}.
\end{equation}
The total performance of the wind farm can be quantified using the averaged power density over the whole wind farm:
\begin{equation}
    \mathcal{P}_T = \frac{1}{N_{row}N_{col}}\sum_{i=1}^{N_{row}}\sum_{j=1}^{N_{col}}\mathcal{P}_{ij}.
\end{equation}
The total power density of the farm increases with decreasing streamwise turbine spacing (i.e., for a denser layout). The power density is normalized with the velocity at the top of the boundary layer $U_H$ to facilitate comparison with the phase-resolved simulations \cite{Yang2014OffshoreW}. The LES with the Dyn-WaSp model accurately predicts the total normalized power density  and is quantified in Table \ref{tab:power_fric_vel}. 
\begin{center}
\begin{table}
\caption{\label{tab:power_fric_vel}Friction velocity and total normalized power from the current LES with the Dyn-WaSp model and the phase resolved simulations from \citet{Yang2014OffshoreW}}
\centering
\begin{ruledtabular}
    
\begin{tabular}{c c c c c c c }

Case &$u^{LES}_{*,bot}/u_*$&$u^{Yang}_{*,bot}/u_*$ & $\mathcal{P}_T^{LES}/U_H^3$ & $\mathcal{P}_T^{Yang}/U_H^3$ &$\langle \tau_{total}^{LES}\rangle_t/U_{H}^2$ & $\langle \tau_{total}^{Yang}\rangle_t/U_{H}^2$\\
\hline
SPWJ2& $0.447$ & $0.44$ & $1.57\times 10^{-3}$ & $1.59\times 10^{-3}$ & $8.18\times 10^{-4}$ & $7.2 \times 10^{-4}$ \\
SPWJ4& $0.36$ & $0.38$ & $2.5\times 10^{-3}$& $2.45\times 10^{-3}$ &$6.33\times 10^{-4}$ & $6.85\times 10^{-4}$\\

\end{tabular}
\end{ruledtabular}

\end{table}
\end{center}
\begin{figure}
    \centering
    \includegraphics[width = 0.8\textwidth]{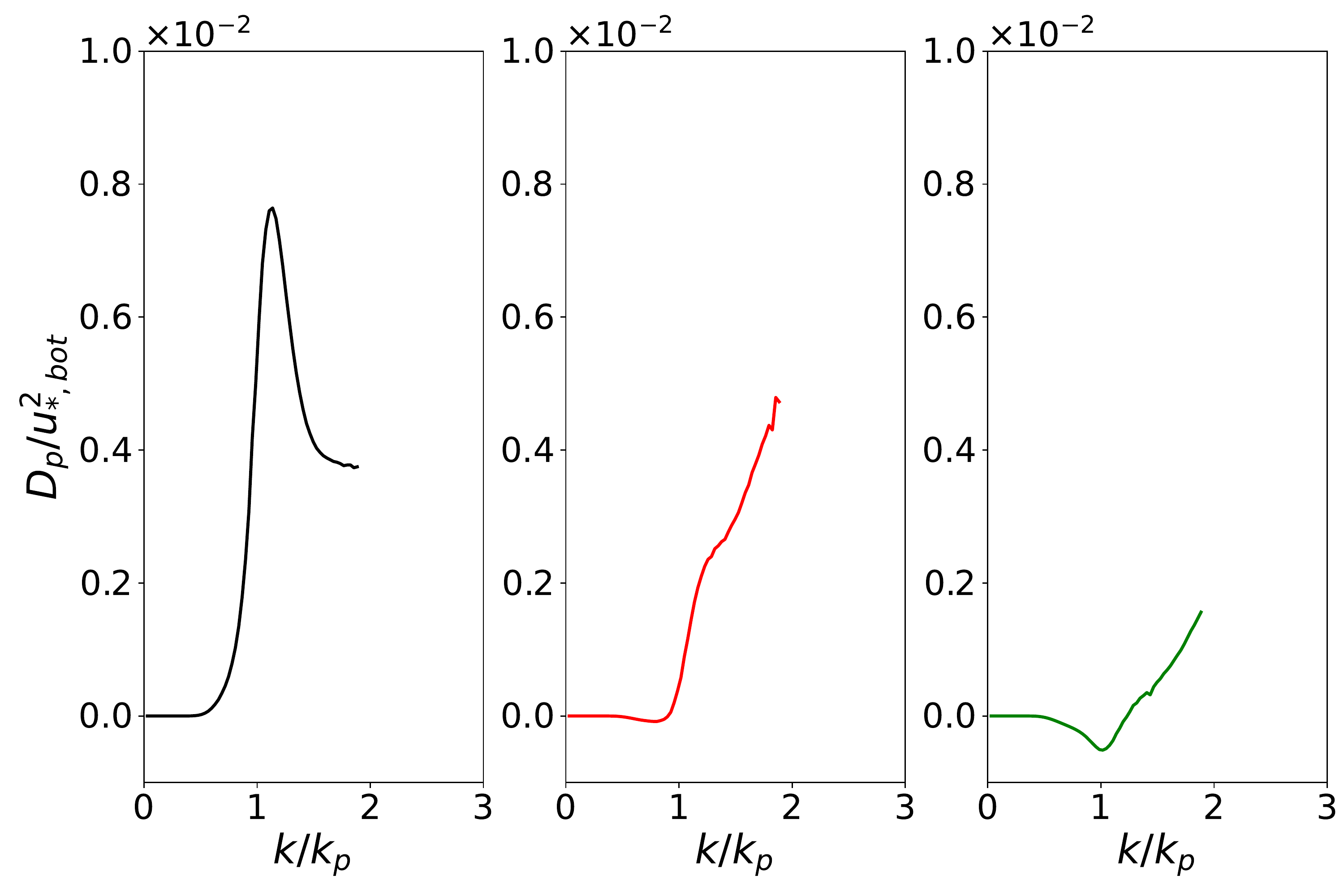}
        \caption{Normalized  wave form stress $D_p = f_d\Delta_z$ as a function of normalized wavenumber for a) airflow above a spectrum of moving waves without wind turbines and flow over a spectrum of waves with wind turbine configuration: b) $N_x\times N_y=2\times3$  and c) $N_x\times N_y = 4\times3$.}
    \label{fig:wind_turb_stress}
\end{figure}

Finally, the vertical transport of kinetic energy  for the temporally and spatially- averaged flow is quantified by multiplying the mean streamwise momentum equation by the spatially averaged streamwise velocity to obtain \cite{Calaf2010}

\begin{equation}
    -\frac{1}{\rho}\frac{dp_{\infty}}{dx}\langle \overline{u}\rangle + \frac{\partial}{\partial z}\left[\left(-\langle\overline{u'w'}\rangle - \langle\overline{u}^{\prime\prime}\overline{w}^{\prime\prime}\rangle\right)\langle \overline{u}\rangle\right] = \left[-\langle\overline{u'w'}\rangle - \langle\overline{u}^{\prime\prime}\overline{w}^{\prime\prime}\rangle\right]\frac{\partial \langle \overline{u}\rangle}{\partial z} - \langle \overline{f_T}\rangle\langle \overline{u}\rangle.
\end{equation}
The fluctuating velocity due to temporal variation is defined as $u_i^{\prime} = u_i - \overline{u}_i$ where $(\overline{\cdot})$ denotes time averaging and $\langle \cdot \rangle$ denotes spatial averaging. The fluctuating velocity due to both temporal and spatial variation is defined as $u_i^{\prime\prime} = u_i - \langle\overline{u}_i\rangle$.
The left hand side terms represent the work per unit time done by imposed pressure-gradient $\mathcal{W}_P$ that drives the flow and the vertical flux of kinetic energy $\Phi_e$ due to the resolved turbulent stress ($\langle\overline{u'w'}\rangle$) and the dispersive stresses ($\langle\overline{u}^{\prime\prime}\overline{w}^{\prime\prime}\rangle$). The first term on the right hand side is the dissipation of kinetic energy $\epsilon_e$, which is the rate at which kinetic energy of the averaged flow is being lost to production of turbulent kinetic energy, and the second term is the total power available to the wind turbine layer $\mathcal{W}_T$. The viscous dissipation and viscous transport terms are neglected due to the high Reynolds number of the flow. 

The vertical profiles of the terms in the mean kinetic energy budget equation are depicted in Figure \ref{fig:KE_budget}. The vertical flux of mean kinetic energy shown in Figure \ref{fig:ke_flux} is maximum at the upper edge of the wind turbine region and decreases towards the top of the domain. Kinetic energy is extracted from the flow by the turbines at the rotor height and transported from the faster mean flow above the turbines. This transport of kinetic energy from above is enhanced for the case with the denser layout due to the increase in number of turbines extracting more energy from the flow.
The turbine force $F_T$ is non-zero only within the turbine layer. The turbine-layer mean power $W_T$ (shown in Figure \ref{fig:turb_work}) increases smoothly from zero at the edge of the disks to the maximum values at the center. Turbulence production (mean kinetic energy dissipation) is large both on the upper portion of the turbines in the shear layer between the wakes and the fast flow above and near the wave surface where the velocity gradient is large. Similar to the kinetic energy flux, the dissipation of kinetic energy is larger for the denser layout.

To facilitate comparison with the phase-resolved simulations, the budget for the kinetic energy within the turbine layer spanning $z_{min} = H_{hub} - D/2$ to $z_{max} = H_{hub} + D/2$ is calculated\cite{Calaf2010}:

\begin{equation}
    \mathcal{W}_{P,D} + \Delta \phi_e = \mathcal{D} + \mathcal{W}_{T,D},
\end{equation}
where $\mathcal{W}_{P,D} = -D\rho_a^{-1} (dp_{\infty}/dx)\langle u\rangle_D$ is the forcing power due to the applied pressure gradient in the wind turbine region, where $\langle u\rangle_D = D^{-1}\int \langle\overline{u}\rangle \mathrm{d}z$ is the temporally and horizontally averaged velocity, additionally averaged vertically over the turbine diameter $D$, $\Delta \phi_e = \phi_e(z_{max}) - \phi_e(z_{min})$ is the flux of kinetic energy across the turbines,
\begin{equation}
    \mathcal{W}_{T,D} = \langle u\rangle_D \frac{\overline{F}_T}{\rho_a s_x s_y D^2},
\end{equation}
  where $F_T = 0.25\rho_a \pi D^2\Delta x f_T$  and 
 \begin{equation}
     \mathcal{D} = \int_{z_{min}}^{z_{max}}  \left[-\langle\overline{u'w'}\rangle - \langle\overline{u}^{\prime\prime}\overline{w}^{\prime\prime}\rangle\right]\frac{\partial \langle \overline{u}\rangle}{\partial z}\mathrm{d}z,
     \end{equation}
     is the dissipation term integrated over the wind turbine area.
     The terms in the total mean kinetic energy budget are quantified in Table \ref{tab:ke_budget}. Consistent with the trends from other studies \cite{Calaf2010,Yang2014OffshoreW}, the budget is dominated by $\Delta \phi_e$ and $\mathcal{W}_T$, while the dissipation and the forcing pressure work are an order of magnitude smaller. Furthermore, the magnitudes of the different terms show excellent agreement with the phase-resolved simulations. 
\begin{figure}
    \centering
    \subfloat[Kinetic Energy Flux\label{fig:ke_flux}]{\includegraphics[width=0.33\textwidth]{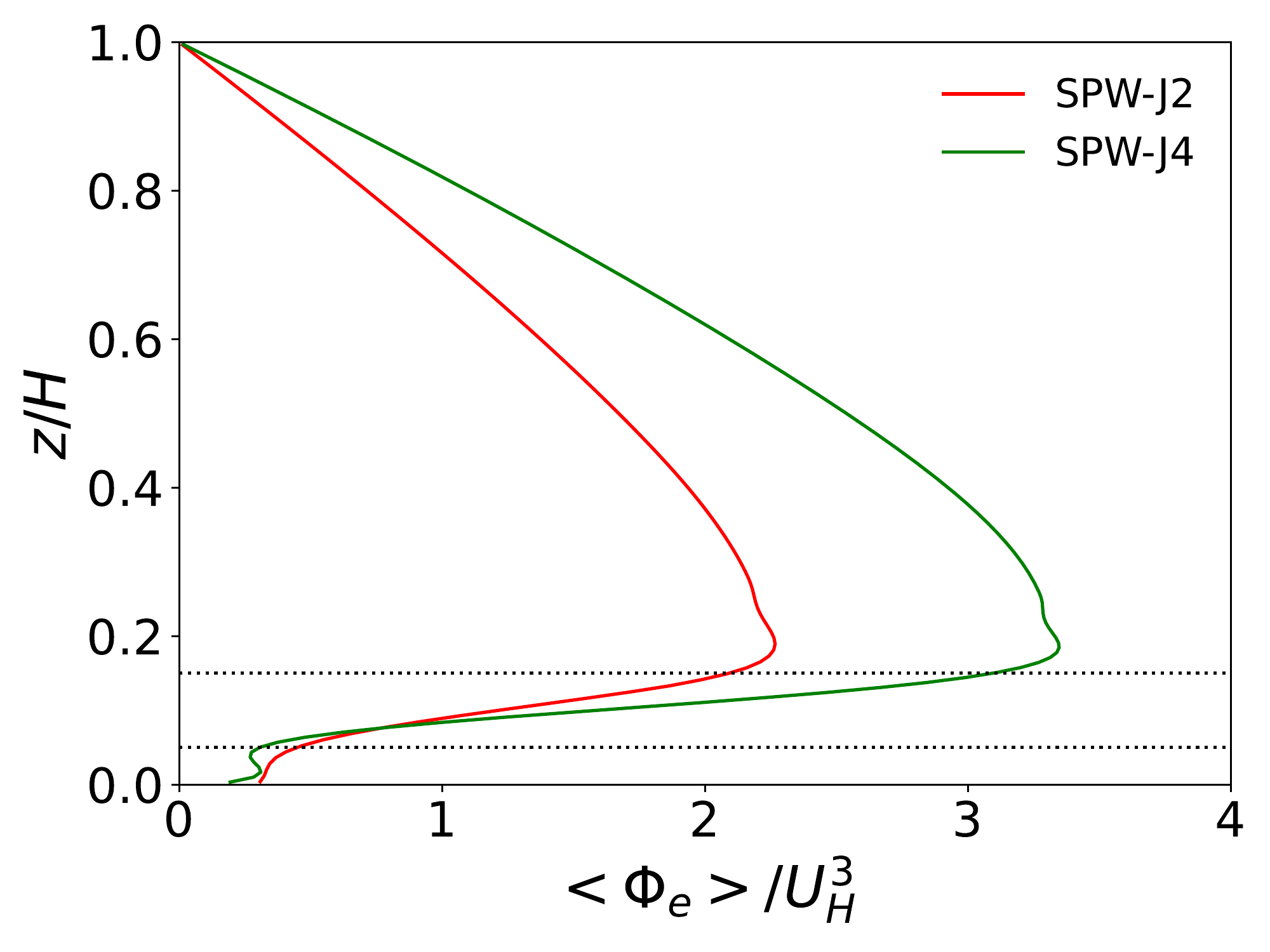}}
    \hfill
    \subfloat[\label{fig:ke_dissip}Kinetic Energy Dissipation]{\includegraphics[width=0.33\textwidth]{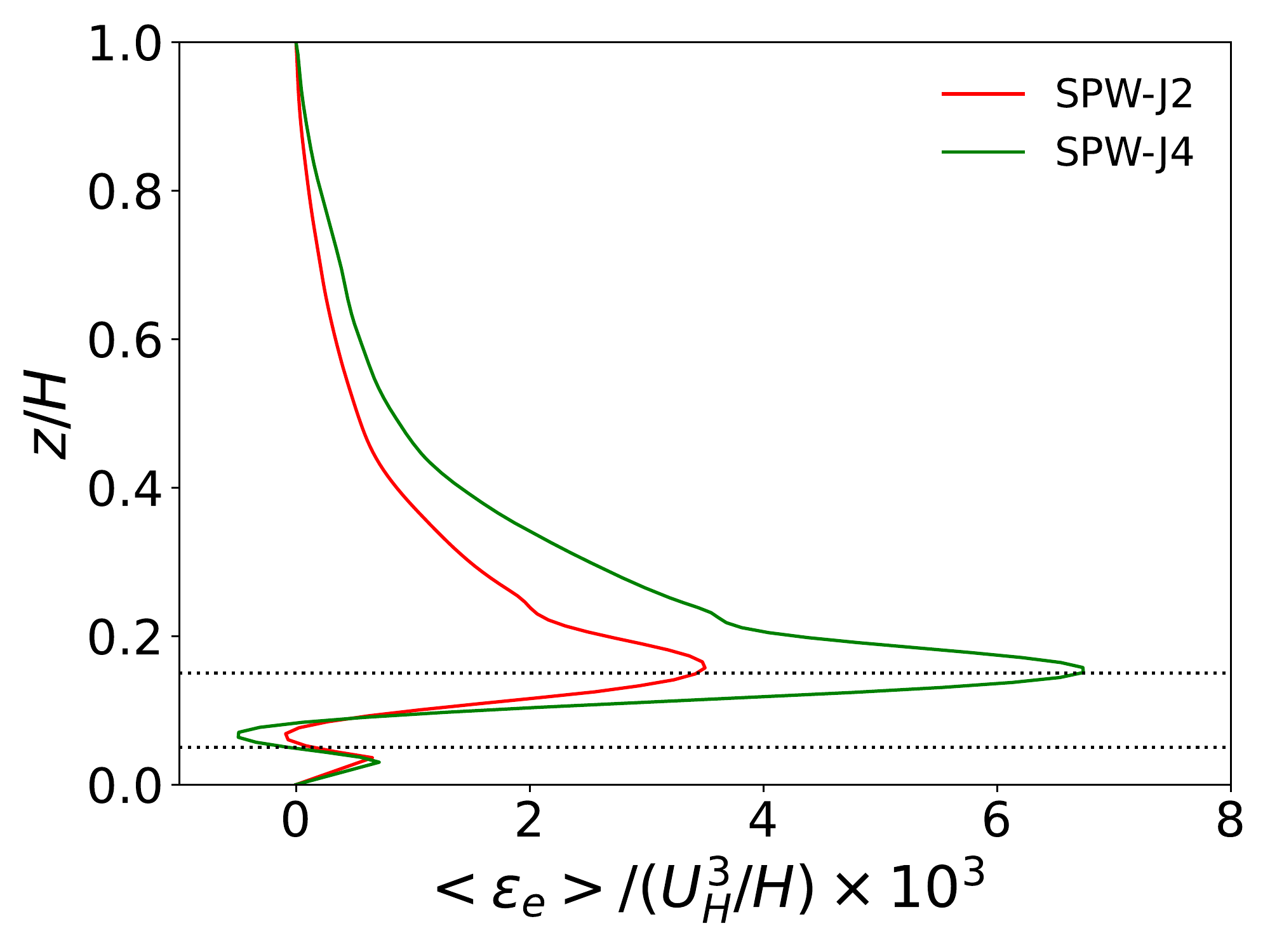}}
    \hfill
    \subfloat[Turbine Work\label{fig:turb_work}]{\includegraphics[width=0.33\textwidth]{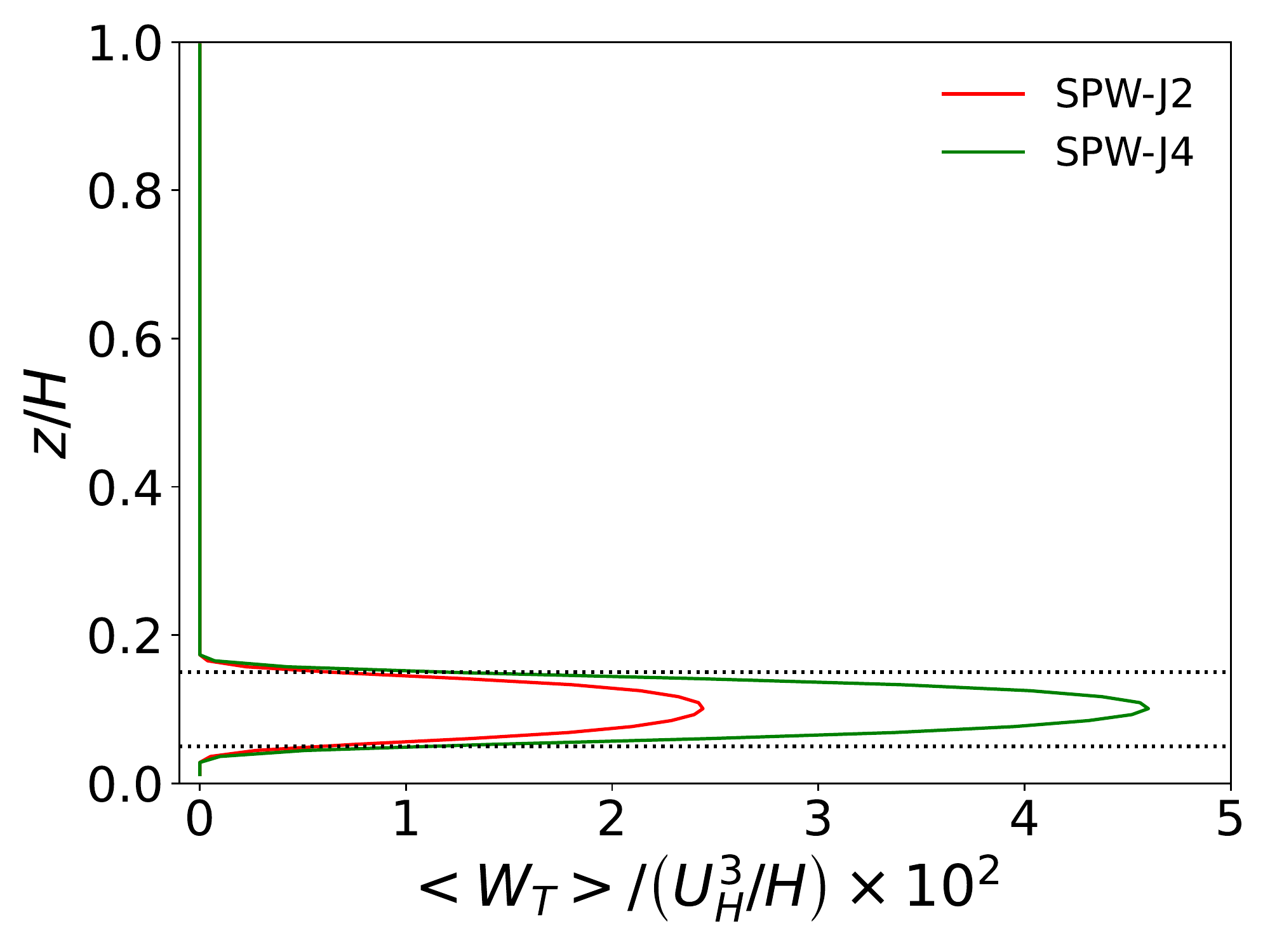}}
   
    \caption{Vertical profiles of normalized \protect\subref{fig:ke_flux}) mean kinetic energy flux, \protect\subref{fig:ke_dissip}) dissipation of mean kinetic energy, and \protect\subref{fig:turb_work}) mean turbine work for SPWJ2 (solid line) and SPWJ4 (dashed line). The wind turbine layer is depicted by the dotted horizontal lines.}
    \label{fig:KE_budget}
\end{figure}
\begin{center}
\begin{table}
\caption{\label{tab:ke_budget}Mean kinetic energy budget in the turbine region. All values are normalized with the velocity at the top of the domain $U_H^3$ and multiplied by $10^{3}$.  }
\centering
\begin{ruledtabular}
    
\begin{tabular}{c c c c c c c c c}

Case & $\mathcal{W}_P^{LES}$ & $\mathcal{W}_P^{Yang}$ & $\Delta \phi_e ^{LES}$ & $\Delta \phi_e ^{Yang}$ & $\mathcal{D}^{LES}$ &$\mathcal{D}^{Yang}$  & $\mathcal{W}_T^{LES}$& $\mathcal{W}_T^{Yang}$ \\
\hline
SPWJ2& $0.247$ & $0.253$ & $1.74$ & $1.76$ & $0.124$& $0.109$ & $1.94$ & $1.88$\\
SPWJ4& $0.347$ & $0.364$ & $2.81$ & $2.78$ & $0.218$& $0.177$  & $3.07$ & $2.92$

\end{tabular}
\end{ruledtabular}
\end{table}
\end{center}

The estimated reduction in computational cost can be quantified by considering the grid resolution and time steps from Ref. \cite{Yang2014OffshoreW}. In the current simulation, the grid resolution is about 2.5 times coarser in the horizontal directions and 1.5 times coarser in the vertical direction compared to Ref. \cite{Yang2014OffshoreW}. Additionally, a separate solver for the wave field is not used for the current approach. The simulations are run with a time step of $0.6$ s corresponding to a CFL number of $0.4$ as the simulation time step is not constrained by a wave solver. A $70$ times lower computational cost is achieved by the current modeling approach due to coarser grids and larger time steps ($0.6$ s compared to $0.08$ s in Ref. \cite{Yang2014OffshoreW}) without even considering the cost of a separate wave model, which would increase the relative computational cost savings with the current approach.


\section{Conclusions}
\label{sec:concl}

Offshore wind farms operate in a complex environment where the interaction between the waves, atmospheric boundary layer, and the wind turbines control the dynamics of the system. Realistic oceanic conditions comprise of multiscale and multi-modal waves, and accounting for their effects in a computationally efficient manner is important to accurately model the flow through offshore wind farms. 

In this work, a wave spectrum drag model applicable for Large Eddy Simulations is developed to calculate the wave form drag for a multiscale moving surface. The model is based on the incoming momentum flux, the relative wind-wave velocity, and the wave surface gradient. Furthermore, a parameterization to account for fast swell waves where the momentum transfer is from the waves to the airflow is proposed. To ensure accurate results for coarse grids where the entire wave spectrum cannot be resolved, the dynamic model proposed by \citet{Anderson2011} and \citet{Yang2013D}  is adapted to develop a  Dynamic Wave Spectrum drag model (Dyn-WaSp). The Dyn-WaSp model is used to simulate airflow over a spectrum of moving waves generated using the JONSWAP spectrum. The wave height distribution is calculated as a superposition of individual wave modes using a random phase model. The effect of resolved wave modes is applied as a bodyforce by adding the contribution due to each wave. The drag from subfilter waves is calculated dynamically using a model based on the subfilter r.m.s of the wave height distribution.  The model produces accurate results for both the mean velocity profiles and the wave stress for different wind and wave conditions. Further, the growth rate, a gauge used to quantify the pressure-based wave stress, was compared to existing experiments and simulations and showed good agreement. 

The Dyn-WaSp model is then applied to simulate flow through an offshore wind farm for two different streamwise turbine spacings, with the bottom boundary bounded by multiscale waves. The time-averaged mean velocity profiles show excellent agreement with phase-resolved simulations. Furthermore, compared to only using a monochromatic wave train, the full spectrum formulation provides superior predictions of the bottom friction velocity and the surface roughness. There is a factor of two difference between the averaged streamwise velocity for the monochromatic and full spectrum case as can be seen in Figure \ref{fig:wind_mean}.
However, for the case of finite length wind farms this effect will be more dramatic, and representing the full spectrum will be crucial. Finally, the total power density from the farm, and the mean kinetic energy budgets are calculated and compared to the phase-resolved counterparts with good agreement.

The dynamic parameter $\alpha_w$ adjusts to the changing flow field as the flow transitions from airflow over a spectrum of waves to a fully developed wind turbine boundary layer.
In all the cases considered here, the grid resolution was sufficient to model the most energetic modes. The Dyn-WaSp model adds minimal computational cost and implementation complexity and is free of system-dependent parameters. Accurately characterizing the wave field  can be used to inform the design and placement of wind turbines and quantify unsteady power generation due to platform motions \cite{Wei2022}. This includes understanding how the waves affect the substructures, such as the foundations and anchors, which can affect the stability of the wind turbines \cite{Zheng2023}.

Another key advantage of the current approach is the low computational cost compared to phase-resolved simulations. The model can be applied as a boundary condition and accurately models wave effects without the need of higher grid resolutions or a separate wave solver. The reduction in computational cost is O(100) based on grid resolution and time steps compared to phase-resolved simulations for the case without turbines. For the turbine simulations the current modeling approach achieves a O(10-100) times lower computational cost due to coarser grids, and larger time steps (compared to Ref. \cite{Yang2014OffshoreW}) without considering the cost of a separate wave model. 
Future studies will focus on the application of the dynamic approach to study wind-wake-wave couplings for different wave characteristics including the effects of swell waves in a finite wind farm setting.

\begin{acknowledgments}
The authors gratefully acknowledge financial support from the Princeton University Andlinger Center for Energy and the Environment and High Meadows Environmental Institute. The simulations presented in this article were performed on computational resources supported by the Princeton Institute for Computational Science and Engineering (PICSciE) and the Office of Information Technology's High Performance Computing Center and Visualization Laboratory at Princeton University.
\end{acknowledgments}

\section*{Data Availability Statement}

Data can be made available upon request.

\bibliography{references,references_2}

\begin{thebibliography}{68}%
\makeatletter
\providecommand \@ifxundefined [1]{%
 \@ifx{#1\undefined}
}%
\providecommand \@ifnum [1]{%
 \ifnum #1\expandafter \@firstoftwo
 \else \expandafter \@secondoftwo
 \fi
}%
\providecommand \@ifx [1]{%
 \ifx #1\expandafter \@firstoftwo
 \else \expandafter \@secondoftwo
 \fi
}%
\providecommand \natexlab [1]{#1}%
\providecommand \enquote  [1]{``#1''}%
\providecommand \bibnamefont  [1]{#1}%
\providecommand \bibfnamefont [1]{#1}%
\providecommand \citenamefont [1]{#1}%
\providecommand \href@noop [0]{\@secondoftwo}%
\providecommand \href [0]{\begingroup \@sanitize@url \@href}%
\providecommand \@href[1]{\@@startlink{#1}\@@href}%
\providecommand \@@href[1]{\endgroup#1\@@endlink}%
\providecommand \@sanitize@url [0]{\catcode `\\12\catcode `\$12\catcode
  `\&12\catcode `\#12\catcode `\^12\catcode `\_12\catcode `\%12\relax}%
\providecommand \@@startlink[1]{}%
\providecommand \@@endlink[0]{}%
\providecommand \url  [0]{\begingroup\@sanitize@url \@url }%
\providecommand \@url [1]{\endgroup\@href {#1}{\urlprefix }}%
\providecommand \urlprefix  [0]{URL }%
\providecommand \Eprint [0]{\href }%
\providecommand \doibase [0]{http://dx.doi.org/}%
\providecommand \selectlanguage [0]{\@gobble}%
\providecommand \bibinfo  [0]{\@secondoftwo}%
\providecommand \bibfield  [0]{\@secondoftwo}%
\providecommand \translation [1]{[#1]}%
\providecommand \BibitemOpen [0]{}%
\providecommand \bibitemStop [0]{}%
\providecommand \bibitemNoStop [0]{.\EOS\space}%
\providecommand \EOS [0]{\spacefactor3000\relax}%
\providecommand \BibitemShut  [1]{\csname bibitem#1\endcsname}%
\let\auto@bib@innerbib\@empty
\bibitem [{\citenamefont {Abkar}\ and\ \citenamefont {Moin}(2017)}]{Akbar2017}%
  \BibitemOpen
  \bibfield  {author} {\bibinfo {author} {\bibnamefont {Abkar}, \bibfnamefont
  {M.}}\ and\ \bibinfo {author} {\bibnamefont {Moin}, \bibfnamefont {P.}},\
  }\bibfield  {title} {\enquote {\bibinfo {title} {Large-eddy simulation of
  thermally stratified atmospheric boundary-layer flow using a minimum
  dissipation model},}\ }\href {\doibase 10.1007/s10546-017-0288-4} {\bibfield
  {journal} {\bibinfo  {journal} {Boundary-Layer Meteorology}\ }\textbf
  {\bibinfo {volume} {165}},\ \bibinfo {pages} {405--419} (\bibinfo {year}
  {2017})}\BibitemShut {NoStop}%
\bibitem [{\citenamefont {Aiyer}, \citenamefont {Deike},\ and\ \citenamefont
  {Mueller}(2022{\natexlab{a}})}]{Aiyer2022}%
  \BibitemOpen
  \bibfield  {author} {\bibinfo {author} {\bibnamefont {Aiyer}, \bibfnamefont
  {A.~K.}}, \bibinfo {author} {\bibnamefont {Deike}, \bibfnamefont {L.}}, \
  and\ \bibinfo {author} {\bibnamefont {Mueller}, \bibfnamefont {M.~E.}},\
  }\bibfield  {title} {\enquote {\bibinfo {title} {{A sea surface-based drag
  model for Large Eddy Simulation of wind-wave interaction}},}\ }\href
  {\doibase 10.1175/jas-d-21-0329.1} {\bibfield  {journal} {\bibinfo  {journal}
  {Journal of the Atmospheric Sciences}\ ,\ \bibinfo {pages} {49--62}}
  (\bibinfo {year} {2022}{\natexlab{a}})}\BibitemShut {NoStop}%
\bibitem [{\citenamefont {Aiyer}, \citenamefont {Deike},\ and\ \citenamefont
  {Mueller}(2022{\natexlab{b}})}]{Aiyer2022T}%
  \BibitemOpen
  \bibfield  {author} {\bibinfo {author} {\bibnamefont {Aiyer}, \bibfnamefont
  {A.~K.}}, \bibinfo {author} {\bibnamefont {Deike}, \bibfnamefont {L.}}, \
  and\ \bibinfo {author} {\bibnamefont {Mueller}, \bibfnamefont {M.~E.}},\
  }\bibfield  {title} {\enquote {\bibinfo {title} {{A wall-modeled approach
  accounting for wave stress in Large Eddy Simulations of offshore wind
  farms}},}\ }in\ \href {\doibase 10.1088/1742-6596/2265/2/022013} {\emph
  {\bibinfo {booktitle} {Journal of Physics: Conference Series}}},\ Vol.\
  \bibinfo {volume} {2265}\ (\bibinfo  {publisher} {Institute of Physics},\
  \bibinfo {year} {2022})\BibitemShut {NoStop}%
\bibitem [{\citenamefont {Alsam}, \citenamefont {Szasz},\ and\ \citenamefont
  {Revstedt}(2015)}]{Alsam2015}%
  \BibitemOpen
  \bibfield  {author} {\bibinfo {author} {\bibnamefont {Alsam}, \bibfnamefont
  {A.}}, \bibinfo {author} {\bibnamefont {Szasz}, \bibfnamefont {R.}}, \ and\
  \bibinfo {author} {\bibnamefont {Revstedt}, \bibfnamefont {J.}},\ }\bibfield
  {title} {\enquote {\bibinfo {title} {{The influence of sea waves on offshore
  wind turbine aerodynamics}},}\ }\href {\doibase 10.1115/1.4031005} {\bibfield
   {journal} {\bibinfo  {journal} {Journal of Energy Resources Technology,
  Transactions of the ASME}\ }\textbf {\bibinfo {volume} {137}},\ \bibinfo
  {pages} {1--10} (\bibinfo {year} {2015})}\BibitemShut {NoStop}%
\bibitem [{\citenamefont {Anderson}\ and\ \citenamefont
  {Meneveau}(2011)}]{Anderson2011}%
  \BibitemOpen
  \bibfield  {author} {\bibinfo {author} {\bibnamefont {Anderson},
  \bibfnamefont {W.}}\ and\ \bibinfo {author} {\bibnamefont {Meneveau},
  \bibfnamefont {C.}},\ }\bibfield  {title} {\enquote {\bibinfo {title}
  {{Dynamic roughness model for large-eddy simulation of turbulent flow over
  multiscale, fractal-like rough surfaces}},}\ }\href {\doibase
  10.1017/jfm.2011.137} {\bibfield  {journal} {\bibinfo  {journal} {Journal of
  Fluid Mechanics}\ }\textbf {\bibinfo {volume} {679}},\ \bibinfo {pages}
  {288--314} (\bibinfo {year} {2011})}\BibitemShut {NoStop}%
\bibitem [{\citenamefont {Barthelmie}\ \emph {et~al.}(2009)\citenamefont
  {Barthelmie}, \citenamefont {Hansen}, \citenamefont {Frandsen}, \citenamefont
  {Rathmann}, \citenamefont {Schepers}, \citenamefont {Schlez}, \citenamefont
  {Phillips}, \citenamefont {Rados}, \citenamefont {Zervos}, \citenamefont
  {Politis},\ and\ \citenamefont {Chaviaropoulos}}]{Barthelmie2009}%
  \BibitemOpen
  \bibfield  {author} {\bibinfo {author} {\bibnamefont {Barthelmie},
  \bibfnamefont {R.~J.}}, \bibinfo {author} {\bibnamefont {Hansen},
  \bibfnamefont {K.}}, \bibinfo {author} {\bibnamefont {Frandsen},
  \bibfnamefont {S.~T.}}, \bibinfo {author} {\bibnamefont {Rathmann},
  \bibfnamefont {O.}}, \bibinfo {author} {\bibnamefont {Schepers},
  \bibfnamefont {J.~G.}}, \bibinfo {author} {\bibnamefont {Schlez},
  \bibfnamefont {W.}}, \bibinfo {author} {\bibnamefont {Phillips},
  \bibfnamefont {J.}}, \bibinfo {author} {\bibnamefont {Rados}, \bibfnamefont
  {K.}}, \bibinfo {author} {\bibnamefont {Zervos}, \bibfnamefont {A.}},
  \bibinfo {author} {\bibnamefont {Politis}, \bibfnamefont {E.~S.}}, \ and\
  \bibinfo {author} {\bibnamefont {Chaviaropoulos}, \bibfnamefont {P.~K.}},\
  }\bibfield  {title} {\enquote {\bibinfo {title} {{Modelling and measuring
  flow and wind turbine wakes in large wind farms offshore}},}\ }\href
  {\doibase 10.1002/we.348} {\bibfield  {journal} {\bibinfo  {journal} {Wind
  Energy}\ }\textbf {\bibinfo {volume} {12}},\ \bibinfo {pages} {431--444}
  (\bibinfo {year} {2009})}\BibitemShut {NoStop}%
\bibitem [{\citenamefont {Belcher}\ and\ \citenamefont
  {Hunt}(1998)}]{Belcher1998}%
  \BibitemOpen
  \bibfield  {author} {\bibinfo {author} {\bibnamefont {Belcher}, \bibfnamefont
  {S.~E.}}\ and\ \bibinfo {author} {\bibnamefont {Hunt}, \bibfnamefont
  {J.~C.~R.}},\ }\bibfield  {title} {\enquote {\bibinfo {title} {Turbulent flow
  over hills and waves},}\ }\href {\doibase 10.1146/annurev.fluid.30.1.507}
  {\bibfield  {journal} {\bibinfo  {journal} {Annual Review of Fluid
  Mechanics}\ }\textbf {\bibinfo {volume} {30}},\ \bibinfo {pages} {507--538}
  (\bibinfo {year} {1998})},\ \Eprint
  {http://arxiv.org/abs/https://doi.org/10.1146/annurev.fluid.30.1.507}
  {https://doi.org/10.1146/annurev.fluid.30.1.507} \BibitemShut {NoStop}%
\bibitem [{\citenamefont {Buckley}, \citenamefont {Veron},\ and\ \citenamefont
  {Yousefi}(2020)}]{Buckley2020}%
  \BibitemOpen
  \bibfield  {author} {\bibinfo {author} {\bibnamefont {Buckley}, \bibfnamefont
  {M.~P.}}, \bibinfo {author} {\bibnamefont {Veron}, \bibfnamefont {F.}}, \
  and\ \bibinfo {author} {\bibnamefont {Yousefi}, \bibfnamefont {K.}},\
  }\bibfield  {title} {\enquote {\bibinfo {title} {{Surface viscous stress over
  wind-driven waves with intermittent airflow separation}},}\ }\href {\doibase
  10.1017/jfm.2020.760} {\bibfield  {journal} {\bibinfo  {journal} {Journal of
  Fluid Mechanics}\ }\textbf {\bibinfo {volume} {905}} (\bibinfo {year}
  {2020}),\ 10.1017/jfm.2020.760}\BibitemShut {NoStop}%
\bibitem [{\citenamefont {Burton}\ \emph {et~al.}(2011)\citenamefont {Burton},
  \citenamefont {Jenkins}, \citenamefont {Sharpe},\ and\ \citenamefont
  {Bossanyi}}]{Burton}%
  \BibitemOpen
  \bibfield  {author} {\bibinfo {author} {\bibnamefont {Burton}, \bibfnamefont
  {T.}}, \bibinfo {author} {\bibnamefont {Jenkins}, \bibfnamefont {N.}},
  \bibinfo {author} {\bibnamefont {Sharpe}, \bibfnamefont {D.}}, \ and\
  \bibinfo {author} {\bibnamefont {Bossanyi}, \bibfnamefont {E.}},\ }\href@noop
  {} {\emph {\bibinfo {title} {Wind energy handbook}}}\ (\bibinfo  {publisher}
  {John Wiley \& Sons},\ \bibinfo {year} {2011})\BibitemShut {NoStop}%
\bibitem [{\citenamefont {Calaf}, \citenamefont {Meneveau},\ and\ \citenamefont
  {Meyers}(2010)}]{Calaf2010}%
  \BibitemOpen
  \bibfield  {author} {\bibinfo {author} {\bibnamefont {Calaf}, \bibfnamefont
  {M.}}, \bibinfo {author} {\bibnamefont {Meneveau}, \bibfnamefont {C.}}, \
  and\ \bibinfo {author} {\bibnamefont {Meyers}, \bibfnamefont {J.}},\
  }\bibfield  {title} {\enquote {\bibinfo {title} {{Large eddy simulation study
  of fully developed wind-turbine array boundary layers}},}\ }\href {\doibase
  10.1063/1.3291077} {\bibfield  {journal} {\bibinfo  {journal} {Physics of
  Fluids}\ }\textbf {\bibinfo {volume} {22}},\ \bibinfo {pages} {015110}
  (\bibinfo {year} {2010})}\BibitemShut {NoStop}%
\bibitem [{\citenamefont {Cao}, \citenamefont {Deng},\ and\ \citenamefont
  {Shen}(2020)}]{Cao2020}%
  \BibitemOpen
  \bibfield  {author} {\bibinfo {author} {\bibnamefont {Cao}, \bibfnamefont
  {T.}}, \bibinfo {author} {\bibnamefont {Deng}, \bibfnamefont {B.~Q.}}, \ and\
  \bibinfo {author} {\bibnamefont {Shen}, \bibfnamefont {L.}},\ }\bibfield
  {title} {\enquote {\bibinfo {title} {{A simulation-based mechanistic study of
  turbulent wind blowing over opposing water waves}},}\ }\href {\doibase
  10.1017/jfm.2020.591} {\bibfield  {journal} {\bibinfo  {journal} {Journal of
  Fluid Mechanics}\ } (\bibinfo {year} {2020}),\
  10.1017/jfm.2020.591}\BibitemShut {NoStop}%
\bibitem [{\citenamefont {Cao}\ and\ \citenamefont {Shen}(2021)}]{Cao2021}%
  \BibitemOpen
  \bibfield  {author} {\bibinfo {author} {\bibnamefont {Cao}, \bibfnamefont
  {T.}}\ and\ \bibinfo {author} {\bibnamefont {Shen}, \bibfnamefont {L.}},\
  }\bibfield  {title} {\enquote {\bibinfo {title} {{A numerical and theoretical
  study of wind over fast-propagating water waves}},}\ }\href {\doibase
  10.1017/jfm.2021.416} {\bibfield  {journal} {\bibinfo  {journal} {Journal of
  Fluid Mechanics}\ }\textbf {\bibinfo {volume} {919}},\ \bibinfo {pages}
  {1--35} (\bibinfo {year} {2021})}\BibitemShut {NoStop}%
\bibitem [{\citenamefont {Castro-Santos}, \citenamefont {Martins},\ and\
  \citenamefont {Guedes~Soares}(2017)}]{Castro-Santos2017}%
  \BibitemOpen
  \bibfield  {author} {\bibinfo {author} {\bibnamefont {Castro-Santos},
  \bibfnamefont {L.}}, \bibinfo {author} {\bibnamefont {Martins}, \bibfnamefont
  {E.}}, \ and\ \bibinfo {author} {\bibnamefont {Guedes~Soares}, \bibfnamefont
  {C.}},\ }\bibfield  {title} {\enquote {\bibinfo {title} {{Economic comparison
  of technological alternatives to harness offshore wind and wave energies}},}\
  }\href {\doibase 10.1016/j.energy.2017.08.103} {\bibfield  {journal}
  {\bibinfo  {journal} {Energy}\ }\textbf {\bibinfo {volume} {140}},\ \bibinfo
  {pages} {1121--1130} (\bibinfo {year} {2017})}\BibitemShut {NoStop}%
\bibitem [{\citenamefont {Cavaleri}\ \emph {et~al.}(2007)\citenamefont
  {Cavaleri}, \citenamefont {Alves}, \citenamefont {Ardhuin}, \citenamefont
  {Babanin}, \citenamefont {Banner}, \citenamefont {Belibassakis},
  \citenamefont {Benoit}, \citenamefont {Donelan}, \citenamefont {Groeneweg},
  \citenamefont {Herbers}, \citenamefont {Hwang}, \citenamefont {Janssen},
  \citenamefont {Janssen}, \citenamefont {Lavrenov}, \citenamefont {Magne},
  \citenamefont {Monbaliu}, \citenamefont {Onorato}, \citenamefont {Polnikov},
  \citenamefont {Resio}, \citenamefont {Rogers}, \citenamefont {Sheremet},
  \citenamefont {{McKee Smith}}, \citenamefont {Tolman}, \citenamefont {{van
  Vledder}}, \citenamefont {Wolf},\ and\ \citenamefont
  {Young}}]{CAVALERI2007603}%
  \BibitemOpen
  \bibfield  {author} {\bibinfo {author} {\bibnamefont {Cavaleri},
  \bibfnamefont {L.}}, \bibinfo {author} {\bibnamefont {Alves}, \bibfnamefont
  {J.-H.}}, \bibinfo {author} {\bibnamefont {Ardhuin}, \bibfnamefont {F.}},
  \bibinfo {author} {\bibnamefont {Babanin}, \bibfnamefont {A.}}, \bibinfo
  {author} {\bibnamefont {Banner}, \bibfnamefont {M.}}, \bibinfo {author}
  {\bibnamefont {Belibassakis}, \bibfnamefont {K.}}, \bibinfo {author}
  {\bibnamefont {Benoit}, \bibfnamefont {M.}}, \bibinfo {author} {\bibnamefont
  {Donelan}, \bibfnamefont {M.}}, \bibinfo {author} {\bibnamefont {Groeneweg},
  \bibfnamefont {J.}}, \bibinfo {author} {\bibnamefont {Herbers}, \bibfnamefont
  {T.}}, \bibinfo {author} {\bibnamefont {Hwang}, \bibfnamefont {P.}}, \bibinfo
  {author} {\bibnamefont {Janssen}, \bibfnamefont {P.}}, \bibinfo {author}
  {\bibnamefont {Janssen}, \bibfnamefont {T.}}, \bibinfo {author} {\bibnamefont
  {Lavrenov}, \bibfnamefont {I.}}, \bibinfo {author} {\bibnamefont {Magne},
  \bibfnamefont {R.}}, \bibinfo {author} {\bibnamefont {Monbaliu},
  \bibfnamefont {J.}}, \bibinfo {author} {\bibnamefont {Onorato}, \bibfnamefont
  {M.}}, \bibinfo {author} {\bibnamefont {Polnikov}, \bibfnamefont {V.}},
  \bibinfo {author} {\bibnamefont {Resio}, \bibfnamefont {D.}}, \bibinfo
  {author} {\bibnamefont {Rogers}, \bibfnamefont {W.}}, \bibinfo {author}
  {\bibnamefont {Sheremet}, \bibfnamefont {A.}}, \bibinfo {author}
  {\bibnamefont {{McKee Smith}}, \bibfnamefont {J.}}, \bibinfo {author}
  {\bibnamefont {Tolman}, \bibfnamefont {H.}}, \bibinfo {author} {\bibnamefont
  {{van Vledder}}, \bibfnamefont {G.}}, \bibinfo {author} {\bibnamefont {Wolf},
  \bibfnamefont {J.}}, \ and\ \bibinfo {author} {\bibnamefont {Young},
  \bibfnamefont {I.}},\ }\bibfield  {title} {\enquote {\bibinfo {title} {Wave
  modelling -- the state of the art},}\ }\href {\doibase
  https://doi.org/10.1016/j.pocean.2007.05.005} {\bibfield  {journal} {\bibinfo
   {journal} {Progress in Oceanography}\ }\textbf {\bibinfo {volume} {75}},\
  \bibinfo {pages} {603--674} (\bibinfo {year} {2007})}\BibitemShut {NoStop}%
\bibitem [{\citenamefont {Charnock}(1955)}]{Charnock1955}%
  \BibitemOpen
  \bibfield  {author} {\bibinfo {author} {\bibnamefont {Charnock},
  \bibfnamefont {H.}},\ }\bibfield  {title} {\enquote {\bibinfo {title} {Wind
  stress on a water surface},}\ }\href {\doibase
  https://doi.org/10.1002/qj.49708135027} {\bibfield  {journal} {\bibinfo
  {journal} {Quarterly Journal of the Royal Meteorological Society}\ }\textbf
  {\bibinfo {volume} {81}},\ \bibinfo {pages} {639--640} (\bibinfo {year}
  {1955})}\BibitemShut {NoStop}%
\bibitem [{\citenamefont {Christiansen}\ and\ \citenamefont
  {Hasager}(2005)}]{Christiansen2005}%
  \BibitemOpen
  \bibfield  {author} {\bibinfo {author} {\bibnamefont {Christiansen},
  \bibfnamefont {M.~B.}}\ and\ \bibinfo {author} {\bibnamefont {Hasager},
  \bibfnamefont {C.~B.}},\ }\bibfield  {title} {\enquote {\bibinfo {title}
  {{Wake effects of large offshore wind farms identified from satellite
  SAR}},}\ }\href {\doibase 10.1016/j.rse.2005.07.009} {\bibfield  {journal}
  {\bibinfo  {journal} {Remote Sensing of Environment}\ }\textbf {\bibinfo
  {volume} {98}},\ \bibinfo {pages} {251--268} (\bibinfo {year}
  {2005})}\BibitemShut {NoStop}%
\bibitem [{\citenamefont {Desjardins}\ \emph {et~al.}(2008)\citenamefont
  {Desjardins}, \citenamefont {Blanquart}, \citenamefont {Balarac},\ and\
  \citenamefont {Pitsch}}]{desjardins2008}%
  \BibitemOpen
  \bibfield  {author} {\bibinfo {author} {\bibnamefont {Desjardins},
  \bibfnamefont {O.}}, \bibinfo {author} {\bibnamefont {Blanquart},
  \bibfnamefont {G.}}, \bibinfo {author} {\bibnamefont {Balarac}, \bibfnamefont
  {G.}}, \ and\ \bibinfo {author} {\bibnamefont {Pitsch}, \bibfnamefont {H.}},\
  }\bibfield  {title} {\enquote {\bibinfo {title} {{High order conservative
  finite difference scheme for variable density low Mach number turbulent
  flows}},}\ }\href {\doibase 10.1016/j.jcp.2008.03.027} {\bibfield  {journal}
  {\bibinfo  {journal} {Journal of Computational Physics}\ }\textbf {\bibinfo
  {volume} {227}},\ \bibinfo {pages} {7125--7159} (\bibinfo {year}
  {2008})}\BibitemShut {NoStop}%
\bibitem [{\citenamefont {Deskos}, \citenamefont {Ananthan},\ and\
  \citenamefont {Sprague}(2022)}]{DESKOS2022109029}%
  \BibitemOpen
  \bibfield  {author} {\bibinfo {author} {\bibnamefont {Deskos}, \bibfnamefont
  {G.}}, \bibinfo {author} {\bibnamefont {Ananthan}, \bibfnamefont {S.}}, \
  and\ \bibinfo {author} {\bibnamefont {Sprague}, \bibfnamefont {M.~A.}},\
  }\bibfield  {title} {\enquote {\bibinfo {title} {Direct numerical simulations
  of turbulent flow over misaligned traveling waves},}\ }\href {\doibase
  https://doi.org/10.1016/j.ijheatfluidflow.2022.109029} {\bibfield  {journal}
  {\bibinfo  {journal} {International Journal of Heat and Fluid Flow}\ }\textbf
  {\bibinfo {volume} {97}},\ \bibinfo {pages} {109029} (\bibinfo {year}
  {2022})}\BibitemShut {NoStop}%
\bibitem [{\citenamefont {Deskos}\ \emph {et~al.}(2021)\citenamefont {Deskos},
  \citenamefont {Lee}, \citenamefont {Draxl},\ and\ \citenamefont
  {Sprague}}]{Deskos2021ReviewLayer}%
  \BibitemOpen
  \bibfield  {author} {\bibinfo {author} {\bibnamefont {Deskos}, \bibfnamefont
  {G.}}, \bibinfo {author} {\bibnamefont {Lee}, \bibfnamefont {J.~C.~Y.}},
  \bibinfo {author} {\bibnamefont {Draxl}, \bibfnamefont {C.}}, \ and\ \bibinfo
  {author} {\bibnamefont {Sprague}, \bibfnamefont {M.~A.}},\ }\bibfield
  {title} {\enquote {\bibinfo {title} {{Review of wind-wave coupling models for
  large-eddy simulation of the marine atmospheric boundary layer}},}\ }\href
  {\doibase 10.1175/jas-d-21-0003.1} {\bibfield  {journal} {\bibinfo  {journal}
  {Journal of the Atmospheric Sciences}\ ,\ \bibinfo {pages} {1--75}} (\bibinfo
  {year} {2021})}\BibitemShut {NoStop}%
\bibitem [{\citenamefont {Donelan}(1990)}]{Donelan1990}%
  \BibitemOpen
  \bibfield  {author} {\bibinfo {author} {\bibnamefont {Donelan}, \bibfnamefont
  {M.~A.}},\ }\bibfield  {title} {\enquote {\bibinfo {title} {Air-sea
  interaction},}\ }\href@noop {} {\bibfield  {journal} {\bibinfo  {journal}
  {Ocean Engineering Science}\ }\textbf {\bibinfo {volume} {9B}},\ \bibinfo
  {pages} {239--292} (\bibinfo {year} {1990})}\BibitemShut {NoStop}%
\bibitem [{\citenamefont {Donelan}\ \emph {et~al.}(2006)\citenamefont
  {Donelan}, \citenamefont {Babanin}, \citenamefont {Young},\ and\
  \citenamefont {Banner}}]{Donelan2006}%
  \BibitemOpen
  \bibfield  {author} {\bibinfo {author} {\bibnamefont {Donelan}, \bibfnamefont
  {M.~A.}}, \bibinfo {author} {\bibnamefont {Babanin}, \bibfnamefont {A.~V.}},
  \bibinfo {author} {\bibnamefont {Young}, \bibfnamefont {I.~R.}}, \ and\
  \bibinfo {author} {\bibnamefont {Banner}, \bibfnamefont {M.~L.}},\ }\bibfield
   {title} {\enquote {\bibinfo {title} {{Wave-follower field measurements of
  the wind-input spectral function. Part II: Parameterization of the wind
  input}},}\ }\href {\doibase 10.1175/JPO2933.1} {\bibfield  {journal}
  {\bibinfo  {journal} {Journal of Physical Oceanography}\ }\textbf {\bibinfo
  {volume} {36}},\ \bibinfo {pages} {1672--1689} (\bibinfo {year}
  {2006})}\BibitemShut {NoStop}%
\bibitem [{\citenamefont {Donelan}\ and\ \citenamefont
  {Pierson}(1987)}]{Donelan1987}%
  \BibitemOpen
  \bibfield  {author} {\bibinfo {author} {\bibnamefont {Donelan}, \bibfnamefont
  {M.~A.}}\ and\ \bibinfo {author} {\bibnamefont {Pierson}, \bibfnamefont
  {W.~J.}},\ }\bibfield  {title} {\enquote {\bibinfo {title} {Radar scattering
  and equilibrium ranges in wind-generated waves with application to
  scatterometry},}\ }\href {\doibase https://doi.org/10.1029/JC092iC05p04971}
  {\bibfield  {journal} {\bibinfo  {journal} {Journal of Geophysical Research:
  Oceans}\ }\textbf {\bibinfo {volume} {92}},\ \bibinfo {pages} {4971--5029}
  (\bibinfo {year} {1987})},\ \Eprint
  {http://arxiv.org/abs/https://agupubs.onlinelibrary.wiley.com/doi/pdf/10.1029/JC092iC05p04971}
  {https://agupubs.onlinelibrary.wiley.com/doi/pdf/10.1029/JC092iC05p04971}
  \BibitemShut {NoStop}%
\bibitem [{\citenamefont {Drennan}\ \emph {et~al.}(2003)\citenamefont
  {Drennan}, \citenamefont {Graber}, \citenamefont {Hauser},\ and\
  \citenamefont {Quentin}}]{Drennan2003}%
  \BibitemOpen
  \bibfield  {author} {\bibinfo {author} {\bibnamefont {Drennan}, \bibfnamefont
  {W.~M.}}, \bibinfo {author} {\bibnamefont {Graber}, \bibfnamefont {H.~C.}},
  \bibinfo {author} {\bibnamefont {Hauser}, \bibfnamefont {D.}}, \ and\
  \bibinfo {author} {\bibnamefont {Quentin}, \bibfnamefont {C.}},\ }\bibfield
  {title} {\enquote {\bibinfo {title} {On the wave age dependence of wind
  stress over pure wind seas},}\ }\href {\doibase
  https://doi.org/10.1029/2000JC000715} {\bibfield  {journal} {\bibinfo
  {journal} {Journal of Geophysical Research: Oceans}\ }\textbf {\bibinfo
  {volume} {108}} (\bibinfo {year} {2003}),\
  https://doi.org/10.1029/2000JC000715}\BibitemShut {NoStop}%
\bibitem [{\citenamefont {Ducrozet}, \citenamefont {Bonnefoy},\ and\
  \citenamefont {Perignon}(2017)}]{DUCROZET2017233}%
  \BibitemOpen
  \bibfield  {author} {\bibinfo {author} {\bibnamefont {Ducrozet},
  \bibfnamefont {G.}}, \bibinfo {author} {\bibnamefont {Bonnefoy},
  \bibfnamefont {F.}}, \ and\ \bibinfo {author} {\bibnamefont {Perignon},
  \bibfnamefont {Y.}},\ }\bibfield  {title} {\enquote {\bibinfo {title}
  {Applicability and limitations of highly non-linear potential flow solvers in
  the context of water waves},}\ }\href {\doibase
  https://doi.org/10.1016/j.oceaneng.2017.07.003} {\bibfield  {journal}
  {\bibinfo  {journal} {Ocean Engineering}\ }\textbf {\bibinfo {volume}
  {142}},\ \bibinfo {pages} {233--244} (\bibinfo {year} {2017})}\BibitemShut
  {NoStop}%
\bibitem [{\citenamefont {Esteban}\ \emph {et~al.}(2011)\citenamefont
  {Esteban}, \citenamefont {Diez}, \citenamefont {L{\'{o}}pez},\ and\
  \citenamefont {Negro}}]{Esteban2011}%
  \BibitemOpen
  \bibfield  {author} {\bibinfo {author} {\bibnamefont {Esteban}, \bibfnamefont
  {M.~D.}}, \bibinfo {author} {\bibnamefont {Diez}, \bibfnamefont {J.~J.}},
  \bibinfo {author} {\bibnamefont {L{\'{o}}pez}, \bibfnamefont {J.~S.}}, \ and\
  \bibinfo {author} {\bibnamefont {Negro}, \bibfnamefont {V.}},\ }\bibfield
  {title} {\enquote {\bibinfo {title} {{Why offshore wind energy?}}}\ }\href
  {\doibase 10.1016/j.renene.2010.07.009} {\bibfield  {journal} {\bibinfo
  {journal} {Renewable Energy}\ }\textbf {\bibinfo {volume} {36}},\ \bibinfo
  {pages} {444--450} (\bibinfo {year} {2011})}\BibitemShut {NoStop}%
\bibitem [{\citenamefont {Fairall}\ \emph {et~al.}(1996)\citenamefont
  {Fairall}, \citenamefont {Bradley}, \citenamefont {Rogers}, \citenamefont
  {Edson},\ and\ \citenamefont {Young}}]{Fairall1996}%
  \BibitemOpen
  \bibfield  {author} {\bibinfo {author} {\bibnamefont {Fairall}, \bibfnamefont
  {C.~W.}}, \bibinfo {author} {\bibnamefont {Bradley}, \bibfnamefont {E.~F.}},
  \bibinfo {author} {\bibnamefont {Rogers}, \bibfnamefont {D.~P.}}, \bibinfo
  {author} {\bibnamefont {Edson}, \bibfnamefont {J.~B.}}, \ and\ \bibinfo
  {author} {\bibnamefont {Young}, \bibfnamefont {G.~S.}},\ }\bibfield  {title}
  {\enquote {\bibinfo {title} {Bulk parameterization of air-sea fluxes for
  tropical ocean-global atmosphere coupled-ocean atmosphere response
  experiment},}\ }\href {\doibase https://doi.org/10.1029/95JC03205} {\bibfield
   {journal} {\bibinfo  {journal} {Journal of Geophysical Research: Oceans}\
  }\textbf {\bibinfo {volume} {101}},\ \bibinfo {pages} {3747--3764} (\bibinfo
  {year} {1996})}\BibitemShut {NoStop}%
\bibitem [{\citenamefont {Fer{\v{c}}{\'{a}}k}\ \emph
  {et~al.}(2022)\citenamefont {Fer{\v{c}}{\'{a}}k}, \citenamefont {Bossuyt},
  \citenamefont {Ali},\ and\ \citenamefont {Cal}}]{Fercak2022}%
  \BibitemOpen
  \bibfield  {author} {\bibinfo {author} {\bibnamefont {Fer{\v{c}}{\'{a}}k},
  \bibfnamefont {O.}}, \bibinfo {author} {\bibnamefont {Bossuyt}, \bibfnamefont
  {J.}}, \bibinfo {author} {\bibnamefont {Ali}, \bibfnamefont {N.}}, \ and\
  \bibinfo {author} {\bibnamefont {Cal}, \bibfnamefont {R.~B.}},\ }\bibfield
  {title} {\enquote {\bibinfo {title} {{Decoupling wind–wave–wake
  interactions in a fixed-bottom offshore wind turbine}},}\ }\href {\doibase
  10.1016/j.apenergy.2021.118358} {\bibfield  {journal} {\bibinfo  {journal}
  {Applied Energy}\ }\textbf {\bibinfo {volume} {309}},\ \bibinfo {pages}
  {118358} (\bibinfo {year} {2022})}\BibitemShut {NoStop}%
\bibitem [{\citenamefont {Fois}\ \emph {et~al.}(2014)\citenamefont {Fois},
  \citenamefont {Hoogeboom}, \citenamefont {Le~Chevalier},\ and\ \citenamefont
  {Stoffelen}}]{fois2014investigation}%
  \BibitemOpen
  \bibfield  {author} {\bibinfo {author} {\bibnamefont {Fois}, \bibfnamefont
  {F.}}, \bibinfo {author} {\bibnamefont {Hoogeboom}, \bibfnamefont {P.}},
  \bibinfo {author} {\bibnamefont {Le~Chevalier}, \bibfnamefont {F.}}, \ and\
  \bibinfo {author} {\bibnamefont {Stoffelen}, \bibfnamefont {A.}},\ }\bibfield
   {title} {\enquote {\bibinfo {title} {An investigation on sea surface wave
  spectra and approximate scattering theories},}\ }in\ \href@noop {} {\emph
  {\bibinfo {booktitle} {2014 IEEE Geoscience and Remote Sensing Symposium}}}\
  (\bibinfo {organization} {IEEE},\ \bibinfo {year} {2014})\ pp.\ \bibinfo
  {pages} {4366--4369}\BibitemShut {NoStop}%
\bibitem [{\citenamefont {Frandsen}\ \emph {et~al.}(2006)\citenamefont
  {Frandsen}, \citenamefont {Barthelmie}, \citenamefont {Pryor}, \citenamefont
  {Rathmann}, \citenamefont {Larsen}, \citenamefont {H{\o}jstrup},\ and\
  \citenamefont {Th{\o}gersen}}]{Frandsen2006}%
  \BibitemOpen
  \bibfield  {author} {\bibinfo {author} {\bibnamefont {Frandsen},
  \bibfnamefont {S.}}, \bibinfo {author} {\bibnamefont {Barthelmie},
  \bibfnamefont {R.}}, \bibinfo {author} {\bibnamefont {Pryor}, \bibfnamefont
  {S.}}, \bibinfo {author} {\bibnamefont {Rathmann}, \bibfnamefont {O.}},
  \bibinfo {author} {\bibnamefont {Larsen}, \bibfnamefont {S.}}, \bibinfo
  {author} {\bibnamefont {H{\o}jstrup}, \bibfnamefont {J.}}, \ and\ \bibinfo
  {author} {\bibnamefont {Th{\o}gersen}, \bibfnamefont {M.}},\ }\bibfield
  {title} {\enquote {\bibinfo {title} {Analytical modelling of wind speed
  deficit in large offshore wind farms},}\ }\href {\doibase
  https://doi.org/10.1002/we.189} {\bibfield  {journal} {\bibinfo  {journal}
  {Wind Energy}\ }\textbf {\bibinfo {volume} {9}},\ \bibinfo {pages} {39--53}
  (\bibinfo {year} {2006})},\ \Eprint
  {http://arxiv.org/abs/https://onlinelibrary.wiley.com/doi/pdf/10.1002/we.189}
  {https://onlinelibrary.wiley.com/doi/pdf/10.1002/we.189} \BibitemShut
  {NoStop}%
\bibitem [{\citenamefont {Goit}\ and\ \citenamefont {Önder}(2022)}]{Goit2022}%
  \BibitemOpen
  \bibfield  {author} {\bibinfo {author} {\bibnamefont {Goit}, \bibfnamefont
  {J.~P.}}\ and\ \bibinfo {author} {\bibnamefont {Önder}, \bibfnamefont
  {A.}},\ }\bibfield  {title} {\enquote {\bibinfo {title} {{The effect of
  coastal terrain on nearshore offshore wind farms: A large-eddy simulation
  study}},}\ }\href {\doibase 10.1063/5.0094476} {\bibfield  {journal}
  {\bibinfo  {journal} {Journal of Renewable and Sustainable Energy}\ }\textbf
  {\bibinfo {volume} {14}} (\bibinfo {year} {2022}),\ 10.1063/5.0094476},\
  \bibinfo {note} {043304},\ \Eprint
  {http://arxiv.org/abs/https://pubs.aip.org/aip/jrse/article-pdf/doi/10.1063/5.0094476/16573559/043304\_1\_online.pdf}
  {https://pubs.aip.org/aip/jrse/article-pdf/doi/10.1063/5.0094476/16573559/043304\_1\_online.pdf}
  \BibitemShut {NoStop}%
\bibitem [{\citenamefont {Grare}\ \emph {et~al.}(2013)\citenamefont {Grare},
  \citenamefont {Peirson}, \citenamefont {Branger}, \citenamefont {Walker},
  \citenamefont {Giovanangeli},\ and\ \citenamefont {Makin}}]{Grare2013}%
  \BibitemOpen
  \bibfield  {author} {\bibinfo {author} {\bibnamefont {Grare}, \bibfnamefont
  {L.}}, \bibinfo {author} {\bibnamefont {Peirson}, \bibfnamefont {W.~L.}},
  \bibinfo {author} {\bibnamefont {Branger}, \bibfnamefont {H.}}, \bibinfo
  {author} {\bibnamefont {Walker}, \bibfnamefont {J.~W.}}, \bibinfo {author}
  {\bibnamefont {Giovanangeli}, \bibfnamefont {J.~P.}}, \ and\ \bibinfo
  {author} {\bibnamefont {Makin}, \bibfnamefont {V.}},\ }\bibfield  {title}
  {\enquote {\bibinfo {title} {{Growth and dissipation of wind-forced,
  deep-water waves}},}\ }\href {\doibase 10.1017/jfm.2013.88} {\bibfield
  {journal} {\bibinfo  {journal} {Journal of Fluid Mechanics}\ }\textbf
  {\bibinfo {volume} {722}},\ \bibinfo {pages} {5--50} (\bibinfo {year}
  {2013})}\BibitemShut {NoStop}%
\bibitem [{\citenamefont {Hao}, \citenamefont {Nagata},\ and\ \citenamefont
  {Zhou}(2020)}]{Hao2020}%
  \BibitemOpen
  \bibfield  {author} {\bibinfo {author} {\bibnamefont {Hao}, \bibfnamefont
  {K.}}, \bibinfo {author} {\bibnamefont {Nagata}, \bibfnamefont {K.}}, \ and\
  \bibinfo {author} {\bibnamefont {Zhou}, \bibfnamefont {Y.}},\ }\bibfield
  {title} {\enquote {\bibinfo {title} {{Scale-by-scale energy transfer in a
  dual-plane jet flow}},}\ }\href {\doibase 10.1063/5.0022103} {\bibfield
  {journal} {\bibinfo  {journal} {Physics of Fluids}\ }\textbf {\bibinfo
  {volume} {32}} (\bibinfo {year} {2020}),\ 10.1063/5.0022103}\BibitemShut
  {NoStop}%
\bibitem [{\citenamefont {Hao}\ and\ \citenamefont {Shen}(2019)}]{Hao2019}%
  \BibitemOpen
  \bibfield  {author} {\bibinfo {author} {\bibnamefont {Hao}, \bibfnamefont
  {X.}}\ and\ \bibinfo {author} {\bibnamefont {Shen}, \bibfnamefont {L.}},\
  }\bibfield  {title} {\enquote {\bibinfo {title} {{Wind-wave coupling study
  using les of wind and phase-resolved simulation of nonlinear waves}},}\
  }\href {\doibase 10.1017/jfm.2019.444} {\bibfield  {journal} {\bibinfo
  {journal} {Journal of Fluid Mechanics}\ ,\ \bibinfo {pages} {391--425}}
  (\bibinfo {year} {2019})}\BibitemShut {NoStop}%
\bibitem [{\citenamefont {Hasselmann}\ \emph {et~al.}(1973)\citenamefont
  {Hasselmann}, \citenamefont {Barnett}, \citenamefont {Bouws}, \citenamefont
  {Carlson}, \citenamefont {Cartwright}, \citenamefont {Enke}, \citenamefont
  {Ewing}, \citenamefont {Gienapp}, \citenamefont {Hasselmann}, \citenamefont
  {Kruseman} \emph {et~al.}}]{hasselmann1973measurements}%
  \BibitemOpen
  \bibfield  {author} {\bibinfo {author} {\bibnamefont {Hasselmann},
  \bibfnamefont {K.}}, \bibinfo {author} {\bibnamefont {Barnett}, \bibfnamefont
  {T.~P.}}, \bibinfo {author} {\bibnamefont {Bouws}, \bibfnamefont {E.}},
  \bibinfo {author} {\bibnamefont {Carlson}, \bibfnamefont {H.}}, \bibinfo
  {author} {\bibnamefont {Cartwright}, \bibfnamefont {D.~E.}}, \bibinfo
  {author} {\bibnamefont {Enke}, \bibfnamefont {K.}}, \bibinfo {author}
  {\bibnamefont {Ewing}, \bibfnamefont {J.}}, \bibinfo {author} {\bibnamefont
  {Gienapp}, \bibfnamefont {A.}}, \bibinfo {author} {\bibnamefont {Hasselmann},
  \bibfnamefont {D.}}, \bibinfo {author} {\bibnamefont {Kruseman},
  \bibfnamefont {P.}},  \emph {et~al.},\ }\bibfield  {title} {\enquote
  {\bibinfo {title} {Measurements of wind-wave growth and swell decay during
  the joint north sea wave project (jonswap).}}\ }\href@noop {} {\bibfield
  {journal} {\bibinfo  {journal} {Ergaenzungsheft zur Deutschen
  Hydrographischen Zeitschrift, Reihe A}\ } (\bibinfo {year}
  {1973})}\BibitemShut {NoStop}%
\bibitem [{\citenamefont {Hwang}(2005)}]{Hwang2005WaveWaves}%
  \BibitemOpen
  \bibfield  {author} {\bibinfo {author} {\bibnamefont {Hwang}, \bibfnamefont
  {P.~A.}},\ }\bibfield  {title} {\enquote {\bibinfo {title} {{Wave number
  spectrum and mean square slope of intermediate-scale ocean surface waves}},}\
  }\href {\doibase 10.1029/2005JC003002} {\bibfield  {journal} {\bibinfo
  {journal} {Journal of Geophysical Research: Oceans}\ }\textbf {\bibinfo
  {volume} {110}},\ \bibinfo {pages} {1--7} (\bibinfo {year}
  {2005})}\BibitemShut {NoStop}%
\bibitem [{\citenamefont {Kawai}\ and\ \citenamefont
  {Larsson}(2012)}]{kawai2012}%
  \BibitemOpen
  \bibfield  {author} {\bibinfo {author} {\bibnamefont {Kawai}, \bibfnamefont
  {S.}}\ and\ \bibinfo {author} {\bibnamefont {Larsson}, \bibfnamefont {J.}},\
  }\bibfield  {title} {\enquote {\bibinfo {title} {Wall-modeling in large eddy
  simulation: Length scales, grid resolution, and accuracy},}\ }\href {\doibase
  10.1063/1.3678331} {\bibfield  {journal} {\bibinfo  {journal} {Physics of
  Fluids}\ }\textbf {\bibinfo {volume} {24}},\ \bibinfo {pages} {015105}
  (\bibinfo {year} {2012})},\ \Eprint
  {http://arxiv.org/abs/https://doi.org/10.1063/1.3678331}
  {https://doi.org/10.1063/1.3678331} \BibitemShut {NoStop}%
\bibitem [{\citenamefont {Kihara}\ \emph {et~al.}(2007)\citenamefont {Kihara},
  \citenamefont {Hanazaki}, \citenamefont {Mizuya},\ and\ \citenamefont
  {Ueda}}]{Kihara2007}%
  \BibitemOpen
  \bibfield  {author} {\bibinfo {author} {\bibnamefont {Kihara}, \bibfnamefont
  {N.}}, \bibinfo {author} {\bibnamefont {Hanazaki}, \bibfnamefont {H.}},
  \bibinfo {author} {\bibnamefont {Mizuya}, \bibfnamefont {T.}}, \ and\
  \bibinfo {author} {\bibnamefont {Ueda}, \bibfnamefont {H.}},\ }\bibfield
  {title} {\enquote {\bibinfo {title} {Relationship between airflow at the
  critical height and momentum transfer to the traveling waves},}\ }\href
  {\doibase 10.1063/1.2409736} {\bibfield  {journal} {\bibinfo  {journal}
  {Physics of Fluids}\ }\textbf {\bibinfo {volume} {19}},\ \bibinfo {pages}
  {015102} (\bibinfo {year} {2007})},\ \Eprint
  {http://arxiv.org/abs/https://doi.org/10.1063/1.2409736}
  {https://doi.org/10.1063/1.2409736} \BibitemShut {NoStop}%
\bibitem [{\citenamefont {Lin}\ and\ \citenamefont {Hasan}(2022)}]{Lin2022}%
  \BibitemOpen
  \bibfield  {author} {\bibinfo {author} {\bibnamefont {Lin}, \bibfnamefont
  {Y.-H.}}\ and\ \bibinfo {author} {\bibnamefont {Hasan}, \bibfnamefont
  {A.~D.}},\ }\bibfield  {title} {\enquote {\bibinfo {title} {{Transient
  analysis of the slamming wave load on an offshore wind turbine foundation
  generated by different types of breaking waves}},}\ }\href {\doibase
  10.1063/5.0107679} {\bibfield  {journal} {\bibinfo  {journal} {Journal of
  Renewable and Sustainable Energy}\ }\textbf {\bibinfo {volume} {14}}
  (\bibinfo {year} {2022}),\ 10.1063/5.0107679},\ \bibinfo {note} {053302},\
  \Eprint
  {http://arxiv.org/abs/https://pubs.aip.org/aip/jrse/article-pdf/doi/10.1063/5.0107679/16637134/053302\_1\_online.pdf}
  {https://pubs.aip.org/aip/jrse/article-pdf/doi/10.1063/5.0107679/16637134/053302\_1\_online.pdf}
  \BibitemShut {NoStop}%
\bibitem [{\citenamefont {Liu}\ \emph {et~al.}(2010)\citenamefont {Liu},
  \citenamefont {Yang}, \citenamefont {Guo},\ and\ \citenamefont
  {Shen}}]{Liu2010}%
  \BibitemOpen
  \bibfield  {author} {\bibinfo {author} {\bibnamefont {Liu}, \bibfnamefont
  {Y.}}, \bibinfo {author} {\bibnamefont {Yang}, \bibfnamefont {D.}}, \bibinfo
  {author} {\bibnamefont {Guo}, \bibfnamefont {X.}}, \ and\ \bibinfo {author}
  {\bibnamefont {Shen}, \bibfnamefont {L.}},\ }\bibfield  {title} {\enquote
  {\bibinfo {title} {{Numerical study of pressure forcing of wind on
  dynamically evolving water waves}},}\ }\href {\doibase 10.1063/1.3414832}
  {\bibfield  {journal} {\bibinfo  {journal} {Physics of Fluids}\ }\textbf
  {\bibinfo {volume} {22}},\ \bibinfo {pages} {1--4} (\bibinfo {year}
  {2010})}\BibitemShut {NoStop}%
\bibitem [{\citenamefont {MacArt}\ and\ \citenamefont
  {Mueller}(2016)}]{macart2016}%
  \BibitemOpen
  \bibfield  {author} {\bibinfo {author} {\bibnamefont {MacArt}, \bibfnamefont
  {J.~F.}}\ and\ \bibinfo {author} {\bibnamefont {Mueller}, \bibfnamefont
  {M.~E.}},\ }\bibfield  {title} {\enquote {\bibinfo {title} {Semi-implicit
  iterative methods for low mach number turbulent reacting flows: Operator
  splitting versus approximate factorization},}\ }\href {\doibase
  https://doi.org/10.1016/j.jcp.2016.09.016} {\bibfield  {journal} {\bibinfo
  {journal} {Journal of Computational Physics}\ }\textbf {\bibinfo {volume}
  {326}},\ \bibinfo {pages} {569--595} (\bibinfo {year} {2016})}\BibitemShut
  {NoStop}%
\bibitem [{\citenamefont {Makin}, \citenamefont {Kudryavtsev},\ and\
  \citenamefont {Mastenbroek}(1995)}]{Makin1995}%
  \BibitemOpen
  \bibfield  {author} {\bibinfo {author} {\bibnamefont {Makin}, \bibfnamefont
  {V.~K.}}, \bibinfo {author} {\bibnamefont {Kudryavtsev}, \bibfnamefont
  {V.~N.}}, \ and\ \bibinfo {author} {\bibnamefont {Mastenbroek}, \bibfnamefont
  {C.}},\ }\bibfield  {title} {\enquote {\bibinfo {title} {Drag of the sea
  surface},}\ }\href {\doibase 10.1007/BF00708935} {\bibfield  {journal}
  {\bibinfo  {journal} {Boundary-Layer Meteorology}\ }\textbf {\bibinfo
  {volume} {73}},\ \bibinfo {pages} {159--182} (\bibinfo {year}
  {1995})}\BibitemShut {NoStop}%
\bibitem [{\citenamefont {Miles}(1993)}]{Miles1993}%
  \BibitemOpen
  \bibfield  {author} {\bibinfo {author} {\bibnamefont {Miles}, \bibfnamefont
  {J.}},\ }\bibfield  {title} {\enquote {\bibinfo {title} {{Surface-Wave
  Generation Revisited}},}\ }\href {\doibase 10.1017/S0022112093002836}
  {\bibfield  {journal} {\bibinfo  {journal} {Journal of Fluid Mechanics}\
  }\textbf {\bibinfo {volume} {256}},\ \bibinfo {pages} {427--441} (\bibinfo
  {year} {1993})}\BibitemShut {NoStop}%
\bibitem [{\citenamefont {Miles}(1957)}]{miles_1957}%
  \BibitemOpen
  \bibfield  {author} {\bibinfo {author} {\bibnamefont {Miles}, \bibfnamefont
  {J.~W.}},\ }\bibfield  {title} {\enquote {\bibinfo {title} {On the generation
  of surface waves by shear flows},}\ }\href {\doibase
  10.1017/S0022112057000567} {\bibfield  {journal} {\bibinfo  {journal}
  {Journal of Fluid Mechanics}\ }\textbf {\bibinfo {volume} {3}},\ \bibinfo
  {pages} {185–204} (\bibinfo {year} {1957})}\BibitemShut {NoStop}%
\bibitem [{\citenamefont {Musial}\ \emph {et~al.}(2019)\citenamefont {Musial},
  \citenamefont {Beiter}, \citenamefont {Spitsen}, \citenamefont {Nunemaker},\
  and\ \citenamefont {Gevorgian}}]{osti_1572771}%
  \BibitemOpen
  \bibfield  {author} {\bibinfo {author} {\bibnamefont {Musial}, \bibfnamefont
  {W.~D.}}, \bibinfo {author} {\bibnamefont {Beiter}, \bibfnamefont {P.~C.}},
  \bibinfo {author} {\bibnamefont {Spitsen}, \bibfnamefont {P.}}, \bibinfo
  {author} {\bibnamefont {Nunemaker}, \bibfnamefont {J.}}, \ and\ \bibinfo
  {author} {\bibnamefont {Gevorgian}, \bibfnamefont {V.}},\ }\bibfield  {title}
  {\enquote {\bibinfo {title} {2018 offshore wind technologies market
  report},}\ }\href {\doibase 10.2172/1572771} {\  (\bibinfo {year} {2019}),\
  10.2172/1572771}\BibitemShut {NoStop}%
\bibitem [{\citenamefont {Phillips}(1985)}]{phillips_1985}%
  \BibitemOpen
  \bibfield  {author} {\bibinfo {author} {\bibnamefont {Phillips},
  \bibfnamefont {O.~M.}},\ }\bibfield  {title} {\enquote {\bibinfo {title}
  {Spectral and statistical properties of the equilibrium range in
  wind-generated gravity waves},}\ }\href {\doibase 10.1017/S0022112085002221}
  {\bibfield  {journal} {\bibinfo  {journal} {Journal of Fluid Mechanics}\
  }\textbf {\bibinfo {volume} {156}},\ \bibinfo {pages} {505–531} (\bibinfo
  {year} {1985})}\BibitemShut {NoStop}%
\bibitem [{\citenamefont {Piomelli}\ and\ \citenamefont
  {Balaras}(2002)}]{Piomelli2002}%
  \BibitemOpen
  \bibfield  {author} {\bibinfo {author} {\bibnamefont {Piomelli},
  \bibfnamefont {U.}}\ and\ \bibinfo {author} {\bibnamefont {Balaras},
  \bibfnamefont {E.}},\ }\bibfield  {title} {\enquote {\bibinfo {title}
  {Wall-layer models for large-eddy simulations},}\ }\href {\doibase
  10.1146/annurev.fluid.34.082901.144919} {\bibfield  {journal} {\bibinfo
  {journal} {Annual Review of Fluid Mechanics}\ }\textbf {\bibinfo {volume}
  {34}},\ \bibinfo {pages} {349--374} (\bibinfo {year} {2002})}\BibitemShut
  {NoStop}%
\bibitem [{\citenamefont {Plant}(1982)}]{plant1982}%
  \BibitemOpen
  \bibfield  {author} {\bibinfo {author} {\bibnamefont {Plant}, \bibfnamefont
  {W.~J.}},\ }\bibfield  {title} {\enquote {\bibinfo {title} {A relationship
  between wind stress and wave slope},}\ }\href {\doibase
  https://doi.org/10.1029/JC087iC03p01961} {\bibfield  {journal} {\bibinfo
  {journal} {Journal of Geophysical Research: Oceans}\ }\textbf {\bibinfo
  {volume} {87}},\ \bibinfo {pages} {1961--1967} (\bibinfo {year} {1982})},\
  \Eprint
  {http://arxiv.org/abs/https://agupubs.onlinelibrary.wiley.com/doi/pdf/10.1029/JC087iC03p01961}
  {https://agupubs.onlinelibrary.wiley.com/doi/pdf/10.1029/JC087iC03p01961}
  \BibitemShut {NoStop}%
\bibitem [{\citenamefont {Port{\'{e}}-Agel}, \citenamefont {Bastankhah},\ and\
  \citenamefont {Shamsoddin}(2020)}]{Porte-Agel2020}%
  \BibitemOpen
  \bibfield  {author} {\bibinfo {author} {\bibnamefont {Port{\'{e}}-Agel},
  \bibfnamefont {F.}}, \bibinfo {author} {\bibnamefont {Bastankhah},
  \bibfnamefont {M.}}, \ and\ \bibinfo {author} {\bibnamefont {Shamsoddin},
  \bibfnamefont {S.}},\ }\href {\doibase 10.1007/s10546-019-00473-0} {\emph
  {\bibinfo {title} {Boundary-Layer Meteorology}}},\ Vol.\ \bibinfo {volume}
  {174}\ (\bibinfo  {publisher} {Springer Netherlands},\ \bibinfo {year}
  {2020})\ pp.\ \bibinfo {pages} {1--59}\BibitemShut {NoStop}%
\bibitem [{\citenamefont {Rozema}\ \emph {et~al.}(2015)\citenamefont {Rozema},
  \citenamefont {Bae}, \citenamefont {Moin},\ and\ \citenamefont
  {Verstappen}}]{Rozema2015}%
  \BibitemOpen
  \bibfield  {author} {\bibinfo {author} {\bibnamefont {Rozema}, \bibfnamefont
  {W.}}, \bibinfo {author} {\bibnamefont {Bae}, \bibfnamefont {H.~J.}},
  \bibinfo {author} {\bibnamefont {Moin}, \bibfnamefont {P.}}, \ and\ \bibinfo
  {author} {\bibnamefont {Verstappen}, \bibfnamefont {R.}},\ }\bibfield
  {title} {\enquote {\bibinfo {title} {{Minimum-dissipation models for
  large-eddy simulation}},}\ }\href {\doibase 10.1063/1.4928700} {\bibfield
  {journal} {\bibinfo  {journal} {Physics of Fluids}\ }\textbf {\bibinfo
  {volume} {27}} (\bibinfo {year} {2015}),\ 10.1063/1.4928700}\BibitemShut
  {NoStop}%
\bibitem [{\citenamefont {Ryabkova}\ \emph {et~al.}(2019)\citenamefont
  {Ryabkova}, \citenamefont {Karaev}, \citenamefont {Guo},\ and\ \citenamefont
  {Titchenko}}]{Ryabkova2019}%
  \BibitemOpen
  \bibfield  {author} {\bibinfo {author} {\bibnamefont {Ryabkova},
  \bibfnamefont {M.}}, \bibinfo {author} {\bibnamefont {Karaev}, \bibfnamefont
  {V.}}, \bibinfo {author} {\bibnamefont {Guo}, \bibfnamefont {J.}}, \ and\
  \bibinfo {author} {\bibnamefont {Titchenko}, \bibfnamefont {Y.}},\ }\bibfield
   {title} {\enquote {\bibinfo {title} {A review of wave spectrum models as
  applied to the problem of radar probing of the sea surface},}\ }\href
  {\doibase https://doi.org/10.1029/2018JC014804} {\bibfield  {journal}
  {\bibinfo  {journal} {Journal of Geophysical Research: Oceans}\ }\textbf
  {\bibinfo {volume} {124}},\ \bibinfo {pages} {7104--7134} (\bibinfo {year}
  {2019})},\ \Eprint
  {http://arxiv.org/abs/https://agupubs.onlinelibrary.wiley.com/doi/pdf/10.1029/2018JC014804}
  {https://agupubs.onlinelibrary.wiley.com/doi/pdf/10.1029/2018JC014804}
  \BibitemShut {NoStop}%
\bibitem [{\citenamefont {Shapiro}, \citenamefont {Gayme},\ and\ \citenamefont
  {Meneveau}(2019)}]{Shapiro2019}%
  \BibitemOpen
  \bibfield  {author} {\bibinfo {author} {\bibnamefont {Shapiro}, \bibfnamefont
  {C.~R.}}, \bibinfo {author} {\bibnamefont {Gayme}, \bibfnamefont {D.~F.}}, \
  and\ \bibinfo {author} {\bibnamefont {Meneveau}, \bibfnamefont {C.}},\
  }\bibfield  {title} {\enquote {\bibinfo {title} {{Filtered actuator disks:
  Theory and application to wind turbine models in large eddy simulation}},}\
  }\href {\doibase 10.1002/we.2376} {\bibfield  {journal} {\bibinfo  {journal}
  {Wind Energy}\ }\textbf {\bibinfo {volume} {22}},\ \bibinfo {pages}
  {1414--1420} (\bibinfo {year} {2019})}\BibitemShut {NoStop}%
\bibitem [{\citenamefont {Sullivan}\ \emph {et~al.}(2008)\citenamefont
  {Sullivan}, \citenamefont {Edson}, \citenamefont {Hristov},\ and\
  \citenamefont {McWilliams}}]{Sullivan2008}%
  \BibitemOpen
  \bibfield  {author} {\bibinfo {author} {\bibnamefont {Sullivan},
  \bibfnamefont {P.~P.}}, \bibinfo {author} {\bibnamefont {Edson},
  \bibfnamefont {J.~B.}}, \bibinfo {author} {\bibnamefont {Hristov},
  \bibfnamefont {T.}}, \ and\ \bibinfo {author} {\bibnamefont {McWilliams},
  \bibfnamefont {J.~C.}},\ }\bibfield  {title} {\enquote {\bibinfo {title}
  {{Large-eddy simulations and observations of atmospheric marine boundary
  layers above nonequilibrium surface waves}},}\ }\href {\doibase
  10.1175/2007JAS2427.1} {\bibfield  {journal} {\bibinfo  {journal} {Journal of
  the Atmospheric Sciences}\ }\textbf {\bibinfo {volume} {65}},\ \bibinfo
  {pages} {1225--1245} (\bibinfo {year} {2008})}\BibitemShut {NoStop}%
\bibitem [{\citenamefont {Sullivan}, \citenamefont {McWilliams},\ and\
  \citenamefont {Moeng}(2000)}]{Sullivan2000}%
  \BibitemOpen
  \bibfield  {author} {\bibinfo {author} {\bibnamefont {Sullivan},
  \bibfnamefont {P.~P.}}, \bibinfo {author} {\bibnamefont {McWilliams},
  \bibfnamefont {J.~C.}}, \ and\ \bibinfo {author} {\bibnamefont {Moeng},
  \bibfnamefont {C.~H.}},\ }\bibfield  {title} {\enquote {\bibinfo {title}
  {{Simulation of turbulent flow over idealized water waves}},}\ }\href
  {\doibase 10.1017/S0022112099006965} {\bibfield  {journal} {\bibinfo
  {journal} {Journal of Fluid Mechanics}\ }\textbf {\bibinfo {volume} {404}},\
  \bibinfo {pages} {47--85} (\bibinfo {year} {2000})}\BibitemShut {NoStop}%
\bibitem [{\citenamefont {Sullivan}, \citenamefont {McWilliams},\ and\
  \citenamefont {Patton}(2014)}]{Sullivan2014}%
  \BibitemOpen
  \bibfield  {author} {\bibinfo {author} {\bibnamefont {Sullivan},
  \bibfnamefont {P.~P.}}, \bibinfo {author} {\bibnamefont {McWilliams},
  \bibfnamefont {J.~C.}}, \ and\ \bibinfo {author} {\bibnamefont {Patton},
  \bibfnamefont {E.~G.}},\ }\bibfield  {title} {\enquote {\bibinfo {title}
  {{Large-eddy simulation of marine atmospheric boundary layers above a
  spectrum of moving waves}},}\ }\href {\doibase 10.1175/JAS-D-14-0095.1}
  {\bibfield  {journal} {\bibinfo  {journal} {Journal of the Atmospheric
  Sciences}\ }\textbf {\bibinfo {volume} {71}},\ \bibinfo {pages} {4001--4027}
  (\bibinfo {year} {2014})}\BibitemShut {NoStop}%
\bibitem [{\citenamefont {Taylor}\ and\ \citenamefont
  {Yelland}(2001)}]{Taylor2001}%
  \BibitemOpen
  \bibfield  {author} {\bibinfo {author} {\bibnamefont {Taylor}, \bibfnamefont
  {P.~K.}}\ and\ \bibinfo {author} {\bibnamefont {Yelland}, \bibfnamefont
  {M.~J.}},\ }\bibfield  {title} {\enquote {\bibinfo {title} {The dependence of
  sea surface roughness on the height and steepness of the waves},}\ }\href
  {\doibase 10.1175/1520-0485(2001)031<0572:TDOSSR>2.0.CO;2} {\bibfield
  {journal} {\bibinfo  {journal} {Journal of Physical Oceanography}\ }\textbf
  {\bibinfo {volume} {31}},\ \bibinfo {pages} {572 -- 590} (\bibinfo {year}
  {2001})}\BibitemShut {NoStop}%
\bibitem [{\citenamefont {Toba}(1973)}]{Toba1973}%
  \BibitemOpen
  \bibfield  {author} {\bibinfo {author} {\bibnamefont {Toba}, \bibfnamefont
  {Y.}},\ }\bibfield  {title} {\enquote {\bibinfo {title} {{Local balance in
  the air-sea boundary processes - II. Partition of wind stress to waves and
  current}},}\ }\href {\doibase 10.1007/BF02109506} {\bibfield  {journal}
  {\bibinfo  {journal} {Journal of the Oceanographical Society of Japan}\
  }\textbf {\bibinfo {volume} {29}},\ \bibinfo {pages} {70--75} (\bibinfo
  {year} {1973})}\BibitemShut {NoStop}%
\bibitem [{\citenamefont {Wei}\ and\ \citenamefont {Dabiri}(2022)}]{Wei2022}%
  \BibitemOpen
  \bibfield  {author} {\bibinfo {author} {\bibnamefont {Wei}, \bibfnamefont
  {N.~J.}}\ and\ \bibinfo {author} {\bibnamefont {Dabiri}, \bibfnamefont
  {J.~O.}},\ }\bibfield  {title} {\enquote {\bibinfo {title} {{Phase-averaged
  dynamics of a periodically surging wind turbine}},}\ }\href {\doibase
  10.1063/5.0076029} {\bibfield  {journal} {\bibinfo  {journal} {Journal of
  Renewable and Sustainable Energy}\ }\textbf {\bibinfo {volume} {14}},\
  \bibinfo {pages} {013305} (\bibinfo {year} {2022})}\BibitemShut {NoStop}%
\bibitem [{\citenamefont {Wu}, \citenamefont {Popinet},\ and\ \citenamefont
  {Deike}(2022)}]{Wu2022}%
  \BibitemOpen
  \bibfield  {author} {\bibinfo {author} {\bibnamefont {Wu}, \bibfnamefont
  {J.}}, \bibinfo {author} {\bibnamefont {Popinet}, \bibfnamefont {S.}}, \ and\
  \bibinfo {author} {\bibnamefont {Deike}, \bibfnamefont {L.}},\ }\bibfield
  {title} {\enquote {\bibinfo {title} {{Revisiting wind wave growth with fully
  coupled direct numerical simulations}},}\ }\href {\doibase
  10.1017/jfm.2022.822} {\bibfield  {journal} {\bibinfo  {journal} {Journal of
  Fluid Mechanics}\ }\textbf {\bibinfo {volume} {951}} (\bibinfo {year}
  {2022}),\ 10.1017/jfm.2022.822}\BibitemShut {NoStop}%
\bibitem [{\citenamefont {Wu}\ and\ \citenamefont
  {Port{\'{e}}-Agel}(2011)}]{Wu2011}%
  \BibitemOpen
  \bibfield  {author} {\bibinfo {author} {\bibnamefont {Wu}, \bibfnamefont
  {Y.~T.}}\ and\ \bibinfo {author} {\bibnamefont {Port{\'{e}}-Agel},
  \bibfnamefont {F.}},\ }\bibfield  {title} {\enquote {\bibinfo {title}
  {{Large-Eddy Simulation of Wind-Turbine Wakes: Evaluation of Turbine
  Parametrisations}},}\ }\href {\doibase 10.1007/s10546-010-9569-x} {\bibfield
  {journal} {\bibinfo  {journal} {Boundary-Layer Meteorology}\ }\textbf
  {\bibinfo {volume} {138}},\ \bibinfo {pages} {345--366} (\bibinfo {year}
  {2011})}\BibitemShut {NoStop}%
\bibitem [{\citenamefont {Xiao}\ and\ \citenamefont {Yang}(2019)}]{Xiao2019}%
  \BibitemOpen
  \bibfield  {author} {\bibinfo {author} {\bibnamefont {Xiao}, \bibfnamefont
  {S.}}\ and\ \bibinfo {author} {\bibnamefont {Yang}, \bibfnamefont {D.}},\
  }\bibfield  {title} {\enquote {\bibinfo {title} {{Large-eddy simulation-based
  study of effect of swell-induced pitch motion on wake-flow statistics and
  power extraction of offshorewind turbines}},}\ }\href {\doibase
  10.3390/en12071246} {\bibfield  {journal} {\bibinfo  {journal} {Energies}\
  }\textbf {\bibinfo {volume} {12}} (\bibinfo {year} {2019}),\
  10.3390/en12071246}\BibitemShut {NoStop}%
\bibitem [{\citenamefont {Yang}, \citenamefont {Meneveau},\ and\ \citenamefont
  {Shen}(2013)}]{Yang2013D}%
  \BibitemOpen
  \bibfield  {author} {\bibinfo {author} {\bibnamefont {Yang}, \bibfnamefont
  {D.}}, \bibinfo {author} {\bibnamefont {Meneveau}, \bibfnamefont {C.}}, \
  and\ \bibinfo {author} {\bibnamefont {Shen}, \bibfnamefont {L.}},\ }\bibfield
   {title} {\enquote {\bibinfo {title} {{Dynamic modelling of sea-surface
  roughness for large-eddy simulation of wind over ocean wavefield}},}\ }\href
  {\doibase 10.1017/jfm.2013.215} {\bibfield  {journal} {\bibinfo  {journal}
  {Journal of Fluid Mechanics}\ }\textbf {\bibinfo {volume} {726}},\ \bibinfo
  {pages} {62--99} (\bibinfo {year} {2013})}\BibitemShut {NoStop}%
\bibitem [{\citenamefont {Yang}, \citenamefont {Meneveau},\ and\ \citenamefont
  {Shen}(2014{\natexlab{a}})}]{Yang2014}%
  \BibitemOpen
  \bibfield  {author} {\bibinfo {author} {\bibnamefont {Yang}, \bibfnamefont
  {D.}}, \bibinfo {author} {\bibnamefont {Meneveau}, \bibfnamefont {C.}}, \
  and\ \bibinfo {author} {\bibnamefont {Shen}, \bibfnamefont {L.}},\ }\bibfield
   {title} {\enquote {\bibinfo {title} {{Effect of downwind swells on offshore
  wind energy harvesting – A large-eddy simulation study}},}\ }\href
  {\doibase 10.1016/J.RENENE.2014.03.069} {\bibfield  {journal} {\bibinfo
  {journal} {Renewable Energy}\ }\textbf {\bibinfo {volume} {70}},\ \bibinfo
  {pages} {11--23} (\bibinfo {year} {2014}{\natexlab{a}})}\BibitemShut
  {NoStop}%
\bibitem [{\citenamefont {Yang}, \citenamefont {Meneveau},\ and\ \citenamefont
  {Shen}(2014{\natexlab{b}})}]{Yang2014OffshoreW}%
  \BibitemOpen
  \bibfield  {author} {\bibinfo {author} {\bibnamefont {Yang}, \bibfnamefont
  {D.}}, \bibinfo {author} {\bibnamefont {Meneveau}, \bibfnamefont {C.}}, \
  and\ \bibinfo {author} {\bibnamefont {Shen}, \bibfnamefont {L.}},\ }\bibfield
   {title} {\enquote {\bibinfo {title} {{Large-eddy simulation of offshore wind
  farm}},}\ }\href {\doibase 10.1063/1.4863096} {\bibfield  {journal} {\bibinfo
   {journal} {Physics of Fluids}\ }\textbf {\bibinfo {volume} {26}} (\bibinfo
  {year} {2014}{\natexlab{b}}),\ 10.1063/1.4863096}\BibitemShut {NoStop}%
\bibitem [{\citenamefont {Yang}\ and\ \citenamefont {Shen}(2010)}]{Yang2010}%
  \BibitemOpen
  \bibfield  {author} {\bibinfo {author} {\bibnamefont {Yang}, \bibfnamefont
  {D.}}\ and\ \bibinfo {author} {\bibnamefont {Shen}, \bibfnamefont {L.}},\
  }\bibfield  {title} {\enquote {\bibinfo {title} {{Direct-simulation-based
  study of turbulent flow over various waving boundaries}},}\ }\href {\doibase
  10.1017/S0022112009993557} {\bibfield  {journal} {\bibinfo  {journal}
  {Journal of Fluid Mechanics}\ }\textbf {\bibinfo {volume} {650}},\ \bibinfo
  {pages} {131--180} (\bibinfo {year} {2010})}\BibitemShut {NoStop}%
\bibitem [{\citenamefont {Yang}, \citenamefont {Shen},\ and\ \citenamefont
  {Meneveau}(2013)}]{Yang2013DA}%
  \BibitemOpen
  \bibfield  {author} {\bibinfo {author} {\bibnamefont {Yang}, \bibfnamefont
  {D.}}, \bibinfo {author} {\bibnamefont {Shen}, \bibfnamefont {L.}}, \ and\
  \bibinfo {author} {\bibnamefont {Meneveau}, \bibfnamefont {C.}},\ }\bibfield
  {title} {\enquote {\bibinfo {title} {{An assessment of dynamic subgrid-scale
  sea-surface roughness models}},}\ }\href {\doibase 10.1007/s10494-013-9459-7}
  {\bibfield  {journal} {\bibinfo  {journal} {Flow, Turbulence and Combustion}\
  }\textbf {\bibinfo {volume} {91}},\ \bibinfo {pages} {541--563} (\bibinfo
  {year} {2013})}\BibitemShut {NoStop}%
\bibitem [{\citenamefont {Yang}\ \emph {et~al.}(2022)\citenamefont {Yang},
  \citenamefont {Ge}, \citenamefont {Abkar},\ and\ \citenamefont
  {Yang}}]{Yang2022x}%
  \BibitemOpen
  \bibfield  {author} {\bibinfo {author} {\bibnamefont {Yang}, \bibfnamefont
  {H.}}, \bibinfo {author} {\bibnamefont {Ge}, \bibfnamefont {M.}}, \bibinfo
  {author} {\bibnamefont {Abkar}, \bibfnamefont {M.}}, \ and\ \bibinfo {author}
  {\bibnamefont {Yang}, \bibfnamefont {X.~I.}},\ }\bibfield  {title} {\enquote
  {\bibinfo {title} {{Large-eddy simulation study of wind turbine array above
  swell sea}},}\ }\href {\doibase 10.1016/j.energy.2022.124674} {\bibfield
  {journal} {\bibinfo  {journal} {Energy}\ }\textbf {\bibinfo {volume} {256}},\
  \bibinfo {pages} {124674} (\bibinfo {year} {2022})}\BibitemShut {NoStop}%
\bibitem [{\citenamefont {Yurovskaya}\ \emph {et~al.}(2013)\citenamefont
  {Yurovskaya}, \citenamefont {Dulov}, \citenamefont {Chapron},\ and\
  \citenamefont {Kudryavtsev}}]{yurovskaya2013directional}%
  \BibitemOpen
  \bibfield  {author} {\bibinfo {author} {\bibnamefont {Yurovskaya},
  \bibfnamefont {M.}}, \bibinfo {author} {\bibnamefont {Dulov}, \bibfnamefont
  {V.}}, \bibinfo {author} {\bibnamefont {Chapron}, \bibfnamefont {B.}}, \ and\
  \bibinfo {author} {\bibnamefont {Kudryavtsev}, \bibfnamefont {V.}},\
  }\bibfield  {title} {\enquote {\bibinfo {title} {Directional short wind wave
  spectra derived from the sea surface photography},}\ }\href@noop {}
  {\bibfield  {journal} {\bibinfo  {journal} {Journal of Geophysical Research:
  Oceans}\ }\textbf {\bibinfo {volume} {118}},\ \bibinfo {pages} {4380--4394}
  (\bibinfo {year} {2013})}\BibitemShut {NoStop}%
\bibitem [{\citenamefont {Zheng}\ \emph {et~al.}(2023)\citenamefont {Zheng},
  \citenamefont {Li}, \citenamefont {Wang}, \citenamefont {Zhou},\ and\
  \citenamefont {Xiao}}]{Zheng2023}%
  \BibitemOpen
  \bibfield  {author} {\bibinfo {author} {\bibnamefont {Zheng}, \bibfnamefont
  {S.}}, \bibinfo {author} {\bibnamefont {Li}, \bibfnamefont {C.}}, \bibinfo
  {author} {\bibnamefont {Wang}, \bibfnamefont {P.}}, \bibinfo {author}
  {\bibnamefont {Zhou}, \bibfnamefont {S.}}, \ and\ \bibinfo {author}
  {\bibnamefont {Xiao}, \bibfnamefont {Y.}},\ }\bibfield  {title} {\enquote
  {\bibinfo {title} {{Wind tunnel and wave flume testing on directionality
  dynamic responses of a 10 MW Y-shaped semi-submersible wind turbine}},}\
  }\href {\doibase 10.1063/5.0131279} {\bibfield  {journal} {\bibinfo
  {journal} {Journal of Renewable and Sustainable Energy}\ }\textbf {\bibinfo
  {volume} {15}} (\bibinfo {year} {2023}),\ 10.1063/5.0131279},\ \bibinfo
  {note} {013305}\BibitemShut {NoStop}%
\end{thebibliography}%

\end{document}